\documentclass[aps,pra,twocolumn,amsmath,amssymb,showpacs]{revtex4-1}
\usepackage{graphicx}
\usepackage{amsmath}
\usepackage{bm}

\newcommand{\be}{\begin{equation}}  
\newcommand{\ee}{\end{equation}}
\newcommand{\ba}{\begin{array}}
\newcommand{\ea}{\end{array}}
\newcommand{\bea}{\begin{eqnarray}}
\newcommand{\eea}{\end{eqnarray}}

\newcommand{\bra}{\langle}
\newcommand{\ket}{\rangle}

\newcommand{\nn}{\nonumber}

\newcommand{\h}{\hbar}

\begin{document}

\title{Thermodynamics from indistinguishability: mitigating and amplifying the effects of the bath \\
} 

\author{C.L. Latune$^{1,2}$, I. Sinayskiy$^{1}$, F. Petruccione$^{1,2,3}$}
\affiliation{$^1$Quantum Research Group, School of Chemistry and Physics, University of
KwaZulu-Natal, Durban, KwaZulu-Natal, 4001, South Africa\\
$^2$National Institute for Theoretical Physics (NITheP), KwaZulu-Natal, 4001, South Africa\\
$^3$School of Electrical Engineering, KAIST, Daejeon, 34141, Republic of Korea}

\date{\today}
\begin{abstract}
Rich quantum effects emerge when several quantum systems are indistinguishable from the point of view of the bath they interact with.
In particular, delocalised excitations corresponding to coherent superposition of excited states (reminiscent of double slit experiments or beam splitters in interferometers) appear and change drastically the dynamics and steady state of the systems. Such phenomena, which are central mechanisms of superradiance, 
present interesting properties 
for thermodynamics and potentially other quantum technologies. Indeed, a recent paper [C.L. Latune, I. Sinayskiy, F. Petruccione, Phys. Rev. A {\bf 99}, 052105 (2019)] studies these properties in a pair of indistinguishable two-level systems and points out surprising effects of mitigation and amplification of the bath's action on the energy and entropy of the pair. Here, we generalise the study to ensembles of arbitrary number of spins of arbitrary size (i.e. dimension). We confirm that the previously uncovered mitigation and amplification effects remain, but also that they become more and more pronounced with growing number of spins and growing spin size. Moreover, we find that the free energy variation and the entropy production associated with the bath-driven dissipation are systematically reduced, formalising the idea of mitigation of the bath's action.
Most remarkably, the combination of mitigation effects from two baths at different temperatures can result in amplifying their action. This is illustrated with cyclic thermal machines, and leads to large power enhancements. 
The reduction of irreversibility is also an interesting aspect since irreversibility is known to limit the performance of thermodynamic tasks. The above findings might also lead to interesting applications in collective work extraction, quantum battery charging, state protection, light harvesting devices, quantum biology, but also for the study of entropy production. Moreover, some experimental realisations and observations suggest that such effects are within reach.
\end{abstract}

\maketitle
\section{Introduction}

Collective interactions of quantum systems with their surrounding environment (bath) generates 
diverse phenomena like superradiance \cite{Dicke,Gross_1982} and entanglement generation \cite{Benatti_2003, Benatti_2010, Passos_2018}. 
It relies on the indistinguishability of the systems from the point of view of the bath \cite{Gross_1982,bathinducedcohTLS}, and as by-product creates coherent superposition of exited states (delocalised excitations). These bath-induced coherences are promising for enhancing thermodynamic tasks 
 (mainly, but not restricted to work extraction and refrigeration) 
\cite{Wang_2009,Scully_2011,Gelbwaser_2015,Uzdin_2016,Mehta_2017,Cakmak_2017,Niedenzu_2018,Holubec_2018,Hewgill_2018,Watanabe_2019}, light harvesting devices \cite{Scully_2010,Svidzinsky_2011,Svidzinsky_2012,Dorfman_2013,Creatore_2013,
 Romero_2014,Killoran_2015,Xu_2015,Su_2016,Chen_2016,Romero_2017,
 Brown_2018}, quantum transport \cite{Robentrost_2009,Ishizaki_2009,Lloyd_2011,Lee_2017}, heat flow reversal \cite{heatflowreversal}, 
and might also be used by living organisms for photosynthesis \cite{Collini_2010,Lloyd_2011,Lambert_2013,Huelga_2013,Dorfman_2013,Chin_2013,Krisnanda_2018} and other vital functions \cite{Lambert_2013,Huelga_2013}. 
%
%
%
%
 %
%

However the lack of consensus on the actual effects of coherences (and entanglement) \cite{Holubec_2018,Brandner_2017,Niedenzu_2018,Farre_2018,
Uzdin_2015, Andolina_2018,Gonzalez_2018,Andolina_2019,Hovhannisyan_2013, Kilgour_2018}
 and the strong model-dependence of some results \cite{Andolina_2018,Farre_2018,Hewgill_2018} emphasise that the underlying mechanisms are still far from fully understood. Moreover, very little is known about the consequences of bath-induced coherences for central quantities like the energy and entropy of the indistinguishable systems. 
Exploring these crucial questions and having in mind the identification of innovative strategies for quantum thermodynamic enhancements, 
we carry out a broad analysis of the thermodynamics effects of bath-induced coherences.

Our results points at diverse and interesting phenomena.  
 First, bath-induced coherences -- generated through collective coupling -- effectively shield the spin ensemble: the impact of the bath's action on the spin ensemble energy and entropy is mitigated. Still, when negative temperatures are considered, an opposite tendency can emerge for the ensemble's energy, amplifying the heat exchanged between the ensemble and the bath. These results extend to ensembles of arbitrary number $n$ of spins of arbitrary size (dimension) the results obtained in \cite{bathinducedcohTLS} for a pair of two-level systems. This is important because not only we show that mitigation and amplification effects remain for larger ensemble and systems, but also that these same effects become more and more pronounced with increasing ensembles -- scaling up with $n$. 

Secondly, we analyse the free energy variation generated by the bath's action and show that it is systematically reduced (in absolute value) in presence of collective bath interactions. This sets a formal ground for the discussion around mitigation and amplification effects. Additionally, 
 we study the entropy production, which has been subject of intense research in a quantum context with pioneering contributions in \cite{Spohn_1978, Spohn_1978b, Alicki_1979}, but also more recently in \cite{Brandao_2013,Muller_2018}, investigating its relation with work extraction \cite{Muller_2018}, irreversibility \cite{Parrondo_2009, Deffner_2011, Santos_2017, Brunelli_2018, Santos_2019}, and its crucial role in non-equilibrium dynamics and in performances of thermal machines \cite{Barato_2015, Gingrich_2016, Pietzonka_2016, Pietzonka_2018, Guarnieri_2019, Timpanaro_2019, Su_2019}. We find a dramatic reduction of entropy production (by a factor up to $1/n$) due to collective bath coupling, opening interesting perspectives in particular with regards to performances of thermodynamic tasks. 
%
More generally, all these phenomena have potential applications (detailed in Section ``Applications and perspectives'') in thermal machines, refrigeration operations, quantum battery charging, state protection, and also contribute to the aforementioned ongoing debate \cite{Uzdin_2015, Andolina_2018,Gonzalez_2018,Niedenzu_2018} 
on genuine quantum effects in thermodynamics. 

Third, we present a remarkable phenomenon in a context of cyclic thermal machines: the combination of mitigation effects from two baths at different temperatures can produce an amplification of their action. We show analytically that this can result in very large power enhancements of thermal machines.

Finally, some experimental realisations of collective bath coupling suggest that the above phenomena might be within reach.  
%
%
%
The underlying mechanism sustaining these effects can be understood in terms of coherences between degenerate energy levels (of the local basis, see also further explanations based on the framework introduced in \cite{apptemp} are detailed in \ref{roleofbathinducedcoh}).
From an alternative point of view, interesting parallels can be drawn with the framework established in \cite{Muller_2018} around catalysis in quantum thermodynamics \cite{Aberg_2014,Ghosh_2017}. 

\section{Collective bath-induced dissipation}\label{seccollective}
We consider an ensemble $A$ of $n$ non-interacting spins $s$ of same Bohr frequency $\omega$ and free Hamiltonian $H_A= \hbar \omega J_z$, where $J_z:= \sum_{k=1}^n j_{z,k}$ is the collective $z$-component of the angular momentum operator (generator of rotation around the $z$-axis), with $j_{z,k}$ the $z$-component of the angular momentum operator associated to the $k^{\rm th}$ spin. In particular, for spin $1/2$, $j_{z,k}$ is one half of the Pauli matrix $\sigma_z$. Note that beyond actual spin $1/2$, any two-level system (like two-level atoms) is isomorphic to a spin $1/2$ so that all the following considerations are also valid for ensembles of two-level systems. We define in the same way the collective angular momentum $J_i:=\sum_{k=1}^n j_{i,k}$ along the direction $i=x,y$ and the local angular momentum $j_{i,k}$ associated to the $k^{\rm th}$ spin with $i=x,y$. 

 We assume that the spin ensemble $A$ interacts collectively with a bath $B$ of inverse temperature $\beta_B$. The collective interaction implicitly requires that the bath does not distinguish the $n$ spins \cite{bathinducedcohTLS}. This can be realised in several platforms \cite{Wood_2014,Wood_2016,Hama_2018} (see also \cite{Niedenzu_2018} for ensemble of two-level atoms) and was experimentally done for instance in \cite{Raimond_1982, Devoe_1996} (see also Section \ref{secsaturation}) and more recently in \cite{Barberena_2019}. The collective coupling to the bath is then of the form $V:= \hbar g J_x O_B$, where $O_B$ is a bath observable, and $g$ characterises the strength of the coupling
  
Assuming that the Born and Markov approximations are valid (namely, the bath correlation time is much smaller than the relaxation time of $A$ \cite{Petruccione_Book, Cohen_Book}), 
the master equation for the reduced density operator $\rho$ of the spin ensemble is (in the interaction picture)
\bea\label{me}
\frac{ d \rho}{dt} &=& \Gamma(\omega) \left(J^{-}\rho J^{+} - J^{+} J^- \rho\right) \nn\\
&&+ \Gamma(-\omega) \left(J^{+}\rho J^{-} - J^{-} J^{+} \rho\right)  + {\rm h.c.},
\eea
where $J^{\pm}:= J_x \pm iJ_y$ are the collective ladder operators of the spin ensemble, $\Gamma(\omega) = \hbar^2g^2\int_0^{\infty} e^{i\omega u} {\rm Tr}\rho_B O_B(u)O_B du$ is the ``half Fourier transform'' of the bath correlation function, $\rho_B$ is the density operator of the bath in the interaction picture (with respect to its free Hamiltonian $H_B$), and $O_B(u)$ denotes the interaction picture of $O_B$. Note that the above master equation \eqref{me} has been derived using the secular approximation (valid when $\omega^{-1}$ is much smaller than the relaxation time of the spin ensemble). Moreover, the master equation \eqref{me} is valid for thermal baths but more generally for stationary baths \cite{Alicki_2014,Alicki_2015} whose apparent temperature can be defined as \cite{Alicki_2014,Alicki_2015,apptemp} ($k_B = 1$)
 \be
T_B:=\hbar\omega \left(\ln \frac{\Gamma(\omega)+\Gamma^{*}(\omega)}{\Gamma(-\omega)+\Gamma^{*}(-\omega)}\right)^{-1}.
\ee
 Importantly, in several usual situations in thermodynamics, like in the context of spin baths \cite{Assis_2018,Kosloff_2019}, thermal machines or more generally when several thermal baths at different temperatures interact with the same system \cite{Brunner_2012,autonomousmachines}, the dissipative dynamics can be described by the interaction with an effective thermal bath at {\it negative} temperature. Therefore, to include such situations relevant for thermodynamics, we consider in the following that the bath interacting with the spin ensemble has a temperature (or apparent temperature) $T_B$ which can be either positive or negative. For convenience, we will prefer to use the inverse temperature $\beta_B=T_B^{-1}$.

\section{Spin ensembles}\label{secspinensembles}
For the $k^{\rm th}$ spin, we denote by $\{|s,m_k\ket_k \}_{-s\leq m_k\leq s}$ the local eigenbasis of $j_{z,k}$, so that $j_{z,k}|s,m_k\ket_k = \hbar m_k |s,m_k\ket_k$. Then, the states of the spin ensemble can be naturally described in the basis
\be\label{localbasis}
|m_1,m_2,...,m_n\ket :=\otimes_{k=1}^n |s,m_k\ket_k
\ee
 obtained from the tensor products of the local eigenbasis. In the following we will refer to this basis as the {\it local} basis.  Alternatively, it is well-known from the theory of addition of angular momenta \cite{Sakurai_Book} that the states of the spin ensemble can be described through another basis made of the eigenvectors of the commuting global observables $J_z$ and ${\cal J}^2:=J_x^2+J_y^2+J_z^2$. Such eigenvectors are denoted by $|J,m\ket$ in reference to their associated eigenvalues, 
 \bea
&&{\cal J}^2|J,m\ket = \hbar J(J+1)|J,m\ket\nn\\
&& J_z|J,m\ket= \hbar m|J,m\ket
\eea
 with $-J\leq m\leq J$ and $J \in [J_0; ns]$, where $J_0 =0$ if $s\geq 1$ and $J_0=1/2$ if $s=1/2$ and $n$ odd. 
A quick calculation shows that the natural basis contains $(2s+1)^n$ elements whereas there are only $(ns+1)^2$ (or $(ns+1/2)(ns+3/2)$ when $s=1/2$ and $n$ is odd) different pairs of eigenvalues $(J,m)$ for $-J\leq m\leq J$ and $J \in [J_0; ns]$. 
Therefore, for $n\geq3$, some eigenspaces must be degenerate.
 We denote by $|J,m\ket_i$ the degenerate eigenstates (of eigenvalue $\hbar J(J+1)$ and $\hbar m$) with the degeneracy index $i$ running from 1 to $l_J$, integer which represents the multiplicity (or degeneracy) of  the associated eigenspace. In the remainder of the paper we call ``eigenspace of total spin $J$'', or simply ``eigenspace $J$'', if no confusion is possible, the eigenspace associated to the eigenvalue $J$ (of the total spin operator ${\cal J}^{2}$). A complete basis is formed by the collection 
 of all eigenvectors $|J,m\ket_i$ (including all the degenerate ones).
In other words, any pure state $|\psi\ket$ of the spin ensemble can be rewritten as 
\be\label{gendecomp}
|\psi\ket = \sum_{J=J_0}^{ns} \sum_{m=-J}^{J}\sum_{i=1}^{l_J} a_{J,m,i}|J,m\ket_{i},
\ee
where $a_{J,m,i}$ are complex coefficients with square module summing up to 1. We will refer to this basis $\{|J,m\ket_i\}$, $J_0\leq J\leq ns$, $-J\leq m\leq J$, $1\leq i\leq l_J$, as the {\it collective} basis. 
Note that one can easily show that the multiplicity $l_J$ is always equal to 1 (no degeneracy) for $J=ns$ and always equal to $n-1$ for $J=ns-1$. However, it is a difficult task to find out the expression of $l_J$ for $J\leq ns-2$ for arbitrary $n$ and $s$. Nevertheless, the theory of addition of angular momenta \cite{Sakurai_Book} guarantees that the above decomposition \eqref{gendecomp} exists. 

It is important to note that the global ladder operators $J^{\pm}$ generate the usual transition between the global eigenstates of $J_z$, namely, $J^{\pm} =\hbar \sqrt{(J\mp m)(J\pm m+1)} |J,m\pm 1\ket_i$.
In the group theory notation the change from local to global basis is often written as
\be
{\cal H}_s^{\otimes^{n}} = \oplus_{J=J_0}^{ns}{\cal H}_J^{\oplus^{l_J}}
\ee
where ${\cal H}_s$ and ${\cal H}_J$ are Hilbert spaces of spin $s$ and $J$, respectively, and $\otimes$ denotes a tensor product whereas $\oplus$ denotes a direct sum. 

This change of basis provides precious information on the spin ensemble evolution under dynamics which preserves the spin-exchange symmetry. 
In particular, the collective dissipation described by the master equation \eqref{me} involves only collective absorptions and collective emissions, represented by the ladder operators $J^{+}$ and $J^{-}$ respectively (preserving therefore the spin-exchange symmetry), so that if the spin ensemble is initialised in the eigenspace of total spin $J$, it remains in it at all times. More generally, if the spin ensemble is initialised in a state $\rho_0$ with a total weight $p_{J,i}:=\sum_{m=-J}^J~_i\bra J,m|\rho_0|J,m\ket_i$ in each eigenspace of total spin $J$ (such that $\sum_{J=J_0}^{ns}\sum_{i=1}^{l_J}p_{J,i}=1$), 
 each component evolves without coupling to the other total spin eigenspaces so that the initial weight $p_{J,i}$ is preserved throughout time. As a consequence, each component $J,i$ thermalises to the thermal state (see Appendix \ref{steadystate})
\be\label{thstji}
\rho_{J,i}^{\rm th}(\beta_B):=Z_{J}(\beta_B)^{-1}\sum_{m=-J}^J e^{- m\hbar\omega\beta_B}|J,m\ket_i\bra J,m|,
\ee
 with 
\bea\label{zj}
Z_J(\beta_B)&:=&\sum_{m=-J}^J e^{-m\hbar\omega\beta_B}\nn\\
&=&e^{J\hbar\omega\beta_B}\frac{1-e^{-(2J+1)\hbar\omega\beta_B}}{1-e^{-\hbar\omega\beta_B}},
\eea
 so that the steady state of the ensemble is
\be\label{ss}
\rho^{\infty}(\beta_B) := \sum_{J=J_0}^{ns}\sum_{i=1}^{l_J} p_{J,i} \rho_{J,i}^{\rm th}(\beta_B).
\ee
 Note that we excluded initial coherences between eigenspaces of different total spin ($_i\bra J,m|\rho_0|J',m'\ket_{i'} =0$ if and only if $J\ne J'$ or $i\ne i'$). We provide in Appendix \ref{geneq7} 
 some arguments to support the claim that for a large class of initial state, in particular for state containing coherences between eigenspaces of different total spin, the steady state is still of the form \eqref{ss}. Nevertheless, we focus in the following on spin ensembles which are initially in arbitrary thermal states -- arguably the most common and experimentally accessible class of states -- having no coherence between different eigenspaces $J,i$ (shown in the following). Therefore, the generalisation of the validity of \eqref{ss} to a larger class of initial states is not necessary here (however we mention it 
 as it might be of interest for other applications). \\

\section{Steady states}\label{sectionsteadys}
As mentioned above we consider a spin ensemble initially in a thermal state at inverse temperature $\beta_0$,
\be\label{thermalstate}
\rho^{\rm th}(\beta_0):= Z(\beta_0)^{-1} e^{-\hbar\omega \beta_0 J_z}
\ee
where 
$Z(\beta_0) := {\rm Tr}e^{-\hbar\omega \beta_0 J_z}$ is the global partition function. Such thermal state can be rewritten as
\bea
\rho^{\rm th}(\beta_0) &=&  \otimes_{k=1}^n\frac{1}{Z_s(\beta_0)}  e^{-\hbar\omega\beta_0 j_z^k}\nn\\
&=&\frac{1}{Z_s(\beta_0)^{n}} \otimes_{k=1}^n \left\{\sum_{m=-s}^s e^{-m\hbar \omega\beta_0}|s,m\ket_k \bra s,m|\right\},\nn\\
\eea
where $Z_s(\beta_0):= \sum_{m=-s}^s e^{-m\hbar \omega \beta_0}$ is the local partition function so that 
\be\label{Z}
Z(\beta_0) = Z_s(\beta_0)^n
\ee
and the states $|s,m\ket_k$ are the eigenstates of $j_{z,k}$ introduced above.
The global thermal state can be re-written as
\bea\label{tslocal}
\rho^{\rm th}(\beta_0) &=& \frac{1}{Z(\beta_0)}\sum_{m=-ns}^{ns} e^{-m\hbar\omega\beta_0}\nn\\
 &&\hspace{-0.5cm}\times \sum_{m_1+ ... + m_n=m} |m_1,...,m_n\ket \bra m_1,...,m_n|.
\eea
 All states $|m_1,...,m_n\ket$ such that $\sum_{k=1}^n m_k = m$ are eigenstates of $J_z$ with the same eigenvalue $\hbar m$. For each $m$ we denote by $I_m$ 
 the number of such eigenstates. They span the subspace associated to the eigenvalue $\hbar m$, that we will refer to in the following as the eigenspace $m$.
We have two orthonormal basis for the eigenspace $m$, 
a collective one $\{|J,m\ket_i\}_{|m| \leq J\leq ns, i\in[1;l_J]}$ and a local one $\{ |m_1,...,m_n\ket\}_{m_1+...+m_n=m}$. This implies in particular the following relation $I_m=\sum_{J=|m|}^{ns} l_J$.
As we saw above, the restriction to the eigenspace $m$ of the thermal state $\rho^{\rm th}(\beta_0)$ is 
\bea
\rho^{\rm th}(\beta_0)_{|_m} &=& \frac{e^{-m\omega\beta_0}}{Z(\beta_0)} \sum_{m_1+ ... + m_n=m} |m_1,...,m_n\ket \bra m_1,...,m_n|\nn\\
&=& \frac{e^{-m\omega\beta_0}}{Z(\beta_0)} {\mathbb I}_m,
\eea
where ${\mathbb I}_m$ denotes the identity of the eigenspace $m$ which can be expressed also in the collective basis as ${\mathbb I}_m = \sum_{J=|m|}^{ns}\sum_{i=1}^{l_J}|J,m\ket_i\bra J,m|$. Therefore, the thermal state can be re-written in the collective basis as
\bea\label{tscollective}
\rho^{\rm th}(\beta_0) &=& \sum_{m=-ns}^{ns} \frac{e^{-m\omega\beta_0}}{Z(\beta_0)} \sum_{J=|m|}^{ns}\sum_{i=1}^{l_J}|J,m\ket_i\bra J,m|\nn\\
&=&  \frac{1}{Z(\beta_0)} \sum_{J=J_0}^{ns}\sum_{i=1}^{l_J}\sum_{m=-J}^{J} e^{-m\omega\beta_0}|J,m\ket_i\bra J,m|\nn\\
&=&   \sum_{J=J_0}^{ns}p_J(\beta_0)\sum_{i=1}^{l_J}\rho_{J,i}^{\rm th}(\beta_0)
\eea
with $p_J(\beta_0) := \frac{Z_J(\beta_0)}{Z(\beta_0)} $. One should note that from the normalisation condition we have automatically the identity $\sum_{J=J_0}^{ns} l_J p_J(\beta_0) =1$ which will be used in the following. Combining \eqref{ss} and \eqref{tscollective} we are now in measure to announce the main result of this paragraph: a state initially in a thermal state at inverse temperature $\beta_0$ tends to the steady state
\be\label{sstate}
\rho^{\infty}_{\beta_0}(\beta_B) := \sum_{J=J_0}^{ns}p_J(\beta_0)\sum_{i=1}^{l_J}\rho_{J,i}^{\rm th}(\beta_B).
\ee
Crucially, $\rho^{\infty}_{\beta_0}(\beta_B) $ is generally not a thermal state. From \eqref{tscollective} we can make a stronger statement: $\rho^{\infty}_{\beta_0}(\beta_B) $ is a thermal state if and only if $\beta_0=\pm\beta_B$ (recovering the fact that the thermal state at inverse temperature $\beta_B$ is a steady state of the dynamics). 

In the following we compare thermodynamic characteristics of $\rho^{\infty}_{\beta_0}(\beta_B)$ with the properties of the thermal equilibrium state $\rho^{\rm th}(\beta_B)$, which is the steady state reached when each spin is distinguishable from the bath's point of view, or equivalently when each spin interacts individually with the bath. We will refer to this distinguishable or individual dissipation as {\it independent} dissipation, by contrast to the {\it collective} dissipation described by \eqref{me}.
Such comparison reveals the energetic and entropic impact of the collective dissipation (or equivalently, indistinguishability) on the spin ensemble.

\section{Steady state energy}\label{mainenergy}
In this Section we look at the energy of the spin ensemble when it reaches its steady state $\rho^{\infty}_{\beta_0}(\beta_B)$. The corresponding energy is defined by
\be\label{energydef}
E^{\infty}_{\beta_0}(\beta_B) := \hbar\omega {\rm Tr} J_z \rho^{\infty}_{\beta_0}(\beta_B) + \h\omega ns.
\ee
 The extra term $\h\omega ns$ is not of fundamental importance, it just means that we are taking the ground state $|J,-J\ket$ as energy reference. In other words, the energy is defined to be proportional to the number of excitations in the spin ensemble. A quick calculation shows that 
\be\label{colen}
E^{\infty}_{\beta_0}(\beta_B) = \sum_{J=J_0}^{ns} p_J(\beta_0)l_Je_J(\beta_B) + \h\omega ns,
\ee 
with 
\bea\label{ej}
e_J(\beta_B)&:=&\hbar\omega {\rm Tr} J_z \rho_{J,i}^{\rm th}(\beta_B)\nn\\
&=& \hbar\omega \sum_{m=-J}^J m \frac{e^{-m\hbar\omega \beta_B}}{Z_J(\beta_B)}\nn\\
&=& -\frac{\partial }{\partial \beta_B} \ln Z_J(\beta_B)\nn\\
&=& \frac{\h\omega}{2}\frac{\cosh (\h\omega\beta_B/2)}{\sinh (\h\omega\beta_B/2)}\nn\\
&& - (2J+1)\frac{\h\omega}{2}\frac{\cosh [(2J+1)\h\omega\beta_B/2]}{\sinh [(2J+1)\h\omega\beta_B/2]}\nn\\
&=& \h\omega \frac{1}{e^{\h\omega\beta_B}-1} -\h\omega\frac{2J+1}{e^{(2J+1)\h\omega\beta_B}-1} -J\h\omega. \nn\\
\eea

\begin{figure}
\centering
\includegraphics[width=7cm, height=4.8cm]{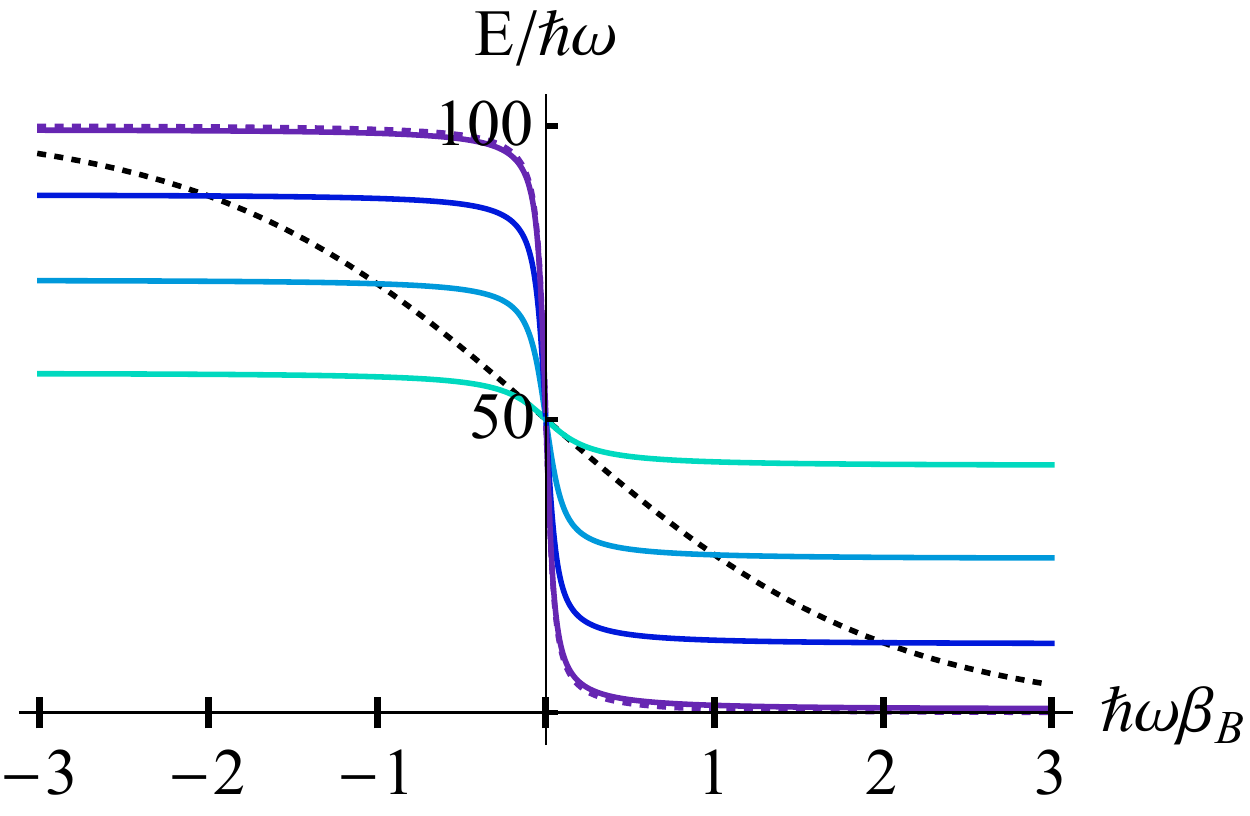}\\
\caption{Plots of $E^{\infty}_{\beta_0}(\beta_B) $ as a  function of $\hbar\omega\beta_B$ for $n=100$ spins $s=1/2$. From the lightest to the darkest blue curves the initial temperature is respectively $\hbar\omega|\beta_0|=0.1$, $\hbar\omega|\beta_0|=1$, $\hbar\omega|\beta_0|=2$, and $\hbar\omega|\beta_0|=5$. The dotted black  curve corresponds to the thermal equilibrium energy $E^{\rm th}(\beta_B)$. The dotted dark blue curve represents the limit $\hbar\omega\beta_0=+\infty$. 
}
\label{half}
\end{figure}

We compare $E^{\infty}_{\beta_0}(\beta_B)$ to the thermal energy $E^{\rm th}(\beta_B):=\h\omega{\rm Tr}J_z\rho^{\rm th}(\beta_B)+\hbar\omega ns$ of the thermal state $\rho^{\rm th}(\beta_B)$ (reached under independent dissipation). One obtains straightforwardly
\be\label{then}
E^{\rm th}(\beta_B) =\sum_{J=J_0}^{ns} p_J(\beta_B)l_Je_J(\beta_B) + \hbar\omega ns.
\ee 
Therefore, the only difference between $E^{\infty}_{\beta_0}(\beta_B) $ and $E^{\rm th}(\beta_B) $ are the weight $p_J(\beta_0)$ which are substituted by $p_J(\beta_B)$ in $E^{\rm th}(\beta_B)$. How does this affect $E^{\infty}_{\beta_0}(\beta_B) $? As a brief preview of the general picture, Fig. \ref{half} contains the plots of $E^{\infty}_{\beta_0}(\beta_B) $ as a  function of $\hbar\omega\beta_B$ for $n=100$ spins $s=1/2$. We used the expressions \eqref{colen} with the degeneracy coefficients 
\be
l_J= (2J+1) \frac{n!}{(\frac{n}{2}+J+1)!(\frac{n}{2}-J)!},
\ee
 obtained -- in the particular case $s=1/2$ -- from $I_m=\frac{n!}{(n/2+m)!(n/2-m)!}$ and $I_m = \sum_{J=|m|}^{n/2} l_J$.
The initial temperature varies from $\hbar\omega|\beta_0|=0.1$ (lightest blue curve), to $\hbar\omega|\beta_0|=5$ (darkest blue curve), with two intermediate values $\hbar\omega|\beta_0|=1$ and  $\hbar\omega|\beta_0|=2$ (intermediate blue curves). The dotted black curve corresponds to the thermal equilibrium energy $E^{\rm th}(\beta_B)$. Finally, the dotted dark blue represents the limit $\hbar\omega\beta_0=+\infty$. 
The dramatic difference between $E^{\infty}_{\beta_0}(\beta_B) $ and $E^{\rm th}(\beta_B)$ is maybe one of the most immediate observation.
Can we draw general tendencies, valid for arbitrary $n$ and $s$?
Since the direct comparison of the analytical expressions of $E^{\infty}_{\beta_0}(\beta_B) $ and $E^{\rm th}(\beta_B) $ is of little help, we use an indirect method.

 In Appendix \ref{derivativeE} we show that 
\bea\label{derivatives}
&&\frac{\partial}{\partial \beta_0}E^{\infty}_{\beta_0}(\beta_B) < 0 \Leftrightarrow \beta_0\beta_B > 0,\nn\\
 &&\frac{\partial}{\partial \beta_0}E^{\infty}_{\beta_0}(\beta_B) > 0  \Leftrightarrow \beta_0\beta_B <0,\nn\\
 &&\frac{\partial}{\partial \beta_0}E^{\infty}_{\beta_0}(\beta_B) =0 \Leftrightarrow \beta_0\beta_B=0.\nn\\
 \eea
  Since $E^{\infty}_{\beta_0=\beta_B}(\beta_B) = E^{\rm th}(\beta_B)$ and $E^{\infty}_{-\beta_0}(\beta_B)=E^{\infty}_{\beta_0}(\beta_B)$ we conclude that, for $\beta_B>0$,
\bea
&&E^{\infty}_{\beta_0}(\beta_B) > E^{\rm th}(\beta_B), {\rm when} ~|\beta_0| < \beta_B,\nn\\
&& E^{\infty}_{\beta_0}(\beta_B) < E^{\rm th}(\beta_B), {\rm when}~ |\beta_0| > \beta_B,
\eea
and for $\beta_B <0$, 
\bea
&&E^{\infty}_{\beta_0}(\beta_B) < E^{\rm th}(\beta_B), {\rm when} ~|\beta_0| < |\beta_B|,\nn\\
&& E^{\infty}_{\beta_0}(\beta_B) > E^{\rm th}(\beta_B), {\rm when}~ |\beta_0| > \beta_B.
\eea
\begin{figure}
\centering
(a)\includegraphics[width=7cm, height=4.8cm]{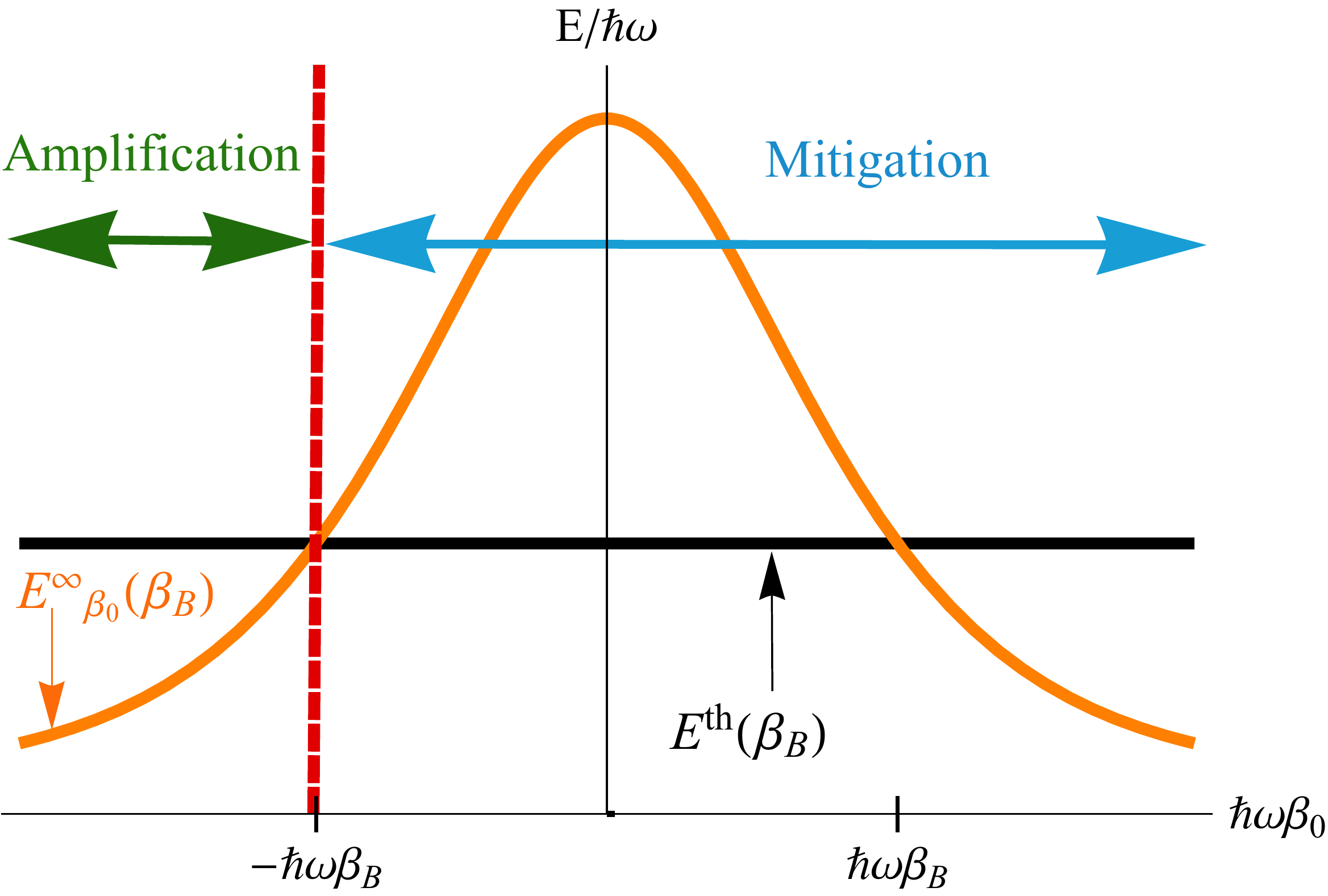}\\
(b)\includegraphics[width=7cm, height=4.8cm]{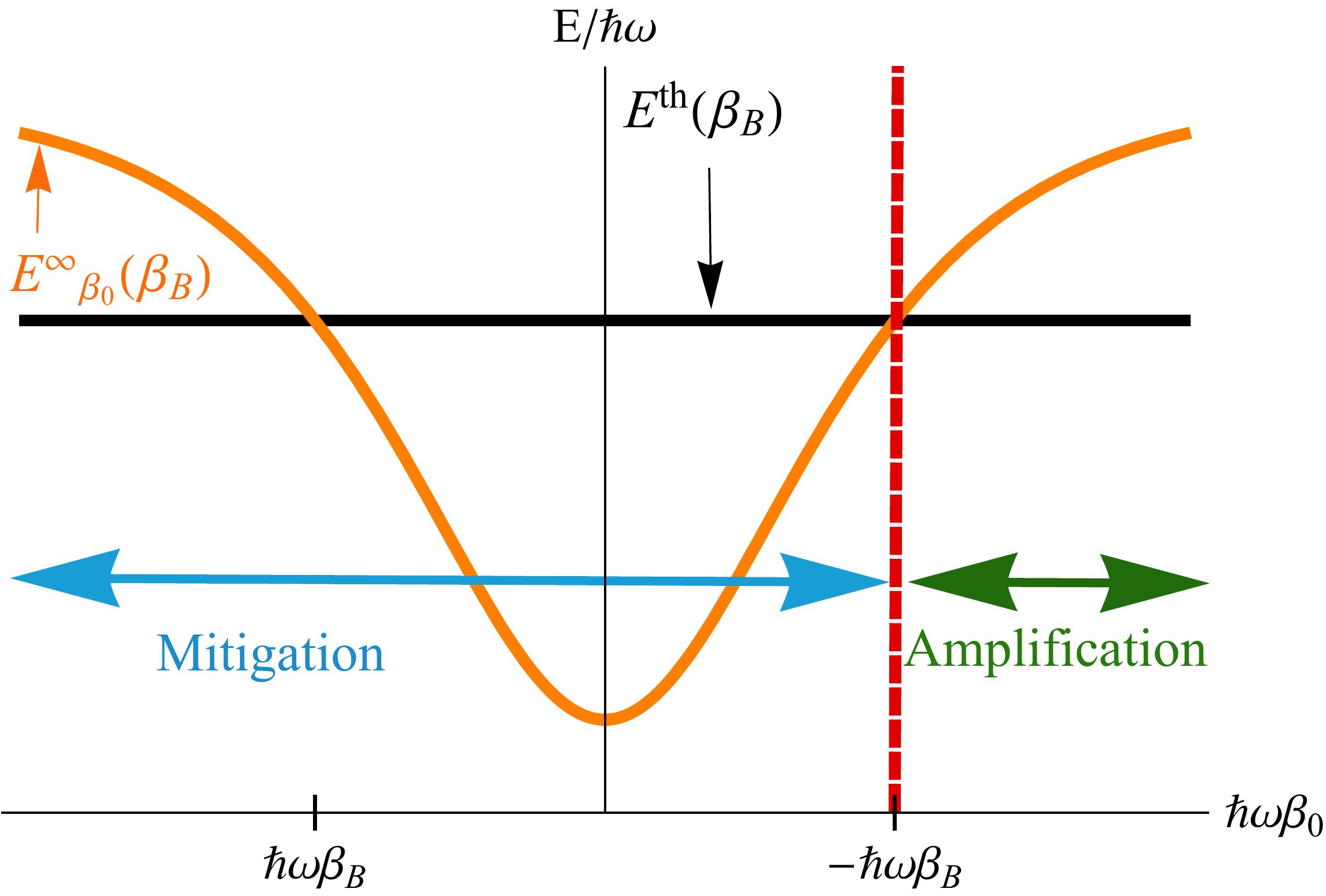}
\caption{Illustration of the general behaviour of the steady state energy $E^{\infty}_{\beta_0}(\beta_B)$ as a function of $\h\omega\beta_0$ (obtained from the particular situation of 10 spins $1/2$). The graph (a) correpsonds to $\beta_B>0$ while the graph (b) to $\beta_B<0$. The value of the thermal energy $E^{\rm th}(\beta_B)$ is indicated by the Black horizontal line.}
\label{enbehaviour}
\end{figure}
The above results are summarised in a more visual way in Fig. \ref{enbehaviour}. 

\subsection{Mitigation and amplification of the bath's action}\label{subsecmitig}
A closer look at these results reveals that the displacement of steady state energy $E^{\infty}_{\beta_0}(\beta_B)$ with respect to the thermal equilibrium energy $E^{\rm th}(\beta_B)$ is not always in the same direction (as appearing also in Fig. \ref{half}). More precisely, in the situation where $\beta_B>0$ and assuming the spin ensemble is initially colder than the bath ($\beta_0>\beta_B$), under independent dissipation the energy of the spin ensemble increases until reaching the thermal equilibrium energy $E^{\rm th}(\beta_B)$. However, the above results show that the collective dissipation limits the steady state energy to a value strictly smaller than the thermal energy $E^{\rm th}(\beta_B)$: the bath's action is mitigated. Similarly, for a spin ensemble initially in a state hotter than the bath such that $-\beta_B<\beta_0<\beta_B$, the energy of the spin ensemble is expected to be reduced to $E^{\rm th}(\beta_B)$ under independent dissipation, but under collective dissipation the reduction is limited to a value of $E^{\infty}_{\beta_0}(\beta_B)$ strictly larger than $E^{\rm th}(\beta_B)$. We have again mitigation of the bath's action. These two situations correspond to the regime designated by ``mitigation'' in Fig. \ref{enbehaviour} (a).

 By contrast, for a spin ensemble initially in a hot state such that $\beta_0<-\beta_B$, the energy of the spin ensemble is brought to lower  levels thanks to the collective dissipation since $E^{\infty}_{\beta_0}(\beta_B)$ is strictly smaller than $E^{\rm th}(\beta_B)$ in such regime. In this situation, corresponding to the region designated by ``amplification'' in Fig. \ref{enbehaviour} panel (a), the bath's action is amplified. This phenomenon resembles the counter-intuitive Mpemba effect \cite{Mpemba_1969,Lasanta_2017,Lu_2017} (under certain conditions, a classical system can be refrigerated faster when it is initially in a hotter state), except that in the present situation an initially hotter system can reach a lower energy. 

Conversely, for an effective bath at negative temperature, similar considerations show that for $\beta_0<|\beta_B|$, the bath's action is mitigated whereas for $\beta_0>|\beta_B|$, the bath's action is amplified, as indicated in  Fig. \ref{enbehaviour} panel (b). 

The above effects on the spin ensemble energy can be recapped in a simple formula: when $\beta_0/\beta_B<-1$, there is amplification of the bath's action, whereas when $\beta_0/\beta_B>-1$ it is substituted by the mitigation of the bath's action.
This brings several potential applications which are detailed in Section \ref{applications}.
These intriguing phenomena can be understood in terms of coherences between degenerate energy levels (of the local basis) which maintain the ensemble in a steady state of energy different from the thermal equilibrium energy. 
 We provide in Appendix \ref{roleofbathinducedcoh} an intuitive explanation in the lights of the framework introduced in \cite{apptemp}.
 In the remainder of this section we present quantitative results on the extent of the amplification and mitigation effects.  \\


\subsection{Extent of the mitigation and amplification effects}
\begin{figure}
\centering
(a)\includegraphics[width=7cm, height=4.5cm]{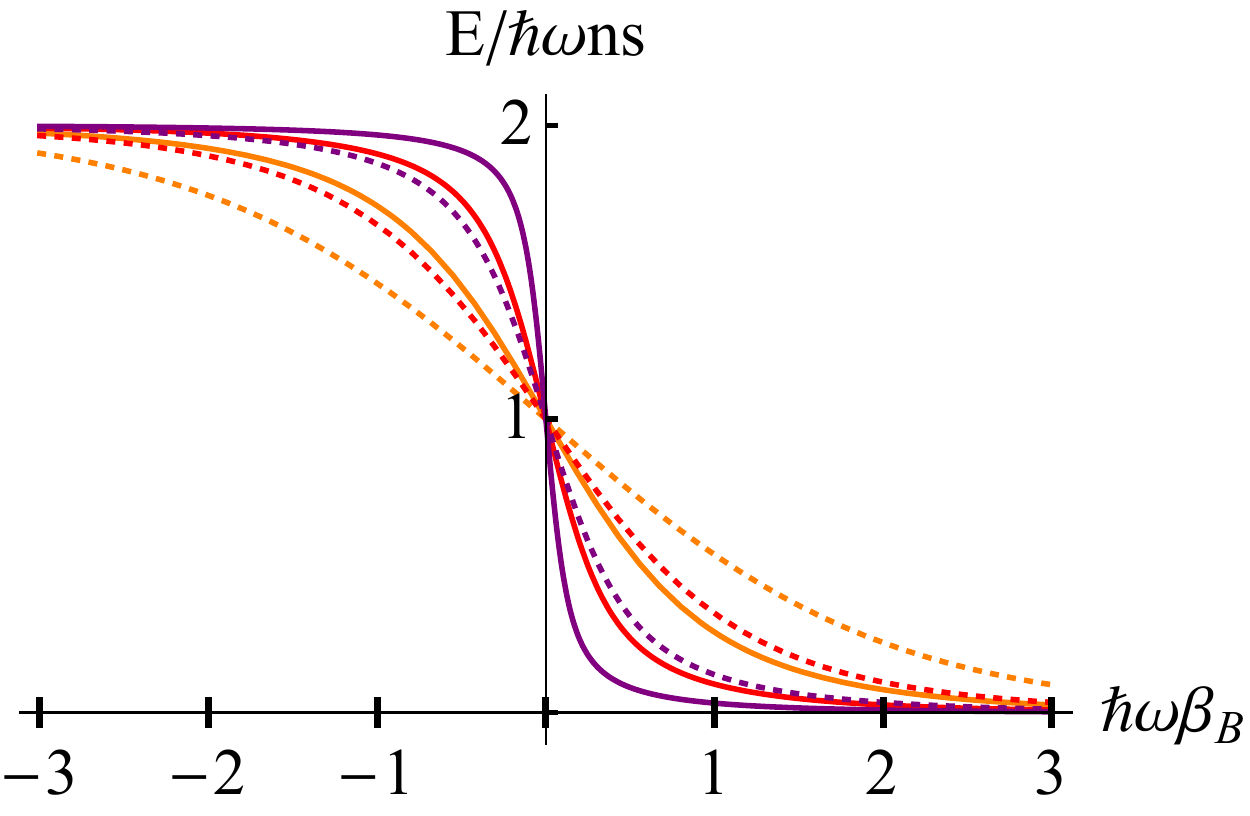}
(b)\includegraphics[width=7cm, height=4.5cm]{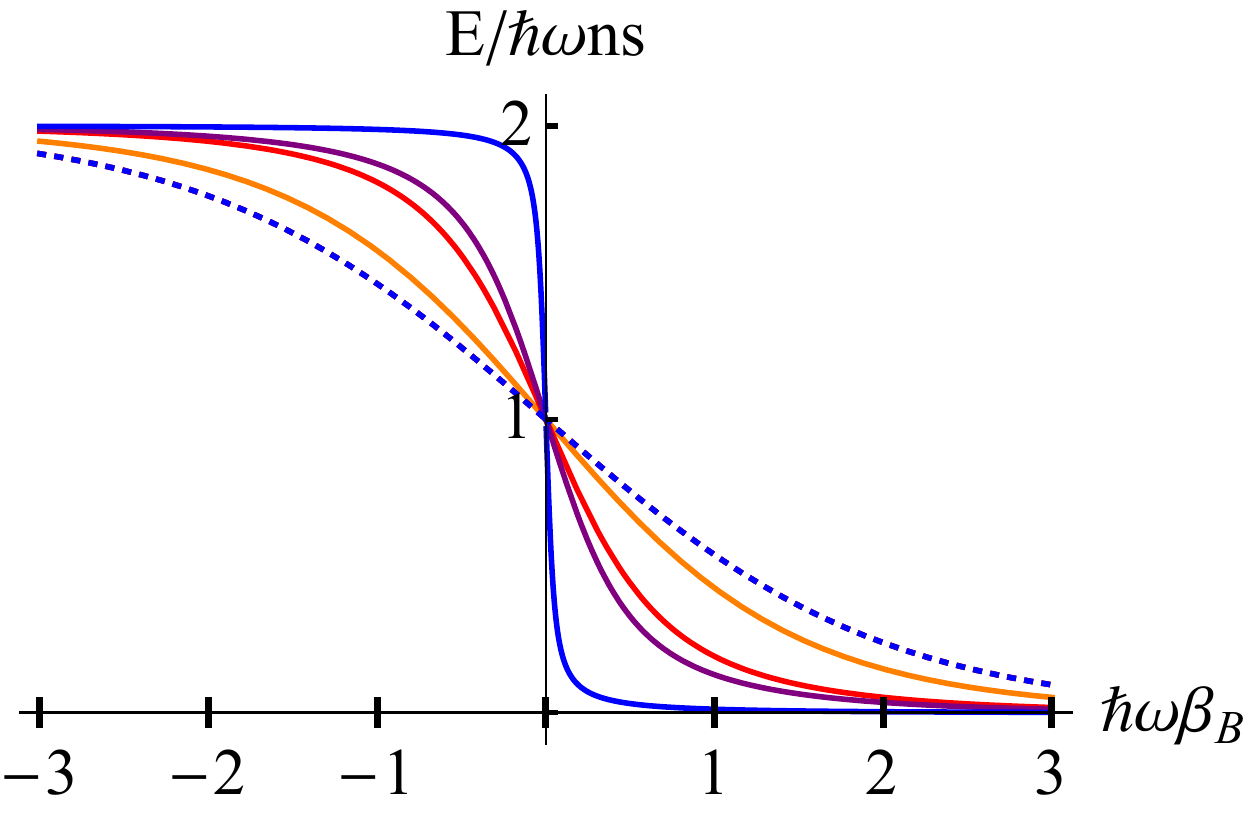}
\caption{(a) Plots of $E_{+}(\beta_B)/ns$ (continuous lines) and $E^{\rm th}(\beta_B)/ns $ (dashed lines) as functions of $\h\omega\beta_B$ for ensembles of $n=4$ spins of size $s=1/2$ (orange curves), $s=3/2$ (red curves), and $s=9/2$ (purple curves).(b) Plots of $E_{+}(\beta_B)/ns$ (continuous lines) and $E^{\rm th}(\beta_B)/ns $ (dashed lines) as functions of $\h\omega\beta_B$ for ensembles containing $n=2$ (orange curves), $n=6$ (red curves), $n=9$ (purple curves), $n=100$ (blue curves) spins of size $s=1/2$. Note that all the four curves $E^{\rm th}(\beta_B)/ns $ are indeed the same.}
\label{enVSns}
\end{figure}

From \eqref{derivatives} and Fig.\ref{enbehaviour} one can conclude that the amplification and mitigation effects are more pronounced for extreme initial inverse temperature $\h\omega|\beta_0|\gg1$.  
 In this limit, one can see from the expressions \eqref{zj} and \eqref{Z} for $Z_J(\beta_0)$ and $Z(\beta_0)$ that $p_J(\beta_0)$ tends to 0 for all $J<ns$ and to $1$ for $J=ns$ (which corresponds to the Dicke subspace \cite{Dicke,Gross_1982}). 
  Then, for $\h\omega|\beta_0|\gg1$, the steady state energy reached by spin ensemble tends to be equal to 
 \bea\label{e+}
E_{+}(\beta_B) &:=& E^{\infty}_{\beta_0=\pm\infty}(\beta_B)\nn\\
&=& l_{ns} e_{ns}(\beta_B) +\h ns\omega \nn\\
&=& e_{ns}(\beta_B) + \h ns\omega\nn\\
&=& \h\omega \frac{1}{e^{\h\omega\beta_B}-1} -\h\omega\frac{2ns+1}{e^{(2ns+1)\h\omega\beta_B}-1},
\eea
obtained using $l_{ns} = 1$ and the expression of $e_{ns}(\beta_B)$ given by \eqref{ej} with the value $J=ns$. Note that $E_{\beta_0}^{\infty}(\beta_B) \simeq E_{+}(\beta_B)$ even for moderate value of $\hbar\omega|\beta_0|$, as one can been seen in Fig. \ref{half} from the curve $\hbar\omega|\beta_0|=5$. 
The thermal equilibrium energy can be obtained 
simply as $n$ times the thermal energy of a spin $s$ (still with the ground state as energy reference),
\bea\label{eth}
E^{\rm th}(\beta_B) &=& n[e_s(\beta_B) + \h\omega s]\nn\\
&=& n\h \omega \frac{1}{e^{\h\omega\beta_B}-1} -n\h\omega\frac{2s+1}{e^{(2s+1)\h\omega\beta_B}-1},\nn\\
\eea 
where we use \eqref{ej} with $J=s$. In order to compare $E_+(\beta_B)$ and $E^{\rm th}(\beta_B)$ we plot several graphs for different value of $n$ and $s$. Fig. \ref{enVSns} presents the plots of $E_{+}(\beta_B)/ns $ and $E^{\rm th}(\beta_B)/ns$ as functions of $\beta_B$ for ensembles of $n=4$ spins of size $s=1/2$, $s=3/2$ and $s=9/2$ (Fig. \ref{enVSns} a), and for ensembles containing $n=2$, $n=6$, $n=9$, and $n=100$ spins $s=1/2$ (Fig. \ref{enVSns} b). One can see that the difference between $E^{+}(\beta_B)$ and $E^{\rm th}(\beta_B)$ becomes larger when $n$  and $s$ increase, even though it is more pronounced with $n$. 
This can also be seen analytically by expanding the expression Eq. \eqref{e+} and \eqref{eth} when $\h\omega |\beta_B|\ll1$. One obtains 
\be\label{enplusexp}
\frac{E_+(\beta_B)}{\h\omega ns} = 1 -\frac{\h\omega\beta_B}{3}(ns+1) + {\cal O}(\h^2\omega^2\beta_B^2)
\ee
and 
\be\label{enthexp}
\frac{E^{\rm th}(\beta_B)}{\h\omega ns} = 1 -\frac{\h\omega\beta_B}{3}(s+1) + {\cal O}(\h^2\omega^2\beta_B^2),
\ee
which shows that the slope around $\beta_B=0$ is almost $n$ times larger for $E_{+}(\beta_B)$, explaining the striking difference between $E_{+}(\beta_B)$ and $E^{\rm th}(\beta_B)$. 

\begin{figure}
\centering
(a)\includegraphics[width=7cm, height=4.5cm]{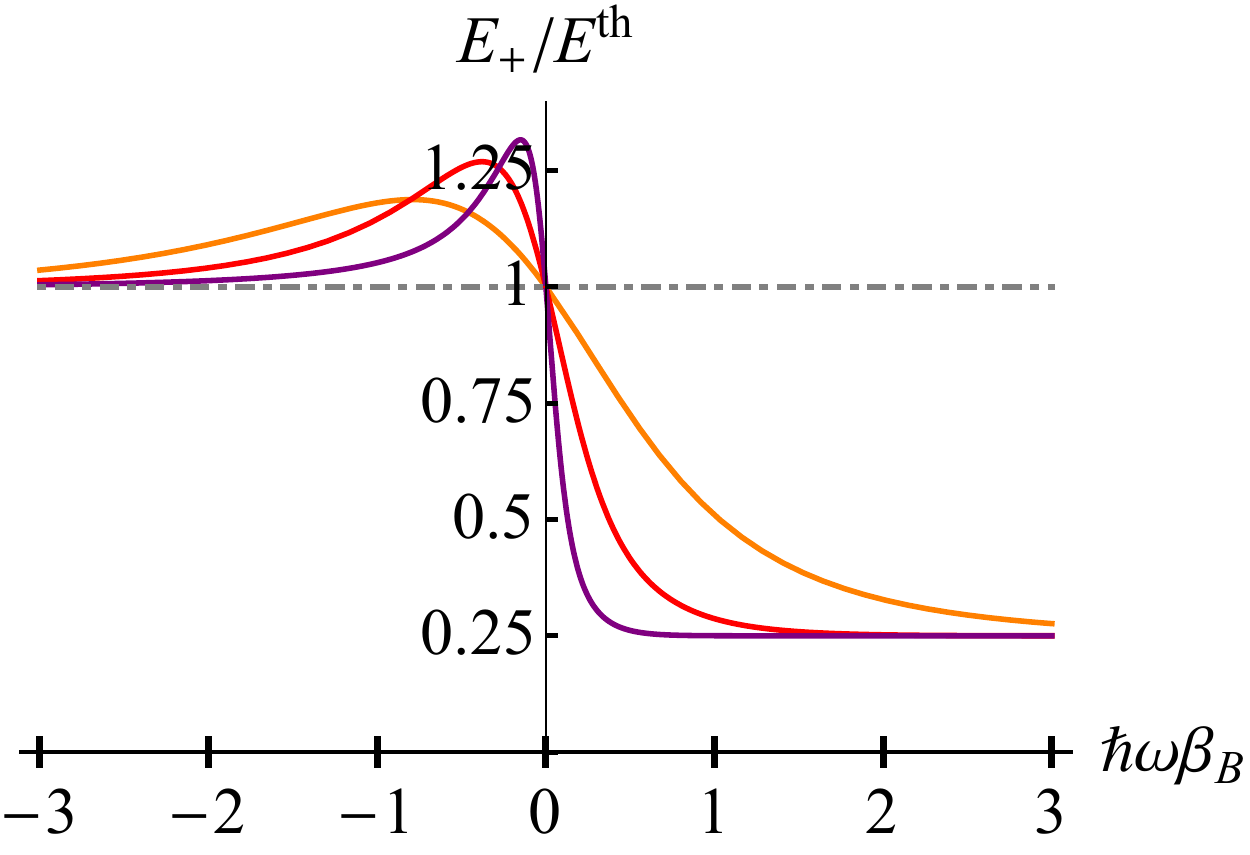}
(b)\includegraphics[width=7cm, height=4.5cm]{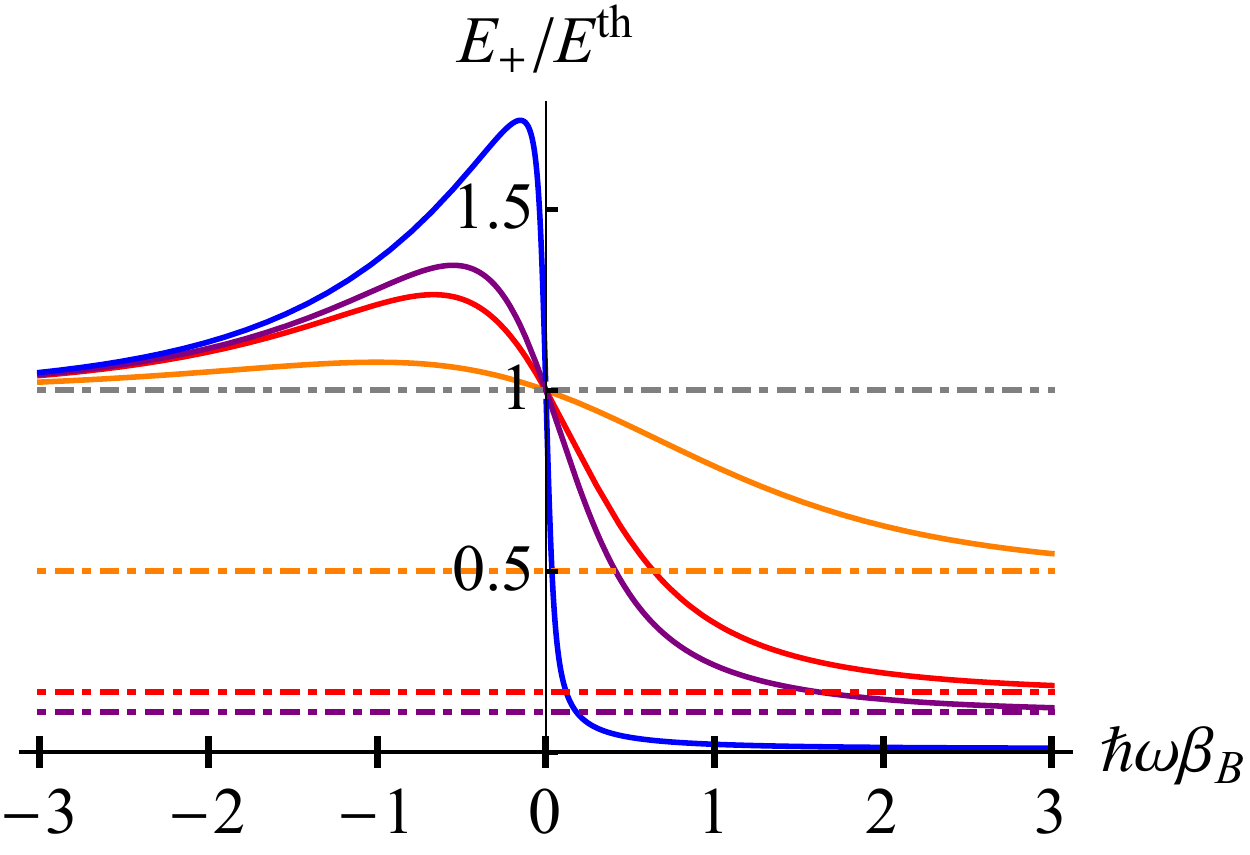}
\caption{(a) Plots of the ratio $E_{+}(\beta_B)/E^{\rm th}(\beta_B) $ as functions of $\h\omega\beta_B$ for an ensemble of 4 spins of size $s=1/2$ (orange curve), $s=3/2$ (red curve), and $s=9/2$ (purple curve). (b) Plots of the ratio $E_{+}(\beta_B)/E^{\rm th}(\beta_B)$ as functions of $\h\omega\beta_B$ for ensembles containing $n=2$ (orange curve), $n=6$ (red curve), $n=9$ (purple curve), and $n=100$ (blue curve) spins $s=1/2$. The gray, orange, red, and purple dot-dashed lines represent the values 1, 1/2, 1/6, and 1/9, respectively. }
\label{energyratiocomp}
\end{figure}

Fig. \ref{energyratiocomp} shows the graphs of the ratio $E_{+}(\beta_B)/E^{\rm th}(\beta_B) $ as a function of $\h\omega\beta_B$ for ensembles of $n=4$ spins of size $s=1/2$, $s=3/2$ and $s=9/2$ (Fig. \ref{energyratiocomp} a) and for ensembles containing $n=2$, $n=6$, $n=9$, and $n=100$ spins $s=1/2$ (Fig. \ref{energyratiocomp} b). One can see that $E_{+}(\beta_B)/E^{\rm th}(\beta_B) $ tends to $1/n$ for $\h\omega\beta_B \gg 1$, which can also be shown analytically from \eqref{e+} and \eqref{eth}, 
\be\label{enplusexp2}
E_{+}(\beta_B)  \underset{\h\omega \beta_B \gg 1}{\simeq} \h\omega \frac{1}{e^{\h\omega\beta_B}-1}\underset{\h\omega \beta_B \gg 1}{\simeq} E^{\rm th}(\beta_B)/n.
\ee

In terms of mitigation of the bath's effects it means for instance that if the spin ensemble is initially in a cold state ($\h\omega \beta_0 \gg 1$) the collective interaction reduces the heating up due to the interaction with a hotter bath by a factor up to $n$. For the sake of completeness we mention an other mitigation effect when the effective bath is in a negative temperature and the spin ensemble is initially close to an inverted population state ($-\h\omega \beta_0 \gg 1$). Then, in such situation the collective interactions keep the spin ensemble in a state of energy up to twice (in the limit of large $ns$) the thermal energy it would reach under independent dissipation. 

In terms of the amplification of the bath's effects, a spin ensemble initially close to an inverted population state ($-\h\omega \beta_0 \gg 1$) interacting with a cold bath can be super refrigerated by a factor close to $n$ (reaching an energy $n$ times smaller) thanks to collective interactions.
Additionally, the amplification of the bath's effects means an extra energy charging when the effective bath is in a negative temperature and the spin ensemble is initially close to the ground state ($\h\omega \beta_0 \gg 1$). Such extra energy charging can go up to twice (in the limit of large $ns$) the energy charged via independent dissipation.\\

%
%

\subsection{Saturation effect and relation with experimental observations}\label{secsaturation}
In addition to the above effects, the collective dissipation can result in a saturation effect. Comparing the expressions Eq. \eqref{e+} of the steady state energy and Eq.\eqref{eth} of the thermal energy one can see that while $E^{\rm th}(\beta_B)$ increases linearly with the number of spins $n$, as expected, the steady state energy $E_+(\beta_B)$ achieved under collective dissipation saturates for growing $n$.
 This curious saturation phenomenon can be related to the experimental observation made on atomic clouds interacting collectively with a thermalised cavity field \cite{Raimond_1982}. The authors observed that when the number of atoms in the cloud is increased, the number of excited atoms after equilibration with the thermal cavity field was saturating instead of increasing linearly with the size of the atomic cloud as one could expect. 
This experimental observation is the nothing but the above saturation effect translated in terms of mean number of excited atoms,
confirming the tendency expected from our predictions, and showing that the effects described throughout this paper should be achievable experimentally.  \\

\subsection{Local state}\label{subseclocstate}
It is also interesting to look at the local state of each spins. From symmetry reason, each spin carries an energy $E^{\infty}_{\beta_0}(\beta_B)/n$ and each spin has the same local state $\rho_{\rm Loc}$. It is interesting to note that for spin $1/2$ (or equivalently for two-level systems), the local state (which is a thermal state) has an inverse temperature $\beta_{\rm Loc}$ different from the bath inverse temperature $\beta_B$. Indeed, the local inverse temperature $\beta_{\rm Loc}$ is a simple  function of the steady state energy $E^{\infty}_{\beta_0}(\beta_B)$ so that $\beta_{\rm Loc}$ reflects the amplification and mitigation of the bath effects described above. In particular, the largest effects happen for $\h\omega\beta_0\gg1$. For the sake of completeness, we give in the following the asymptotic behaviour of $\beta_{\rm Loc}$,
  \be\label{templocll}
\h\omega \beta_{\rm Loc} \underset{\h\omega|\beta_B|\ll1}{=} \h\omega\beta_B\frac{n+2}{3}
\ee
and 
 \be\label{templocgg}
\h\omega \beta_{\rm Loc} \underset{\h\omega|\beta_B|\gg1}{=}\ln n + \h\omega\beta_B.
\ee
The above equations \eqref{templocll} and \eqref{templocgg} show that the amplification of the bath effects (when $\beta_0/\beta_B<-1$) and mitigation of the bath effects (when $\beta_0/\beta_B>-1$) grow with $n$. 

By contrast, for ensemble of spins $s\geq 1$, we show in Appendix \ref{applocstate} that the local state is not a thermal state. This is also interesting since non-thermality was shown to be a useful resource \cite{Brandao_2013} which can be harnessed, for instance, to boost the performances of autonomous thermal machines \cite{autonomousmachines}. 

Note that the interpretation of the mitigation effects in terms of dark states made in \cite{bathinducedcohTLS} for a pair of spins $s=1/2$ would be still valid here in the sense that the variation of $E^{\infty}_{\beta_0}(\beta_B)$ as a function of $\beta_0$ can be seen as an interplay between the weight of dark and bright states. However, a more quantitative description based on dark states is out of reach in general (the structure of dark and bright states becomes too complex for increasing $n$ and $s$). 
As a conclusion we mention that the properties described throughout this Section are all fruit of collective dissipation which is itself rooted in the indistinguishability of each spins. These properties have promising applications detailed in Section ``Applications''. 
In the next Section we look at an other central property, the entropy.

\section{Steady state entropy}\label{secentropy}
Entropy is an other fundamental property of quantum systems, and we shall see in this Section that it is also dramatically affected by the collective 
character of the interaction with the bath. The von Neumann entropy of the steady state $\rho^{\infty}_{\beta_0}(\beta_B)$ is given by 
$S[\rho^{\infty}_{\beta_0}(\beta_B)]= - {\rm Tr} \rho^{\infty}_{\beta_0}(\beta_B) \ln \rho^{\infty}_{\beta_0}(\beta_B)$. Since the states $\rho^{\rm th}_{J,i}(\beta_B)$ have support on orthogonal subspaces, the following identity holds \cite{Nielsen_Book},
\be\label{ent1}
S[\rho^{\infty}_{\beta_0}(\beta_B)]= \sum_{J=J_0}^{ns} \sum_{i=1}^{l_J} p_{J,i}S[\rho^{\rm th}_{J,i}(\beta_B)] + H(p),
\ee
where $H(p)= -\sum_{J=J_0}^{ns}\sum_{i=1}^{l_J} p_{J,i}\ln p_{J,i}$ is the Shannon entropy of the distribution probability $p_{J,i}$. For an ensemble initially in a thermal state at inverse temperature $\beta_0$, $p_{J,i}=p_J(\beta_0)$ (independent of $i$) so that 
\be\label{ent2}
H(p) = - \sum_{J=J_0}^{ns}l_J p_J(\beta_0)\ln p_J(\beta_0).
\ee 
One can also verifies that the von Neumann entropy of $\rho^{\rm th}_{J,i}(\beta_B)$ takes the usual expression of any thermal state,
\bea\label{ent3}
S[\rho^{\rm th}_{J,i}(\beta_B)] &=& \ln Z_J(\beta_B) +  \hbar\omega \beta_B{\rm Tr} J_z \rho^{\rm th}_{J,i}(\beta_B) \nn\\
&=& \ln Z_J(\beta_B) +  \beta_Be_{J}(\beta_B). 
\eea
 Combining \eqref{ent1}, \eqref{ent2}, and \eqref{ent3} we obtain
\bea\label{entropygen}
S[\rho^{\infty}_{\beta_0}(\beta_B)]&=&\beta_B\Big[E^{\infty}_{\beta_0}(\beta_B) - \h \omega ns\Big] \nn\\
&& + \sum_{J=J_0}^{ns} l_J p_J(\beta_0)\ln \frac{Z_J(\beta_B)}{p_J(\beta_0)}.
\eea
By contrast, the thermal equilibrium entropy $S[\rho^{\rm th}(\beta_B)]$ reached under independent dissipation is equal to 
\bea
S^{\rm th}(\beta_B)&:=&S[\rho^{\rm th}(\beta_B)]\nn\\
&=&nS[\rho^{\rm th}_{J=s}(\beta_B)]\nn\\
&=&n [\ln Z_s(\beta_B) +  \beta_Be_s(\beta_B)]\nn\\
&=&n \ln Z_s(\beta_B) +  \beta_B[E^{\rm th}(\beta_B) - \h \omega ns] \nn\\
&=& \beta_B E^{\rm th}(\beta_B) + n \ln\frac{1-e^{-(2s+1)\omega\beta_B}}{1-e^{-\omega\beta_B}},\nn\\ \label{exprthentropy}
\eea
obtained from \eqref{ent3} or from \eqref{entropygen} with $\beta_0=\beta_B$. The expression of $\ln Z_s(\beta_B)$ presented in the last line was obtained using \eqref{zj}.

As for the steady state energy, it is challenging to compare directly $S[\rho^{\infty}_{\beta_0}(\beta_B)]$ with the thermal entropy $S^{\rm th}(\beta_B)$. Thus, we follow the same strategy as in the previous Section which consists in studying the behaviour of $S[\rho^{\infty}_{\beta_0}(\beta_B)]$ as a function of $\beta_0$. It is shown in Appendix \ref{smentropycomp} that 
 $S[\rho^{\infty}_{\beta_0}(\beta_B)]$ is a monotonic strictly increasing function of $\beta_0$ for $\beta_0 <0$ and strictly decreasing for $\beta_0 >0$.
 Consequently, since $S[\rho^{\infty}_{\beta_0=\pm\beta_B}(\beta_B)]=S^{\rm th}(\beta_B)$, we have
\be\label{ineqentropy}
S[\rho^{\infty}_{\beta_0}(\beta_B)] < S^{\rm th}(\beta_B)
\ee
for all $|\beta_0| > |\beta_B|$, and 
\be\label{ineqentropy2}
S[\rho^{\infty}_{\beta_0}(\beta_B)] > S^{\rm th}(\beta_B)
\ee
for all $|\beta_0| < |\beta_B|$.
In particular, this implies, for all $\beta_0$ and $\beta_B$,
\be
 |S[\rho^{\infty}_{\beta_0}(\beta_B)] -S[\rho^{\rm th}(\beta_0)] |< |S[\rho^{\rm th}(\beta_B)] -S[\rho^{\rm th}(\beta_0)] |.
 \ee
 It means that the variation (in absolute value) of entropy between the initial and final states is always reduced when the dissipation is collective. In other words, the collective interaction always mitigates the bath's action from the point of view of the entropy. We show in the following that the spin ensemble entropy can be reduced by a factor up to $n$.  
  This is an interesting additional properties since for various applications in the regime $|\beta_0|>|\beta_B|$ (like collective work extraction, state protection and cooling operations, see Section \ref{applications}), it is highly desirable that the spin ensemble remains in a low entropy state. \\

 We now analyse, as for the energy, the extent of the mitigation effect for the entropy. 
The mitigation is more pronounced for $\h\omega|\beta_0|\gg1$, which also corresponds to the largest amplification and mitigation effects for the energy. In this regime $\h\omega|\beta_0| \gg 1$ the steady state entropy $S[\rho^{\infty}_{\beta_0}(\beta_B)]$ tends to be equal to
\bea\label{ent+}
S_{+}(\beta_B) &:=& S[\rho^{\infty}_{\beta_0=\pm\infty}(\beta_B)] \nn\\
 &=&\beta_B\Big[E_{+}(\beta_B) - \h ns\omega\Big] +  \ln Z_{ns}(\beta_B)\nn\\
&=& \beta_B E_+(\beta_B) + \ln\frac{1-e^{-(2ns+1)\omega\beta_B}}{1-e^{-\omega\beta_B}},
\eea
where the last line was obtained using \eqref{zj}.
This is to be compared with the thermal entropy given in \eqref{exprthentropy}.

\begin{figure}
\centering
(a)\includegraphics[width=7cm, height=4.5cm]{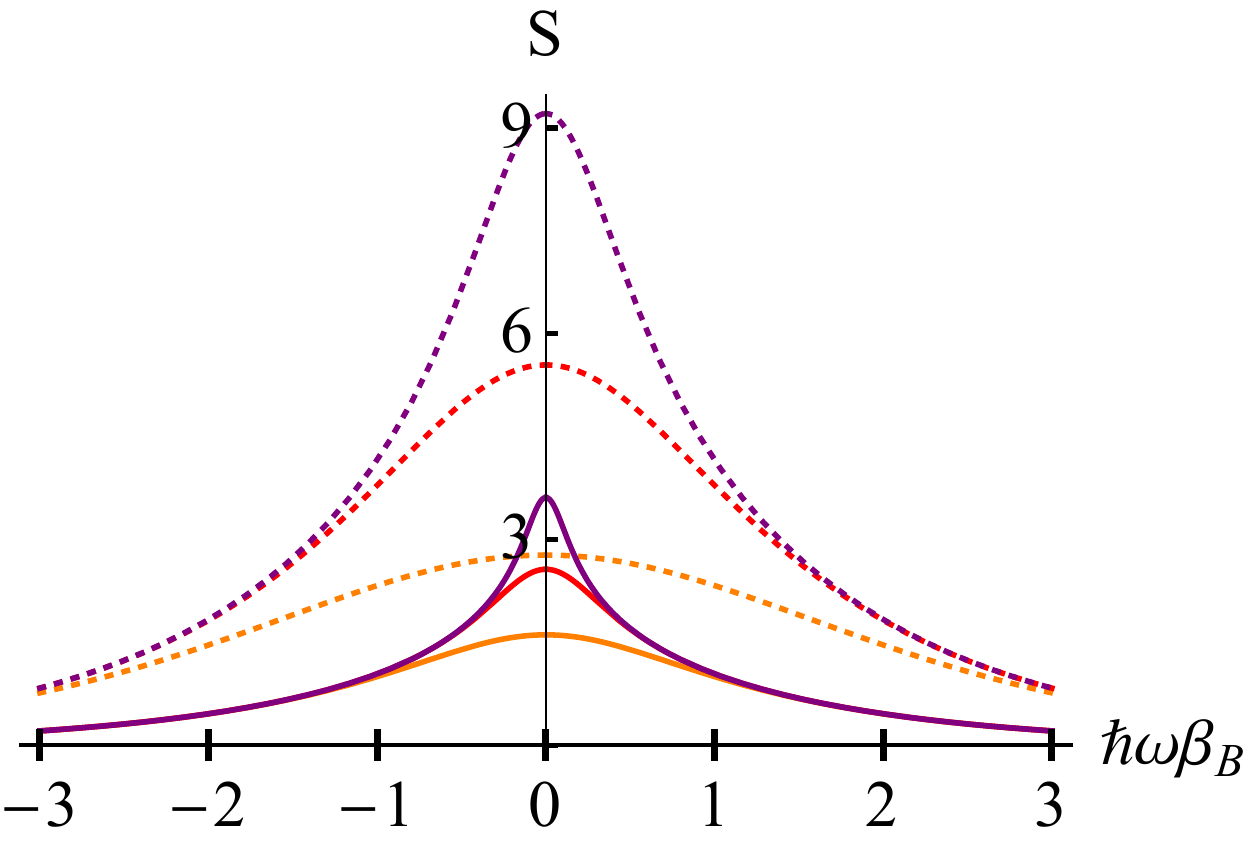}
(b)\includegraphics[width=7cm, height=4.5cm]{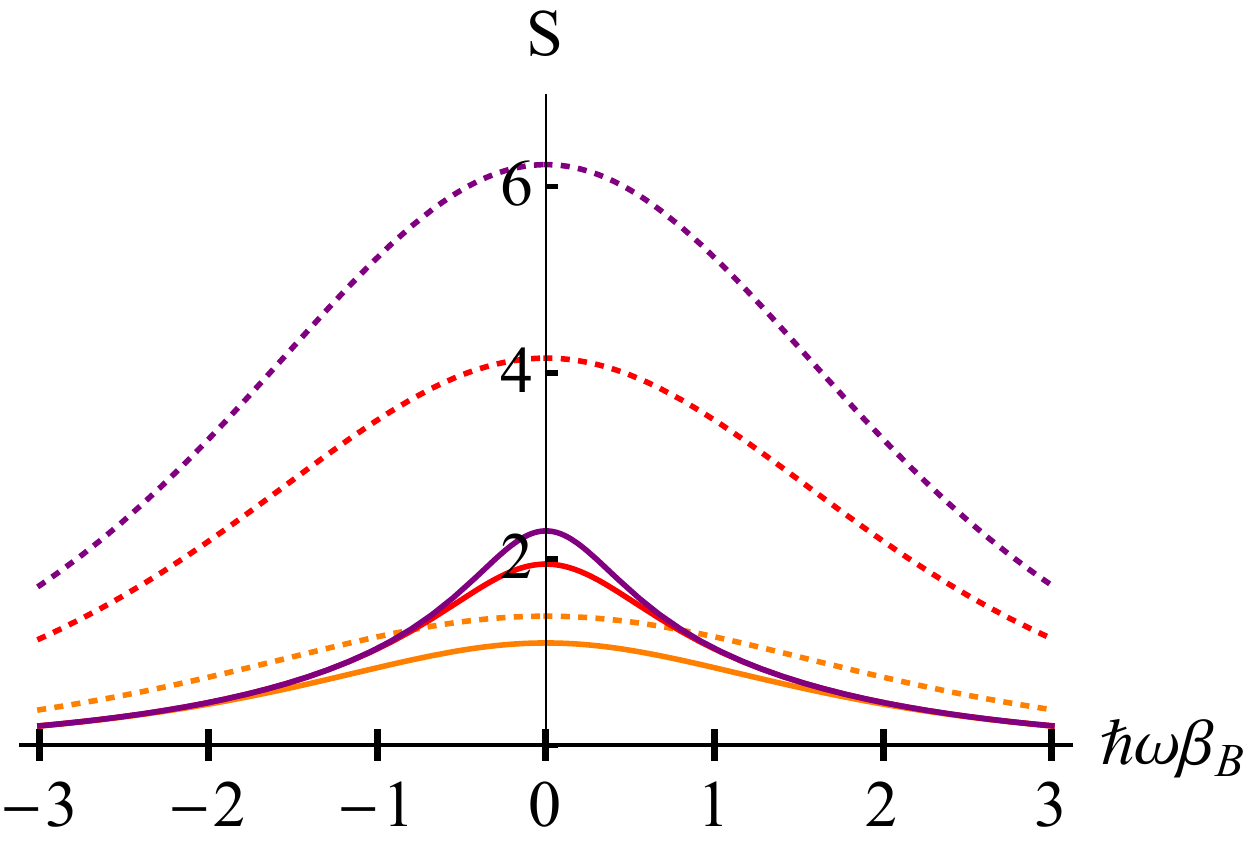}
\caption{(a) Plots of $S_{+}(\beta_B)$ (continuous lines) and $S^{\rm th}(\beta_B)$ and (dashed lines) as functions of $\h\omega\beta_B$ for ensembles of 4 spins of size $s=1/2$ (ornage curves), $s=3/2$ (red curves), and $s=9/2$ (purple curves).(b) Plots of $S_{+}(\beta_B)$ (continuous lines) and $S^{\rm th}(\beta_B) $ (dashed lines) as functions of $\h\omega\beta_B$ for ensembles containing $n=2$ (orange curves), $n=6$ (red curves), and $n=9$ (purple curves) spins of size $s=1/2$. }
\label{entropycomp}
\end{figure}

\begin{figure}
\centering
(a)\includegraphics[width=7cm, height=4.5cm]{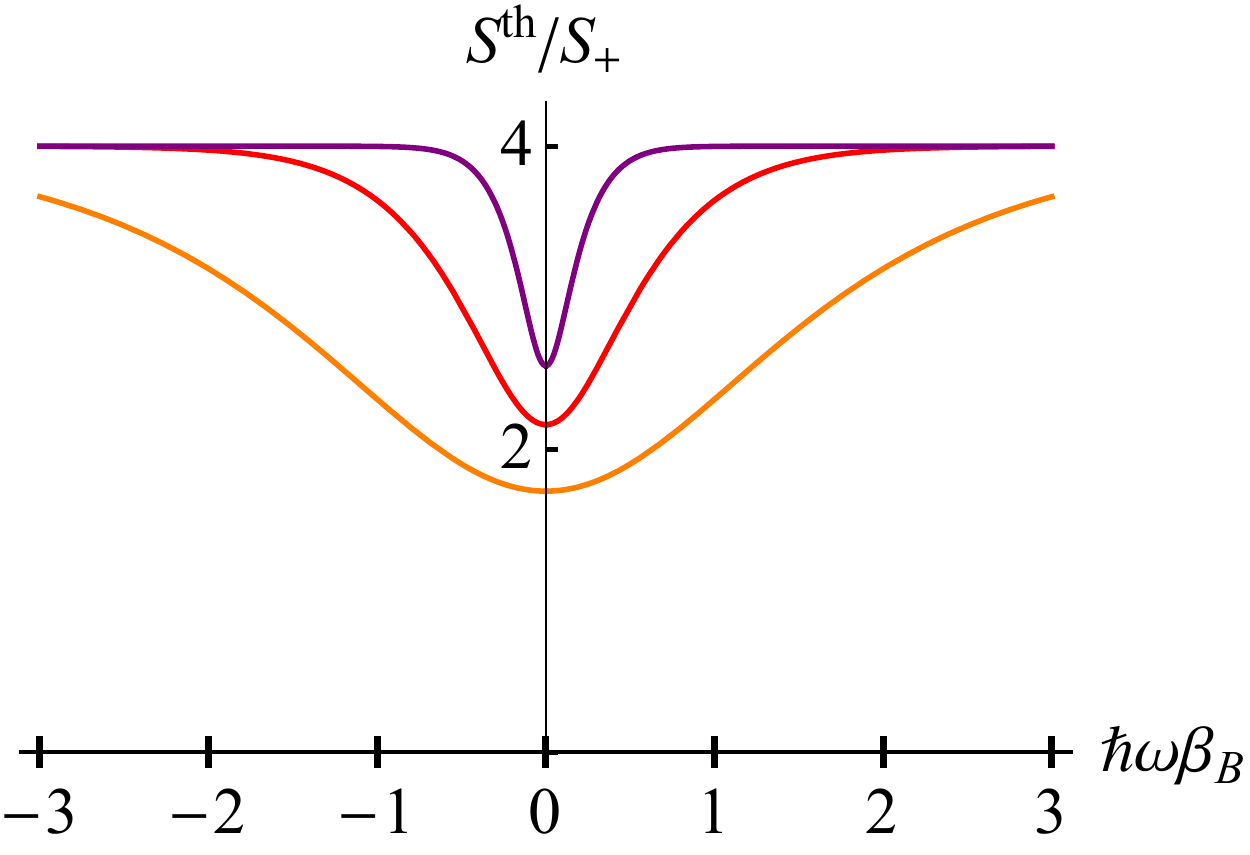}
(b)\includegraphics[width=7cm, height=4.5cm]{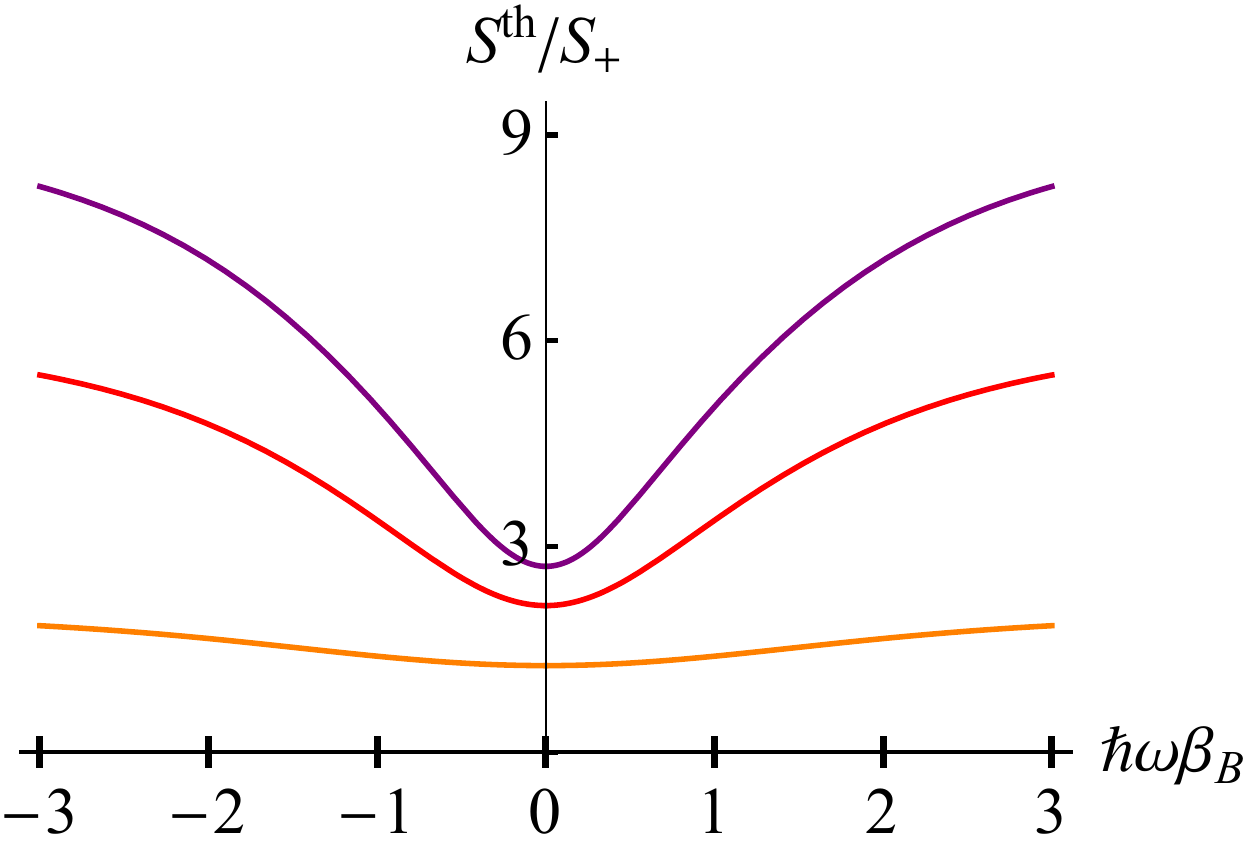}
\caption{(a) Plots of the ratio $S^{\rm th}(\beta_B)/S_{+}(\beta_B)$ as a function of $\h\omega\beta_B$ for ensembles of 4 spins of size $s=1/2$ (orange curves), $s=3/2$ (red curves), and $s=9/2$ (purple curves).(b) Plots of the ratio $S^{\rm th}(\beta_B)/S_{+}(\beta_B)$ again as functions of $\h\omega\beta_B$ for ensembles containing $n=2$ (orange curves), $n=6$ (red curves), and $n=9$ (purple curves) spins of size $s=1/2$.}
\label{entropyratiocomp}
\end{figure}

Fig. \ref{entropycomp} presents the graphs of $S_{+}(\beta_B) $ and $S^{\rm th}(\beta_B)$ as a function of $\h\omega\beta_B$ for ensembles of $n=4$ spins of size $s=1/2$, $s=3/2$, and $s=9/2$ (Fig. \ref{entropycomp} a), and for ensembles containing $n=2$, $n=6$, and $n=9$ spins of size $s=1/2$ (Fig. \ref{entropycomp} b). Fig. \ref{entropyratiocomp} corresponds to the plots of the ratio $S^{\rm th}(\beta_B)/S_{+}(\beta_B) $ again as a function of $\h\omega\beta_B$ for ensembles of $n=4$ spins of size $s=1/2$, $s=3/2$, and $s=9/2$ (Fig. \ref{entropyratiocomp} a), and for ensembles containing $n=2$, $n=6$, and $n=9$ spins of size $s=1/2$ (Fig. \ref{entropyratiocomp} b).
 One can see a very large reduction of entropy over the whole range of values of $\beta_B$. 
  In particular, the entropy tends to be reduced by a factor $n$ for $\h\omega |\beta_B| \gg 1$, which can also be seen analytically from \eqref{exprthentropy} and \eqref{ent+},
\bea\label{entplusexp}
S_{+}(\beta_B)  &\underset{\h\omega |\beta_B| \gg 1}{\simeq}& \frac{\beta_B\h \omega}{e^{\h\omega\beta_B}-1} -\ln(1-e^{-\omega\beta_B})\nn\\
& \underset{\h\omega |\beta_B| \gg 1}{\simeq}& S^{\rm th}(\beta_B)/n.
\eea
One can also note that the difference between the steady state and the thermal equilibrium entropies increases for increasing spins size $s$. The effect can be seen also analytically when taking the limit $\h\omega|\beta_B|\ll1$,
 \bea\label{entplusexp2}
S_{+}(\beta_B) &=&\ln(2ns+1) -\frac{(\hbar\omega\beta_B)^2}{6}ns(ns+1)\nn\\
 &&+{\cal O}[(\hbar\omega\beta_B)^3],
\eea
and
 \bea\label{entthexp}
S^{\rm th}(\beta_B) &=&n\ln(2s+1)-\frac{(\hbar\omega\beta_B)^2}{6}ns(s+1)\nn\\
 &&+{\cal O}[(\hbar\omega\beta_B)^3].
\eea

\section{Free energy and entropy production}\label{secfreeen}
We conclude this overview of the thermodynamic implications of collective dissipation by one other fundamental thermodynamic quantity, the free energy, defined for a state $\rho$ of the spin ensemble by
\be
F(\rho):=\h\omega {\rm Tr}J_z\rho +\h\omega ns - S(\rho)/\beta_B.
\ee
Note that the term $\h\omega ns $ is due to our energy reference. 
 The meaningful quantity is the variation of free energy $\Delta F$, which gives precious information on the irreversibility of the evolution \cite{Batalhao_2018}, but also on the quantity of extractable work \cite{Muller_2018} -- one can think of the free energy as the ``accessible'' energy of the system. As such, when noise is added to the system, its free energy should decrease, which is verified for instance for dissipative evolution
 (at least for Markovian processes \cite{Strasberg_2019}) with $\beta_B>0$.

One can show (Appendix \ref{appfreeenergy}) for $\beta_B>0$ that under collective dissipation the free energy variation $\Delta F^{\infty}_{\beta_0}(\beta_B):=F[\rho^{\infty}_{\beta_0}(\beta_B)]-F[\rho^{\rm th}(\beta_0)]$ is always larger (i.e. smaller in absolute value) than the free energy variation $\Delta F^{\rm th}(\beta_B):=F[\rho^{\rm th}(\beta_B)]-F[\rho^{\rm th}(\beta_0)]$ under independent dissipation. This holds for any $\beta_0$. In this sense, the bath's action is always mitigated by collective coupling (even though we saw in Section \ref{subsecmitig} that, from an energetic point of view, the bath's action is amplified when $\beta_0/\beta_B<-1$). This general statement can be extended to $\beta_B<0$ (see Appendix \ref{appfreeenergy}).

As illustrations, Fig. \ref{freeenergy} shows the plots of $\Delta F^{\infty}_{\beta_0}(\beta_B)$ and $\Delta F^{\rm th}(\beta_B)$ as functions of $\h\omega\beta_B$ for $\h\omega\beta_0\gg1$ (where the collective effects are more pronounced) and for ensembles containing $n=2$, $n=6$, $n=9$, and $n=100$ spins $s=3/2$. 
 Curiously, one can see in Fig. \ref{freeenergy} (a) (full curves) that the free energy variation corresponding to collective dissipation remains almost the same for any $n$. This implies that the variation of free energy per spins is highly increased (decreased in absolute value) as it can be observed in Fig. \ref{freeenergy} (b). The asymptotic behaviour of the variation of free energy can be obtained straightforwardly from \eqref{enplusexp2} and \eqref{entplusexp}, leading to
\be\label{freeenexp}
\Delta F^{\infty}_{\beta_0}(\beta_B)\underset{\h\omega\beta_B\gg1}{\underset{\h\omega\beta_0\gg1}{\simeq}} \frac{\Delta F^{\rm th}(\beta_B)}{n}.
\ee


\begin{figure}
\centering
(a)\includegraphics[width=7cm, height=4.5cm]{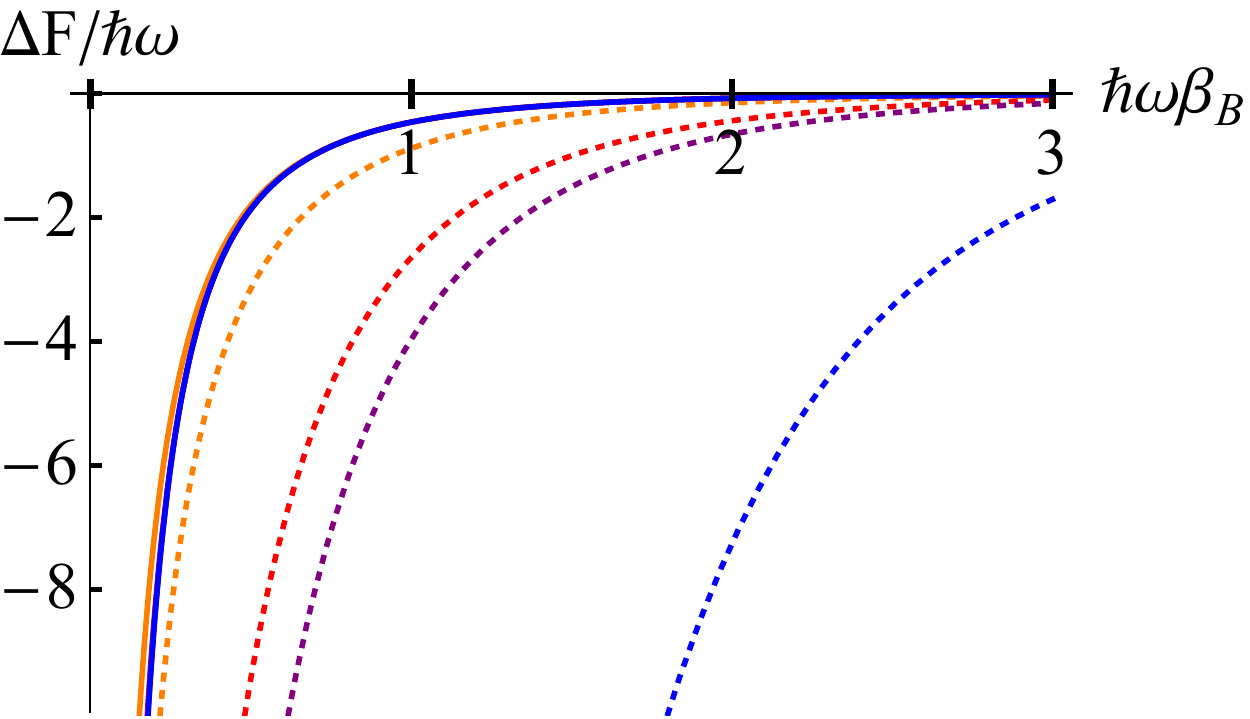}
(b)\includegraphics[width=7cm, height=4.5cm]{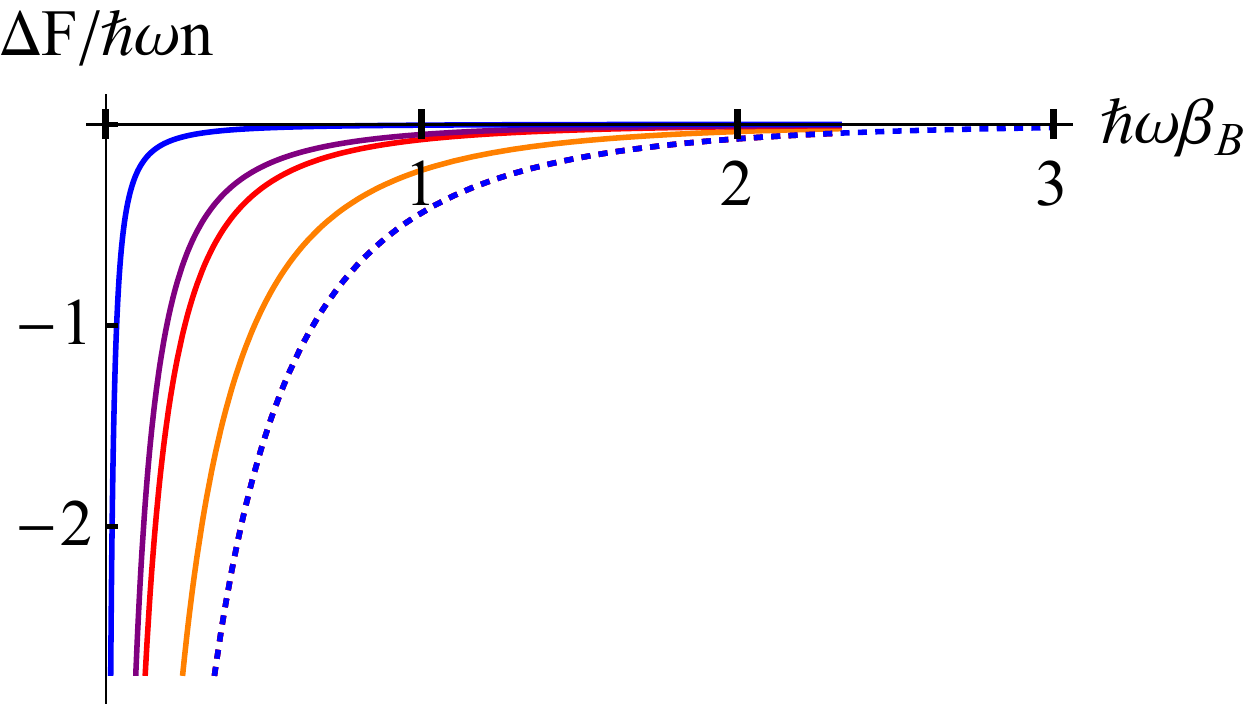}
\caption{(a) Plots of the variation of free energy $\Delta F^{\infty}_{\beta_0}(\beta_B)$ (continuous curves) for $\h\omega\beta_0\gg1$ and $\Delta F^{\rm th}(\beta_B)$ (dashed curves) as functions of $\h\omega\beta_B$ for ensembles containing $n=2$ (orange curves), $n=6$ (red curves), $n=9$ (purple curves), and $n=100$ (blue curves) spins of size $s=3/2$. Note that the curves $\Delta F^{\infty}_{\beta_0}(\beta_B)$ are almost the same and therefore cannot be properly distinguished. (b) Same plots as in the panel (a) but for the variation of free energy per spin, $\Delta F^{\infty}_{\beta_0}(\beta_B)/n$ and $\Delta F^{\rm th}(\beta_B)/n$. Note that the curves $\Delta F^{\rm th}(\beta_B)/n$ are exactly the same, reason why only one dotted curve appears.}
\label{freeenergy}
\end{figure}

\begin{figure}
\centering
\includegraphics[width=7cm, height=4.5cm]{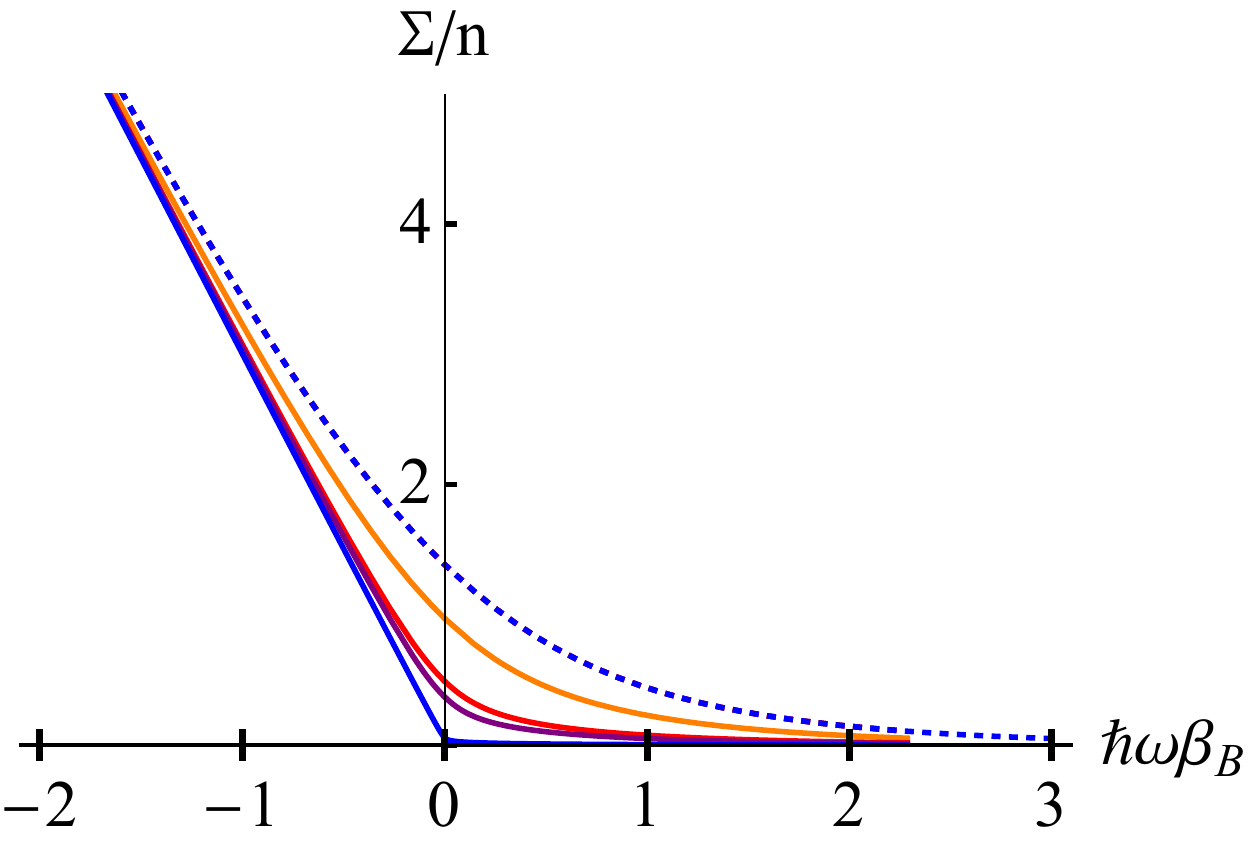}
\caption{Plots of the entropy production per spin $\Sigma_{\beta_0}^{\infty}(\beta_B)/n$ (continous lines) (for $\h\omega\beta_0\gg1$) and $\Sigma^{\rm th}(\beta_B)/n$ (dashed lines) as functions of $\h\omega\beta_B$ for ensembles containing $n=2$ (orange curves), $n=6$ (red curves), $n=9$ (purple curves), and $n=100$ (blue curves) spins os size $s= 3/2$. Note that all curves $\Sigma^{\rm th}(\beta_B)/n$ (dashed lines) are the same.}
\label{entproduction}
\end{figure}

The extension of the plots of Fig. \ref{freeenergy} to negative effective bath temperatures (which can be obtained, for $\hbar\omega\beta_0\gg1$, by $\Delta F(-\beta_B) = 2\hbar\omega ns - \Delta F(\beta_B)$) leads to positive variation of free energy. There is in fact nothing special with that, it is merely an effect of $\beta_B$ assuming negative values. 

Finally, we mention the entropy production -- the core concept of the Second Law of thermodynamics -- which has been object of intense research also due to its relation with irreversibility \cite{Parrondo_2009, Deffner_2011, Santos_2017, Brunelli_2018, Santos_2019}, believed to play a central role in non-equilibrium dynamics but also in the performances of thermal machines and the so-called thermodynamics uncertainty relations \cite{Barato_2015,Gingrich_2016,Pietzonka_2016,Pietzonka_2018, Guarnieri_2019,Timpanaro_2019,Su_2019}. 

The entropy production is simply given here by $\Sigma=-\beta_B\Delta F$ \cite{Deffner_2011,Batalhao_2018,Santos_2019}. As a direct consequence of the general result on the free energy variation, the entropy production is always reduced for collective dissipation, implying a reduction of irreversibility of the dissipation.
 Fig. \ref{entproduction} presents the plot of the entropy production $\Sigma_{\beta_0}^{\infty}(\beta_B)/n$ associated to the process of collective dissipation, and the entropy production $\Sigma^{\rm th}(\beta_B)/n$ associated to the process of independent dissipation for ensembles containing $n=2$, $n=6$, $n=9$, and $n=100$ spins $s=3/2$. One can see the dramatic impact of collective dissipation, turning the process almost a reversible process for $\beta_B>0$. From \eqref{freeenexp} one obtains 
\be
\Sigma_{\beta_0}^{\infty}(\beta_B) \underset{\h\omega\beta_B\gg1}{\underset{\h\omega\beta_0\gg1}{\simeq}} \frac{\Sigma^{\rm th}(\beta_B)}{n}.
\ee
One can question the claim of reduction of entropy production because we are comparing two processes, collective and independent dissipations, which do not yield the same final state. However, what we can do at least is to compare processes yielding states of same final energy. Doing that, one still obtain a dramatic reduction of entropy production. Considering for instance $\hbar\omega\beta_0\gg1$, and choosing a final energy per spin (let us say $0.5 \hbar\omega s$ per spin), one can find on Fig. \ref{enVSns} (b) the bath temperature yielding such final energy  for both collective and independent dissipation. One can see on Fig. \ref{entproduction} that the corresponding entropy production is still much smaller for collective coupling, especially for large $n$ and final energy per spin close to $ \hbar\omega s$.
 
For $\beta_B<0$, the steep increase in the entropy production can be explained by the following relation $\Sigma_{\beta_0}^{\infty}(-\beta_B) = 2\h\omega\beta_Bns + \Sigma_{\beta_0}^{\infty}(\beta_B)$ (for $\hbar\omega\beta_0\gg1$).


\section{Applications and perspectives}\label{applications}
The above effects described in Sections \ref{mainenergy}, \ref{secentropy}, and \ref{secfreeen} have several important applications and consequences.
 The first ones are related to thermal machines. One long standing question in quantum thermodynamics is whether and how quantum effects can enhance the performance of thermodynamic tasks like refrigeration and work or energy extraction from thermal baths. 
In the following we show how the mitigation effects described in Section \ref{mainenergy} can be harnessed to increase the output power of cyclic thermal machines. More applications are briefly mentioned in Section \ref{secmoreapp}.

\subsection{Effective amplification}\label{seceffectiveamp}
We consider a thermal machine undergoing a quantum Otto cycle \cite{Scully_2002,Quan_2007} with a working medium composed of an ensemble of $n$ spins $s$ (or two-level systems). Some designs of thermal machines using many-body working medium have already been studied in \cite{Jaramillo_2016,Gelbwaser_2019} where it was reported that collective effects can be beneficial when using non-adiabatic strokes instead of the usual adiabatic ones. Enhancements using phase-transitions in many-body systems have also been investigated in \cite{Campisi_2016, Holubec_2017, Ma_2017}. Power increase was pointed out for ensembles of spins $1/2$ in \cite {Kloc_2019} where the equilibration speed-up stemming from collective effects allows one to reduce the duration of the cycle, and hence increase the delivered power. Other studies investigate the effect of internal coupling and entanglement between the subsystems constituting the many-body working medium (pair of two-level systems \cite{Wang_2009,Altintas_2014}, pair of degenerate two-level systems \cite{ Mehta_2017}, a two-level system coupled to a harmonic oscillator \cite{Altintas_2015,Barrios_2017}, and ensemble of spins $1/2$ \cite{Hardal_2018}). Additionally, many-body effects have also been investigated in continuous thermal machines \cite{Niedenzu_2018,Watanabe_2019}.

In this section, we suggest an alternative mechanism to increase the output power of many-body thermal machines. The successive mitigation effects of the hot and cold baths can result in an {\it effective amplification} of the baths' action, leading to an increase of the extracted work. The cycle is described by the four usual strokes composing the Otto cycle \cite{Scully_2002, Quan_2007}. The first isochoric stroke is realised through the interaction with a hot bath at inverse temperature $\beta_h$. Crucially, we assume that the hot bath does not distinguish the spins composing the working medium so that the ensemble is dissipated collectively. Considering that the working medium was initially in a thermal state $\rho^{\rm th}(\beta_0)$ {\it before} the machine starts to operate, the steady state reached at the end of this isochoric stroke is $\rho_1:=\rho_{\beta_0}^{\infty}(\beta_h)$ (defined with respect to the initial free Hamiltonian denoted by $H$). The second stroke is adiabatic, preserving the state of the working medium but changing the Hamiltonian, from $H$ to $H'$. Then, follows a second isochoric stroke realised by a cold bath at inverse temperature $\beta_c$, taking the working media from $\rho_1$ to $\rho_2$. Assuming also that the cold bath does not distinguish the spins of the working medium, the steady state reached at the end of the second isochoric stroke is $\rho_2:={\rho'}_{\beta_0}^{\infty}(\beta_c)$ (defined with respect to $H'$). Note that although the working medium was not in the state $\rho^{\rm th}(\beta_0)$ at the beginning of this second isochoric stroke, it still reaches the steady state ${\rho'}_{\beta_0}^{\infty}(\beta_c)$. This is because the steady state is determined by the weights $p_{J,i}$ (see Section \ref{secspinensembles} and Eq. \eqref{ss}), which are unaffected by the dissipation processes. Therefore, the initial weights $p_J(\beta_0)$ are preserved throughout the cycles, determining the properties and performances of the engine as we show in the following.
 Finally, the last stroke is a second adiabatic evolution, preserving the state of the ensemble but taking the Hamiltonian back to its original value, $H$.

 The work $W$ extracted per cycle by the engine is the sum of the work realised during the two adiabatic strokes,
 \be
W = {\rm Tr} \rho_1(H'-H) + {\rm Tr} \rho_2(H-H') 
\ee 
which should be negative (for work extraction). The heat invested is $Q_{h}={\rm Tr} (\rho_1-\rho_2) H$, and the heat dumped into the cold bath is $Q_{c}={\rm Tr} (\rho_2-\rho_1) H'$, verifying the first law, $Q_h+Q_c=-W$. 

Assuming homogeneous adiabatic strokes \cite{Scully_2002,Quan_2007,Gelbwaser_2019}, meaning that the two Hamiltonian are proportional, $H'=\lambda H$, with the ``compression'' $\lambda$ chosen in the interval $\frac{\beta_h}{\beta_c}\leq \lambda\leq1$ (in order to extract energy), one can verify that with this design the efficiency of work extraction, defined by 
\be
\eta:= \frac{-W}{Q_{h}},
\ee
is equal to the usual value \cite{Scully_2002,Quan_2007,Gelbwaser_2019}, namely $\eta = 1-\lambda \leq 1-\frac{T_c}{T_h}$. For indistinguishable spins, the work extracted per cycle, determining the power of the engine, is 
\bea\label{wextcoh}
-W^{\rm coh}&:=&(1-\lambda){\rm Tr}(\rho_1-\rho_2)H\nn\\
&=&(1-\lambda) [E^{\infty}_{\beta_0}(\beta_h)-E^{\infty}_{\beta_0}(\lambda\beta_c)],
\eea
where the factor $\lambda$ in the argument of the second energy is to take into account that $\rho_2:={\rho'}_{\beta_0}^{\infty}(\beta_c)$ is defined with respect to $H'$. 
By comparison, the same thermal engine using distinguishable spins extracts per cycle a work equal to
\bea\label{wextinc}
-W^{\rm inc}&:=&(1-\lambda){\rm Tr}[\rho^{\rm th}(\beta_h)-{\rho'}^{\rm th}(\beta_c)]H\nn\\
&=&(1-\lambda) [E^{\rm th}(\beta_h)-E^{\rm th}(\lambda\beta_c)],
\eea
where, as previously with the collective coupling, ${\rho'}^{\rm th}(\beta_c)$ is defined with respect to $H'$, leading to the factor $\lambda$ in the argument of the second energy.

\begin{figure}
\centering
(a)~\includegraphics[width=7.45cm, height=4.75cm]{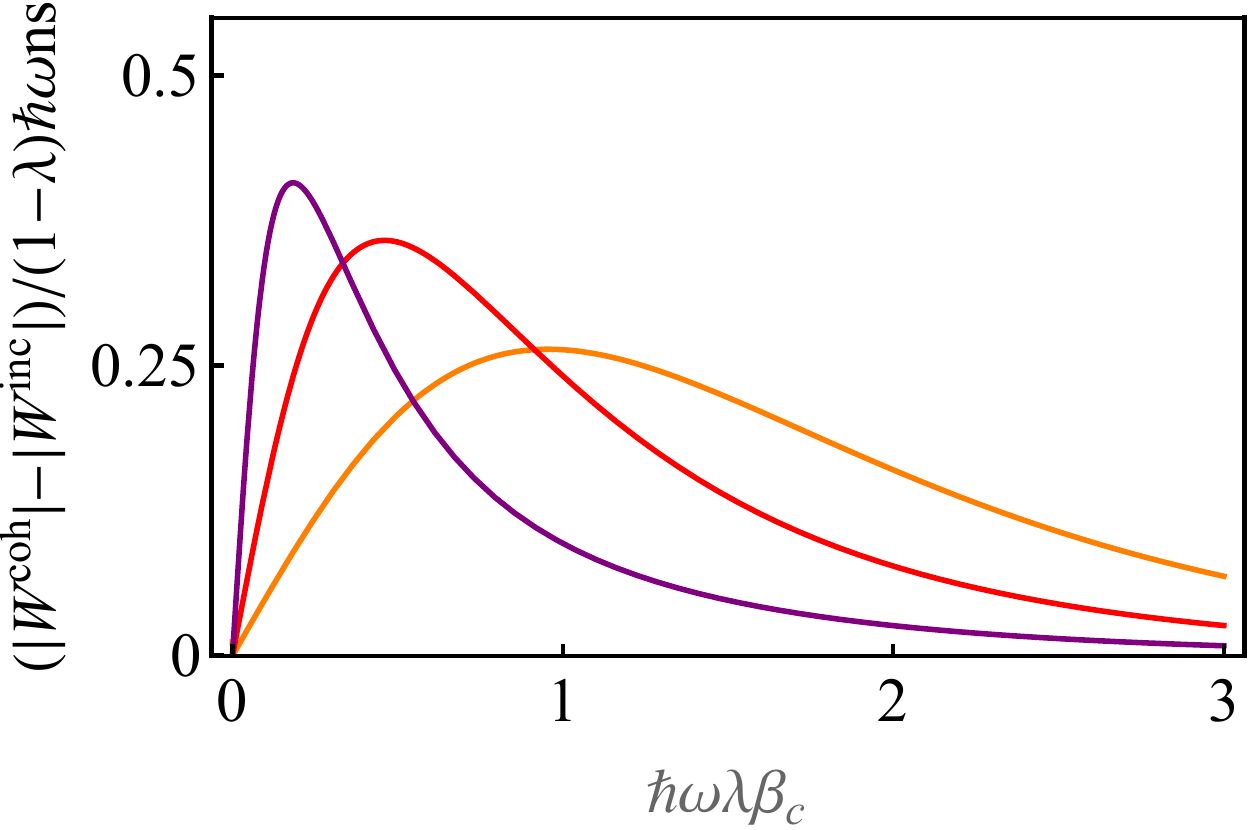}
(b)~~~~~\includegraphics[width=7cm, height=4.5cm]{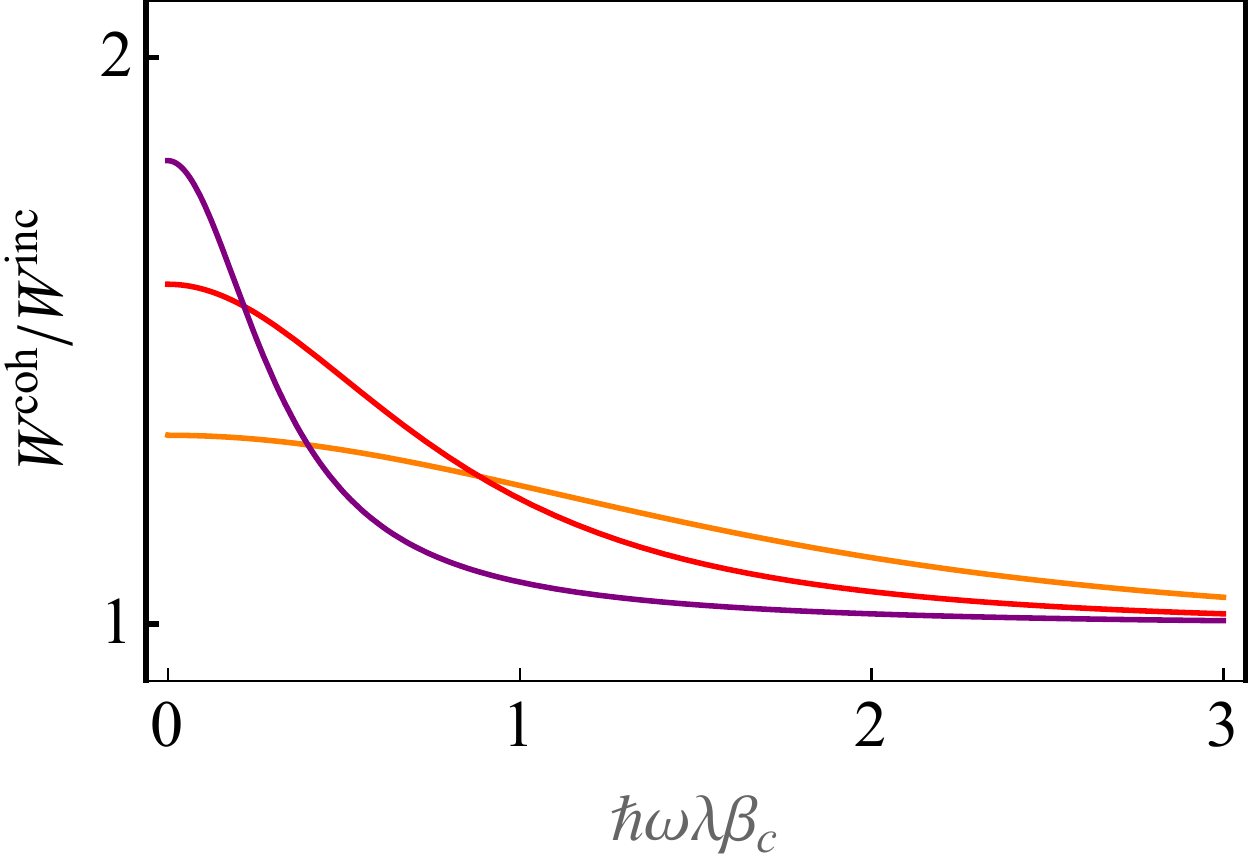}
(c)\includegraphics[width=7.55cm, height=4.79cm]{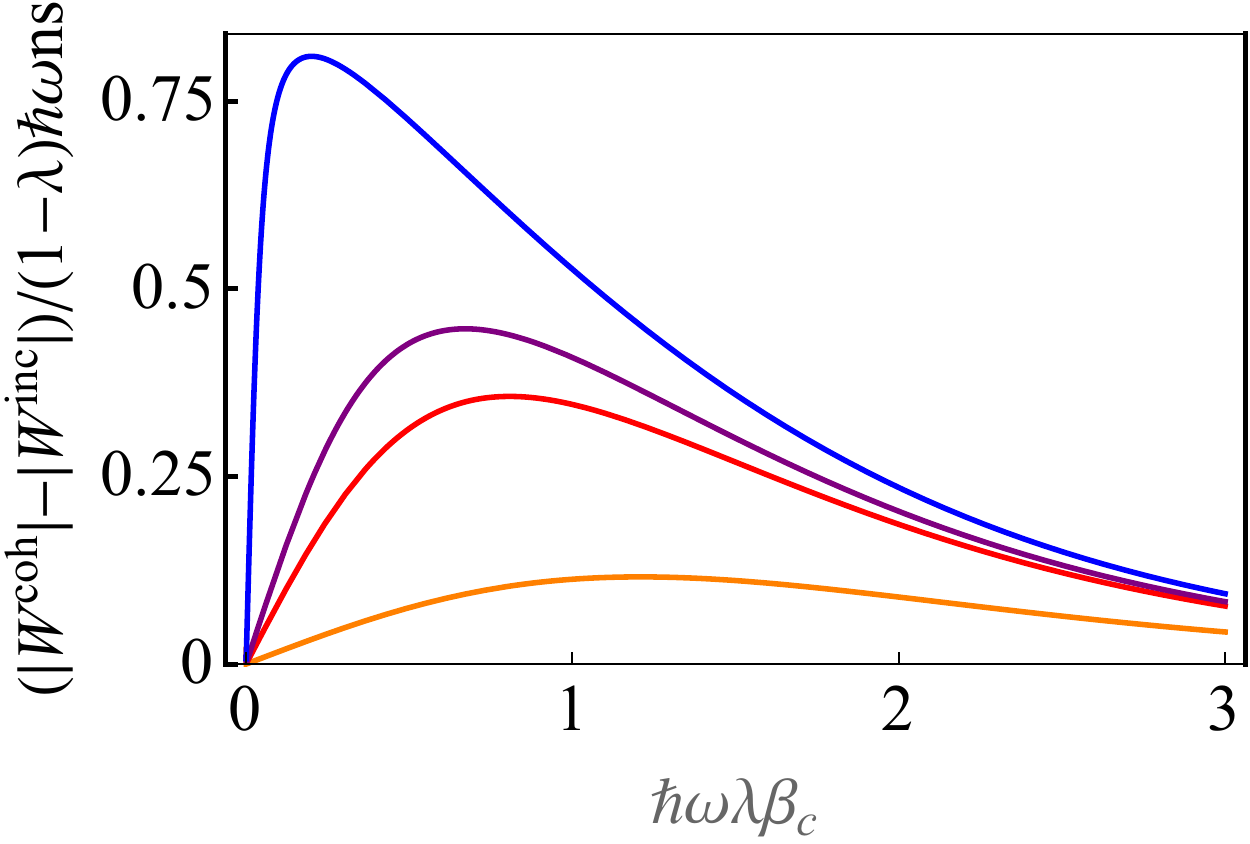}
(d)~~~~~\includegraphics[width=7cm, height=4.5cm]{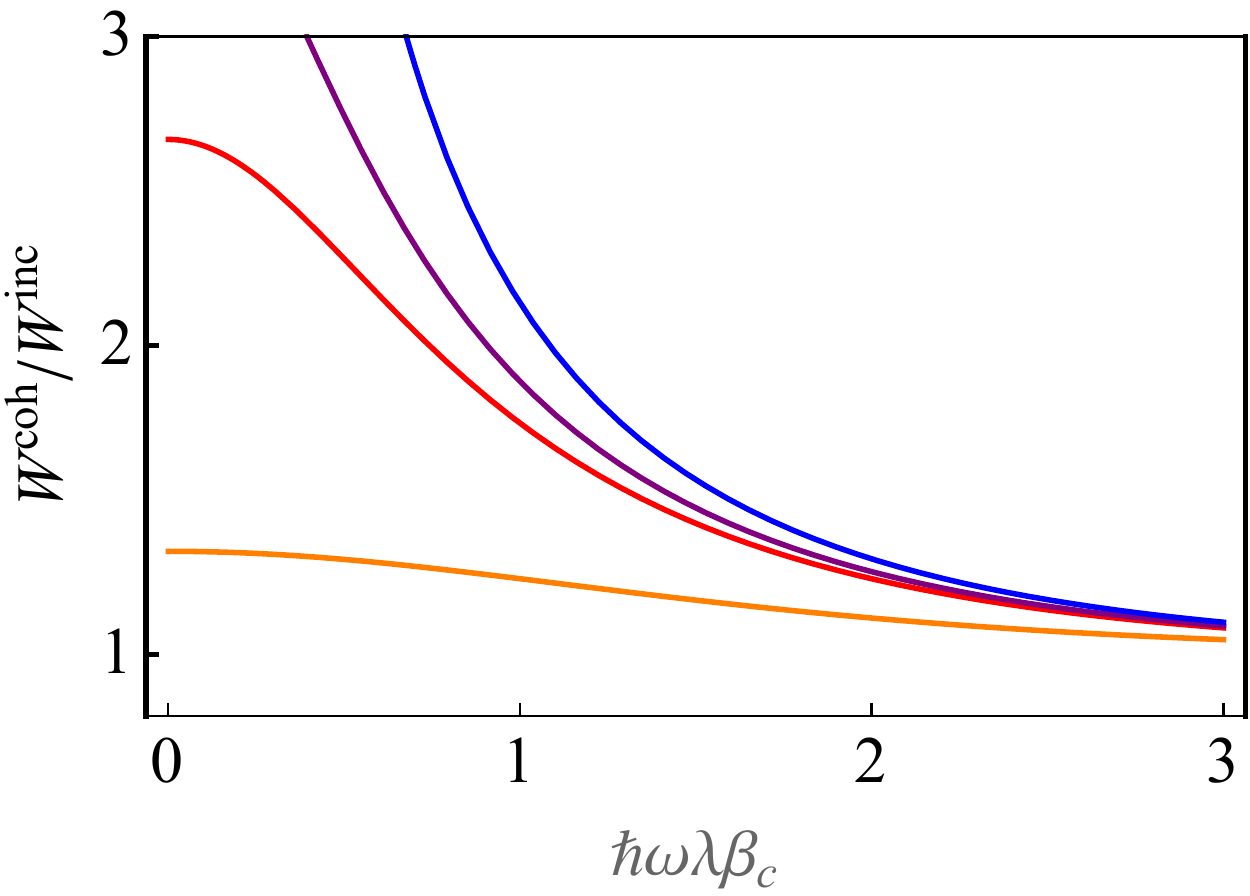}
\caption{Plots of (a) $|W^{\rm coh}| -|W^{\rm inc}|$ (normalised by $(1-\lambda)ns\hbar\omega$) and (b) $W^{\rm coh}/W^{\rm inc}$ as functions of $\hbar\omega\lambda\beta_c$, for $\beta_h=0$ and ensembles of $n=4$ spins of size $s=1/2$ (orange curve), $s=3/2$ (red curve), and $s=9/2$ (purple curve). Plots of (c) $|W^{\rm coh}| -|W^{\rm inc}|$ (normalised by $(1-\lambda)ns\hbar\omega$) and (d) $W^{\rm coh}/W^{\rm inc}$ as functions of $\hbar\omega\lambda\beta_c$, for $\beta_h=0$ and ensembles containing $n=2$ (orange curve), $n=6$ (red curve), $n=9$ (purple curve), and $n=100$ (blue curve), spins of size $s=1/2$. All the curves corresponds to ensembles of spins (or two-level atoms) initially in a thermal state at inverse temperature $\beta_0$ before the engine started to operate.   }
\label{workdiff}
\end{figure}
The central question is therefore: can we have $|W^{\rm coh}|>|W^{\rm inc}|$?
We show analytically in Appendix \ref{appeffamplification} that indeed one can have $|W^{\rm coh}|>|W^{\rm inc}|$ for $\h\omega|\beta_0| \gg 1$, when $\beta_h$ and $\lambda\beta_c$ are chosen in the interval $[0;\beta_l]$ (and such that $\lambda\beta_c>\beta_h$). The limit inverse temperature $\beta_l$ is strictly positive and depends on $n$ and $s$. 
 Alternatively, the analytical proof in Appendix \ref{appeffamplification} can simply be seen graphically in Fig. \ref{enVSns}. Choosing adequately $\lambda\beta_c$ and $\beta_h$, one can see on both curves \ref{enVSns} (a) and (b) that $E^{\infty}_{\beta_0}(\beta_h)-E^{\infty}_{\beta_0}(\lambda\beta_c)> E^{\rm th}(\beta_h)-E^{\rm th}(\lambda\beta_c)$, implying $|W^{\rm coh}|>|W^{\rm inc}|$. It also appears clearly that the range of temperatures leading to an indistinguishability-enhanced work extraction depends on $n$ and $s$. Conversely, a bad choice of $\beta_h$ and $\lambda\beta_c$ leads to a reduction of the work extracted by the indistinguishable spins, illustrating that enhancements are not systematic and require careful analysis. 
Importantly, we also show in Appendix \ref{appeffamplification} that indistinguishability-enhanced work extraction is not limited to $\h\omega|\beta_0| \gg1$. Even for moderate or small value of $|\beta_0|$, indistinguishability-induced enhancements can still be obtained.

As an illustration, we consider the largest 
enhancements, obtained for $\hbar\omega\beta_h\ll1 \ll\h\omega|\beta_0|$. 
 Fig. \ref{workdiff} presents the plots of $|W^{\rm coh}| -|W^{\rm inc}|$ (normalised by $(1-\lambda)ns\hbar\omega$) as a function of $\lambda\beta_c$, for $\beta_h=0$. The plot \ref{workdiff} (a) contains the curves for $n=4$ and $s=1/2$, $s=3/2$, and $s=9/2$. The plot \ref{workdiff} (c) contains the curves for $s=1/2$ and $n=2$, $n=6$, $n=9$, and $n=100$. 
 The maximum of the curves, corresponding to the maximal difference of work extraction between indistinguishable and distinguishable spins, is attained for $\lambda\beta_c=\beta_l$,
and we have
\be
\frac{|W^{\rm coh}|-|W^{\rm inc}|}{ 1-\lambda} \underset{ns\gg1}{\rightarrow}\h\omega ns
\ee
 (see also Fig. \ref{workdiff} (a) and (c)). Interestingly, the trade-off between efficiency (determined by $\lambda$) and the engine's power is also illustrated in Fig. \ref{workdiff} (a) and (c). In particular, for $\lambda$ going to $0$ the engine's power decrease slower for indistinguishable spins, and in the regime of larger $\lambda\beta_c$, increasing the efficiency (i.e. decreasing $\lambda$) can increase the indistinguishability-induced enhancement.   
 
 Furthermore, using \eqref{enplusexp2} 
one can show that the ratio of the two extracted works tends to
\be
 \frac{W^{\rm coh}}{W^{\rm inc}} \underset{\h\omega\beta_c\ll1}{\rightarrow} \frac{ns+1}{s+1},
\ee
 which appears also in Fig. \ref{workdiff} (b) and (d).  
Note that increasing both $n$ or $s$ increase the indistinguishability-induced enhancement.  
Additionally, one can see from Fig. \ref{workdiff} that for $\beta_h=0$, any value of $\beta_c$ leads to indistinguishability-induced enhancements. However, for $\beta_h>0$, the range of $\beta_c$ yielding enhancements is finite. 
 In particular, still for $\beta_h>0$, there is a threshold for the value of $ns$ beyond which indistinguishability leads only to smaller work extractions, showing again that indistinguishability-induced enhancements are not systematic and require a detailed analysis. 
 
 One should note that we did not mention and study equilibration speed-up emerging from collective effects \cite{Kloc_2019}. Taking into consideration the reduction of time period of each cycle one can obtain additional power increases. We also did not consider baths with negative temperatures (as in \cite{Assis_2018}). A hot bath with negative temperature is expected to bring larger enhancements. 
 
 Additionally, we mention that  using the same cyclic machine but with different parameter (namely $\lambda \leq \beta_h/\beta_c$), similar enhancements as the one presented above can be obtained for refrigeration operations. 
 
\begin{figure}
\centering
\includegraphics[width=7cm, height=4.5cm]{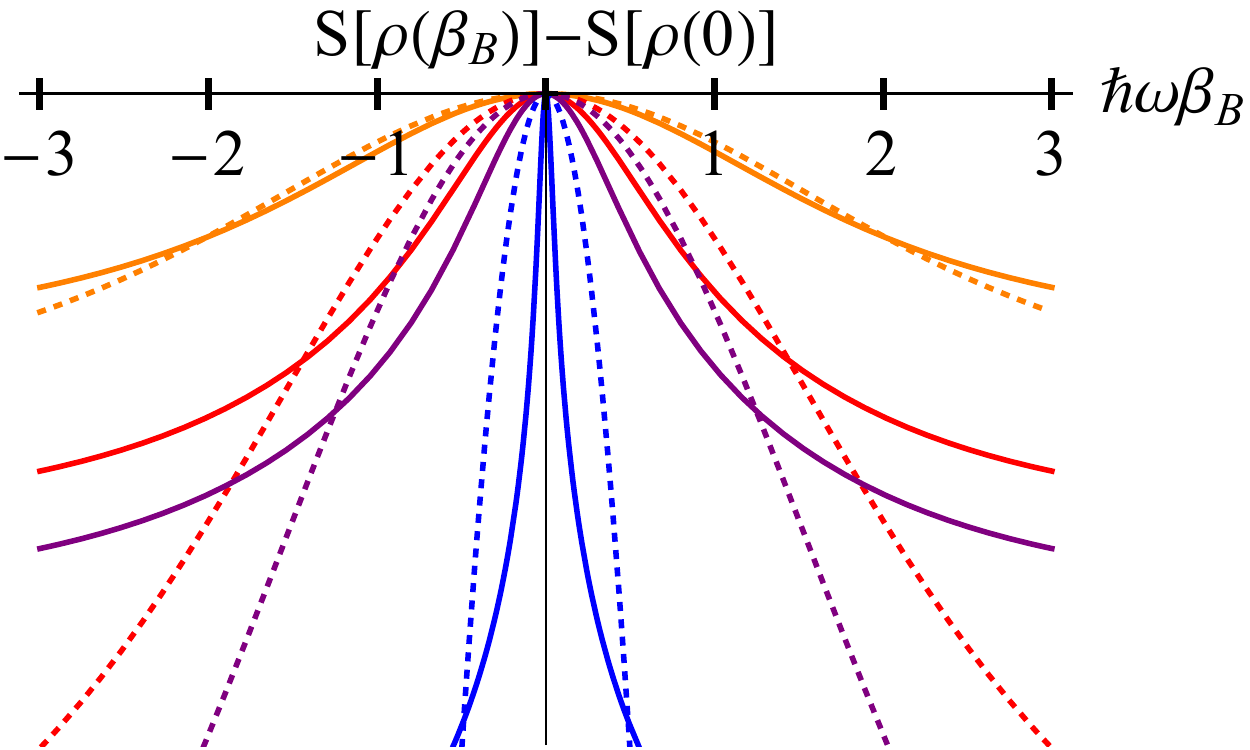}
\caption{Plots of the entropy difference $S[\rho_{\beta_0}^{\infty}(\beta_B)]-S[\rho_{\beta_0}^{\infty}(0)]$ (full lines) and $S[\rho^{\rm th}(\beta_B)]-S[\rho^{\rm th}(0)]$ (dotted lines) as functions of $\hbar\omega\beta_B$ for $\hbar\omega|\beta_0|\gg1$ and for ensembles containing $n=2$ (orange curve), $n=6$ (red curve), $n=9$ (purple curve), and $n=100$ (clue curve) spins $s=1/2$. For a given $n$, the combined mitigation effects lead to an overall amplification of the baths' action when $\beta_h$ and $\beta_c$ are both in the region where the full curve is below the dotted curve.
 }
\label{entdiff}
\end{figure}

 We can formalised the claim of effective amplification of the baths effect emerging from combined mitigation effects. The free energy variation per cycle is 
\be
\Delta F_{\rm cyc}^{\rm coh} = \left(\frac{1}{\beta_c}-\frac{1}{\beta_h}\right)\{S[\rho_{\beta_0}^{\infty}(\beta_h)]-S[\rho_{\beta_0}^{\infty}(\lambda\beta_c)]\}\leq0,
\ee
for collective coupling, and 
\be
\Delta F_{\rm cyc}^{\rm inc} = \left(\frac{1}{\beta_c}-\frac{1}{\beta_h}\right)\{S[\rho^{\rm th}(\beta_h)]-S[\rho^{\rm th}(\lambda\beta_c)]\}\leq0,
\ee
for independent coupling. Then, the combined action of the baths results in an amplified  action when $|\Delta F_{\rm cyc}^{\rm coh} |>|\Delta F_{\rm cyc}^{\rm inc} |$, which happens for 
\be\label{condamplif}
S[\rho_{\beta_0}^{\infty}(\beta_h)]-S[\rho_{\beta_0}^{\infty}(\lambda\beta_c)] >S[\rho^{\rm th}(\beta_h)]-S[\rho^{\rm th}(\lambda\beta_c)].
\ee
 In a similar way as for the steady state energy, one can see analytically through \eqref{entplusexp2} and \eqref{entthexp} or graphically in Fig. \ref{entdiff} that the later condition \eqref{condamplif} is always verified for $\beta_h$ and $\beta_c$ small enough.

The core mechanism of these indistinguishability-induced enhancements stems from the strong dependence of magnitude of the mitigation effects on the bath temperature. Even though the steady state energy of both isochoric strokes is reduced (mitigation effects), the magnitude of the reduction at the end of the second stroke (driven by the cold bath) can be much larger than the magnitude of the reduction of the end of the first isochoric stroke (driven by the hot bath). This  results in an enhanced energy difference yielding an enhanced extracted work. Thus, the fact that the cycle goes through two different steady states $\rho_1$ and $\rho_2$  is essential. It is not obvious how this indistinguishability-induced enhancement would survive in a continuous engine architecture where the working medium interacts simultaneously with both hot and cold bath and tends to a (single) steady state \cite{Niedenzu_2018}. This is an indication that the mechanism presented here is different in nature from the one in place in \cite{Niedenzu_2018}. Moreover, the power enhancement suggested here stems from a steady state effect, which is itself related to bath-induced coherences (in the local basis) as shown in \cite{bathinducedcohTLS} for a pair of two-level systems and extended in Appendixes \ref{roleofbathinducedcoh}, \ref{smapptempss}, and \ref{roleofcoh} for ensemble of $n$ spins of size $s$.
 Thus, it is not obvious whether such phenomena have a classical analogue (see also discussion in the next Section \ref{secmoreapp}). 
By contrast, the result from \cite{Niedenzu_2018} stems from superradiance, which is a dynamical effect. Moreover, classical analogues of superradiance can be found (for instance several classical emitters in phase) \cite{Gross_1982,Nefedkin_2016}. 
In conclusion, this suggests that indistinguishability-induced power enhancement relies on mechanisms unexploited so far and with probably no classical analogue.

\subsection{More applications}\label{secmoreapp}
In addition to the application detailed in the previous section, we mention briefly other operations which might benefit from the mitigation and amplification effects introduced in Sections \ref{mainenergy}, \ref{secentropy}, and \ref{secfreeen}.

First, the energetic amplification effect described in Section \ref{mainenergy} can yield a precious enhancement in a context of storing work in an ensemble of quantum batteries. Indeed, from Fig. \ref{enVSns} (b) and Fig. \ref{energyratiocomp} (b) one can expect an increase of up to $100\%$ of stored energy when using an ensemble of quantum batteries made indistinguishable from the point of view of the baths (typically two baths in order to realise an effective negative temperature bath \cite{Brunner_2012,autonomousmachines}). This phenomenon can be investigated in more details using for instance the versatile framework introduced in \cite{autonomousmachines}. Note that this is different from the design detailed in the previous section where it was the working medium itself which was composed of many subsystems. 
We also expect promising applications in the slightly different context of quantum battery charging \cite{Campaioli_2017, Ferraro_2018,Campaioli_2018} (via unitary operations).


In addition to that, there is an active debate \cite{Uzdin_2015, Andolina_2018,Gonzalez_2018,Niedenzu_2018} around whether performance enhancements stemming from collective effects are genuinely quantum or not. 
We believe our results can give a valuable contribution to this debate.  
In particular, the mitigation and amplification of the bath's effects introduced here rely on bath-induced coherences in the local basis as mentioned in the previous Section \ref{seceffectiveamp}.  
 Moreover, whereas constructive interferences of classical emitters can reproduce some aspects of superradiance \cite{Gross_1982,Nefedkin_2016}, it is not obvious how such interference effects would affect the steady state energy of the emitters.  
Therefore, one crucial question is whether the bath amplification and mitigation can have a classical analogue. 


 Regarding the reduction of irreversibility shown in Section \ref{secfreeen}, it can be of great value to reduce the entropy production of dissipative processes, but also of thermal machines, which is expected to lead to an increase of performances (efficiency and power) \cite{Feldmann_2006,Plastina_2014,Barato_2015,Gingrich_2016,Pietzonka_2016,Pietzonka_2018, Guarnieri_2019,Timpanaro_2019,Su_2019,Holubec_2018b}. 
 Additionally, it suggests that similar benefits stemming from collective effects could also happen for driven systems, opening interesting perspectives to reduce unwanted entropy production and frictions in diverse situations. 
 Moreover, it rises the question of the role of coherences between degenerate energy levels in the production of entropy. It would be interesting also to have a closer look at how this is related to coherence-induced reversibility reported in \cite{Uzdin_2016}.


The amplified cooling happening when the spin ensemble is initially in an inverse temperature $\beta_0< -\beta_B$ (with $\beta_B\geq0$) can lead to a reduction of the steady state energy and entropy by a factor up to $1/n$. 
If ensembles of large number $n$ of spins can be made indistinguishable, this amplified cooling can become a valuable cooling technique. It might also be of interest in algorithmic cooling \cite{Boykin_2002} and recent extensions \cite{Alhambra_2019}.  

Furthermore, we also showed that the mitigation of the bath effects becomes stronger when the number of spins increases. Then, an other interesting application can be to maintain a spin ensemble (much) colder than the bath, essential in many fields like quantum error correction and computation. 
 Assuming for instance that the ensemble is initially in a state colder than the bath, the mitigation effect can 
 maintain it in an energy and entropy up to $n$ times smaller than the thermal equilibrium energy and entropy.

Other applications might come up in other thermodynamical problems, light-harvesting devices \cite{Scully_2010,Svidzinsky_2011,Svidzinsky_2012,Dorfman_2013,Creatore_2013,
 Romero_2014,Killoran_2015,Xu_2015,Su_2016,Chen_2016,Romero_2017} using for instance superabsoption \cite{Brown_2018},
 but also possibly 
in quantum biology \cite{Lambert_2013}, particularly in light-harvesting complexes \cite{Collini_2010,Lloyd_2011,Lambert_2013,Huelga_2013,Dorfman_2013,Chin_2013,Krisnanda_2018}.  
 The non-thermality of the local states (for $s\geq 1$) shown in Section \ref{subseclocstate} is a valuable resource \cite{Brandao_2013,autonomousmachines} which can find interesting applications in some thermodynamic or computational tasks.

 
Finally, one can draw an interesting parallel with the conclusions of \cite{Muller_2018}. Let us consider a system $A$ initially in a state $\rho_A$ that we want to bring to a state $\rho_{A'}$ and having for that access to an ancillary system $M$ and a thermal bath (using the notation of \cite{Muller_2018}). One of the conclusions of \cite{Muller_2018} is that allowing correlations between $A$ and $M$ to build up reduces the constraints on energy (work) that must be invested in order to realise a given transformation. Here, we can see a similar effects. Within the spin ensemble, we single out one spin that we consider our ``main system'' of interest $A$ while the remaining spins are considered as an ancillary system $M$. If, for instance, $A$ is initially in an inverted population state and one is interested in cooling $A$ using a cold thermal bath, the simple fact of allowing correlations between $A$ and $M$ to build up increases the performance of the cooling, or alternatively relaxes the requirements on the cold bath. 
In this sense, the bath-induced coherences can be seen as catalysts \cite{Aberg_2014,Muller_2018}. Moreover, similarly as in \cite{Muller_2018}, the larger the ancillary system $M$, the larger the benefit. It would be interesting to continue this comparison and to see if our results could bring new aspects related to coherences to the results of \cite{Muller_2018}.

\section{Concluding remarks}

The phenomenon of collective dissipation relies on the indistinguishability of the subsystems (here spins or two-level atoms) from the point of view of the bath \cite{bathinducedcohTLS}. 
 Such indistinguishability is in general not present naturally but can be engineered  \cite{Devoe_1996, Hama_2018, Niedenzu_2018}, for instance by introducing an ancillary system between the spin ensemble and the bath (like an optical cavity \cite{Raimond_1982, Wood_2014, Wood_2016, Barberena_2019}), erasing part of the information ``seen'' by the bath.

 The requirement of non-interacting spins can be lifted for a pair of spins 
 as in this situation the interactions do not break the spin-exchange symmetry. For larger ensemble, there is typically a trade-off between non-interaction which can be obtained by dilution (spins far apart) and indistinguishability. As just mentioned, bath engineering, by introducing for instance an ancillary system, can help in overcoming this trade-off. Moreover, the experiments \cite{Raimond_1982, Devoe_1996, Barberena_2019} are indications that the effects reported throughout this study could indeed be achieved experimentally. Alternatively, if the spins are spatially arranged so that the coupling between each pair of spin is the same (for instance in ring configuration \cite{Gross_1982}), there is no symmetry breaking and our results should still hold. Additionally, 
 we show in Appendix \ref{appnoisy} that weak perturbations 
 (spin inhomogeneities or spin-spin interactions much smaller than the bath coupling) do not prevent the emergence of mitigation and amplification effects. However, for long times, such weak perturbations are expected to destroy these effects.  
This is not an issue for most applications envisioned in Section \ref{applications} as long as the mitigation and amplification effects emerge, even if temporarily. 
These predictions on the impact of weak perturbations coincide with the mathematical analysis done in \cite{Merkli_2015}. 
For stronger imperfections, involving energies at least of the order of $g^2\tau_c$ (where $\tau_c$ stands for the bath correlation time), the treatment in Appendix \ref{appnoisy} is not valid. Alternative methods should be used to investigate the survival of the mitigation and amplification effects. This is left for future research.

Note additionally that the above phenomena are not limited to spins. One can expect similar results for ensemble of harmonic oscillators. However, ensembles of indistinguishable harmonic oscillators do not seem to be common, limiting the applications. 

Finally, this study provides an overview of some stunning consequences of collective dissipation, analysing the effects on the main thermodynamical quantities, including energy, entropy, free energy variation and entropy production. The studied systems are ensembles containing an arbitrary number $n$ of spins of arbitrary size $s$ (or of two-level atoms), which are assumed to be initially in a thermal state, arguably the most widespread and experimentally accessible state. From the point of view of the spin ensemble energy, the collective dissipation results in an amplification ($\beta_0/\beta_B<-1$, where $\beta_0$ is the initial inverse temperature of the ensemble, and by $\beta_B$ the bath inverse temperature) or mitigation ($\beta_0/\beta_B>-1$) of the bath's action. This can be understood as a consequence of bath-induced coherences in the local basis (as detailed in Appendix \ref{roleofbathinducedcoh}). 
These amplification and mitigation effects grow with the number of spins in the ensemble, attaining considerable levels (see Fig. \ref{enVSns} and \ref{energyratiocomp}), and the size of the spins boost these effects. 
By contrast, in terms of entropy, free energy variation and entropy production, the action of the bath is always mitigated. Still, the combination of two baths at different temperatures can result in an overall amplification of their action while their individual action is mitigated. The emergence of such effective amplification in cyclic thermal machines can bring very large power enhancements (by a factor up to $(ns+1)/(s+1)$ in ideal conditions, see Section \ref{seceffectiveamp}). 
 %
Potential additional applications (detailed in Section \ref{applications}) stemming from these mitigation and amplification effects  include mainly collective work extraction, quantum batteries charging, cooling operations, and state protection. 
It also rises interesting questions around the production of entropy and its interplay with coherences.

%

 We hope our results will incentivise more research around the consequences and potential beneficial effects of collective interaction and indistinguishability, striving for realisable, scalable and sustainable quantum technologies. \\

\acknowledgements
This  work  is  based  upon  research  supported  by  the
South  African  Research  Chair  Initiative, Grant No. UID 64812 of  the  Department  of  Science  and  Technology of the Republic of South Africa and  National  Research Foundation of the Republic of South Africa.

\appendix
\numberwithin{equation}{section}

\section{Steady state}\label{steadystate}
As mentioned in the main text, each eigenspace of ${\cal J}^2$ is stable under $J^{\pm}$. Consequently, assuming no initial correlation between eigenspaces of ${\cal J}^2$, the dynamics remains confined in each eigenspace.  
  Therefore, within each eigenspace of ${\cal J}^2$, 
   the corresponding dynamics is the same as the relaxation of a non-degenerate system of $2J+1$ levels. The well-known equilibrium state is a thermal distribution (namely weighted by the Boltzmann factors) of the energy (here $J_z$) eigenstates \cite{Petruccione_Book}.
  One can see it directly from the master equation \eqref{me}, which yields for the dynamics of the populations and coherences $\rho_{m,m'}:=\!~_i\bra J,m|\rho|J,m'\ket_i$,   
\begin{widetext}
\bea\label{cohdyn}
\dot \rho_{m,m'} &=& -\Big[\Gamma(\omega) (J+m)(J-m+1) + \Gamma^*(\omega)(J+m')(J-m'+1)\nn\\
&& \hspace{0.5cm}+ \Gamma(-\omega)(J-m)(J+m+1) + \Gamma^*(-\omega)(J-m')(J+m'+1)\Big] \rho_{m,m'}\nn\\
&&+ G(\omega) \sqrt{(J-m)(J+m+1)(J-m')(J+m'+1)}\rho_{m+1,m'+1} \nn\\
&&+ G(-\omega)\sqrt{(J+m)(J-m+1)(J+m')(J-m'+1)}\rho_{m-1,m'-1},
\eea
\end{widetext}
with $G(\omega):= \Gamma(\omega)+\Gamma^*(\omega)$. One can verify that the steady state is given  
\bea
&&\rho_{m,m'}^{\infty} = 0, {\rm if ~ m\ne m'},\nn\\
&&\rho_{m,m}^{\infty} = e^{-m\hbar\omega\beta_B}\rho_{0,0}=e^{-\hbar(J+m)\hbar\omega\beta_B}\rho_{-J,-J},
\eea
where $\beta_B$ is the inverse temperature (or inverse apparent temperature) of the bath, which can be of arbitrary sign as mentioned in the main text after the master equation Eq.\eqref{me}. Since the sum of the populations is constant, $\sum_{m=-J}^{J} \dot \rho_{m,m} = \dot p_{J,i} = 0$ (as one can verifies directly from \eqref{cohdyn}), the steady state populations can be rewritten as, 
\be
\rho_{m,m}^{\infty} = p_{J,i}e^{-m\hbar\omega\beta_B}/Z_J(\beta_B),
\ee 
where
\bea\label{SMzj}
Z_J(\beta_B)&:=&\sum_{m=-J}^J e^{-m\hbar\omega\beta_B}\nn\\
&=&e^{J\hbar\omega\beta}\frac{1-e^{-(2J+1)\hbar\omega\beta}}{1-e^{-\hbar\omega\beta}},
\eea
and $p_{J,i}:= \sum_{m=-J}^J\!\!~_i\bra J,m |\rho_0|J,m\ket_i$. Then, the steady state restricted to the eigenspace $J,i$ is \be
 p_{J,i} \rho_{J,i}^{\rm th}(\beta_B)=p_{J,i}Z_{J}(\beta_B)^{-1}\sum_{m=-J}^J e^{- m\hbar\omega\beta_B}|J,m\ket_i\bra J,m|,
\ee
 as announced in \eqref{thstji} so that the steady state of the spin ensemble is $\rho^{\infty}(\beta_B) := \sum_{J=J_0}^J \sum_{i=1}^{l_J} p_{J,i} \rho_{J,i}^{\rm th}(\beta_B)$ as announced in \eqref{ss} of the main text.

\section{Generalisation of Eq. \eqref{ss}}\label{geneq7}
We assume here that the spin ensemble is initially in an arbitrary state. As mentioned in the main text, this initial state can be decomposed onto the collective basis $\{|J,m\ket_i\}$, $J_0\leq J\leq ns$, $-J\leq m\leq J$, $1\leq i\leq l_J$. The following reasoning is based on an unraveling view of the dissipation process, where the ensemble follows a quantum trajectory composed of jumps corresponding to absorption and emission of excitations. 
Each time there is an excitation absorbed from the bath or emitted to the bath, all components of the initial state gain or lose one excitation. 
Components with $m=\pm J$ cannot absorb or lose excitations and thus disappear. We ``follow'' the trajectory of a coherence between two arbitrary states $|J,m\ket_i$ and $|J',m'\ket_i'$. The coherence is initially preserved for the first absorptions and emissions if $m\ne \pm J$ and $m'\ne \pm J$ since both states gain or lose excitations simultaneously. 
 After a few absorptions/emissions, our pair of states reaches a stage where one of the two states cannot absorb/emit anymore while the other can (if $J\ne J'$). Then, if the absorption/emission happens, one of the two states gains/loses an excitation but the other disappears, and the coherence is destroyed. Similarly, if one considers a coherence between two levels $|J,m\ket$ and $|J,m'\ket$, with both $m$ and $m'$ in the interval $[-J+1;J-1]$ and $m\ne m'$, the same reasoning shows that the coherence will be destroyed by the bath interaction, which coincides with the well-known and established fact that a spin $J$ relaxes to the thermal state when interacting with a thermal bath. 
 However, the above reasoning is not valid for coherences between degenerate spin components like $|J,m\ket_i$ and $|J,m\ket_{i'}$ with $i\ne i'$. It is therefore possible that such kind of coherences survive the dissipation and affect the steady states. Such situation is left for further research.

\section{Derivative of $E_{\beta_0}^{\infty}(\beta_B)$}\label{derivativeE}
 In this Section we determine the sign of the derivative of $E_{\beta_0}^{\infty}(\beta_B)$ with respect to $\beta_0$ depending on the value of $\beta_0$ and $\beta_B$. We have,
\bea
\frac{\partial }{\partial \beta_0} E_{\beta_0}^{\infty}(\beta_B) = \sum_{J=J_0}^{ns} l_J \frac{\partial}{\partial \beta_0} p_J(\beta_0) e_J(\beta_B).
\eea
One can easily verify that 
\be\label{derpj}
\frac{\partial}{\partial \beta_0} p_J(\beta_0) = -p_J(\beta_0)[e_J(\beta_0) - E^{\rm th}(\beta_0)+\hbar\omega ns]
\ee
 so that 
\bea
&&\frac{\partial }{\partial \beta_0} E_{\beta_0}^{\infty}(\beta_B)\nn\\
&& = -\sum_{J=J_0}^{ns} l_J  p_J(\beta_0) [e_J(\beta_0) - E^{\rm th}(\beta_0)+\hbar\omega ns] \nn\\
&&\hspace{0.3cm}\times e_J(\beta_B)\label{c3}\\
&& = -\sum_{J=J_0}^{ns} l_J  p_J(\beta_0) \label{c4}\\
&&\hspace{0.3cm}\times \Big\{e_J(\beta_0) - \sum_{J'=J_0}^{ns} l_{J'}  p_{J'}(\beta_0) e_{J'}(\beta_0)\Big\} e_J(\beta_B)\nn\\
&& = -\sum_{J=J_0}^{ns}\sum_{J'=J_0}^{ns} l_J l_{J'}  p_J(\beta_0)p_{J'}(\beta_0) \label{c5}  \\
&&\hspace{0.3cm}\times \Big[e_J(\beta_0) -    e_{J'}(\beta_0)\Big] e_J(\beta_B)\nn\\
&& = -\sum_{J>J'}^{ns} l_J l_{J'}  p_J(\beta_0)p_{J'}(\beta_0)  \\
&&\hspace{0.3cm}\times [e_J(\beta_0) -    e_{J'}(\beta_0)] [e_J(\beta_B)-e_{J'}(\beta_B)],\nn
\eea
where we used \eqref{then} from \eqref{c3} to \eqref{c4} and the identity $\sum_{J'=J_0}^{ns} l_{J'} p_{J'}(\beta_0) = 1$ from \eqref{c4} to \eqref{c5}.
We need to determine the sign of $[e_J(\beta) -    e_{J'}(\beta)] $ depending on the value of $\beta$. This can be done through the identity
 \be\label{appendiff}
 e_J(\beta) -    e_{J'}(\beta) = -\frac{\partial}{\partial \beta} \ln \frac{Z_J(\beta)}{Z_{J'}(\beta)}.
 \ee
  From the expression \eqref{zj} we obtain $\frac{Z_J(\beta)}{Z_{J'}(\beta)} = \frac{\sinh(J+1/2)\omega\beta}{\sinh(J'+1/2)\omega\beta}$ which, for $J>J'$, is a strictly decreasing function of $\beta$ on the interval $]-\infty;0[$ and strictly increasing function on $]0;+\infty[$ (the derivative canceling at the point $\beta=0$). Therefore, we deduce that $e_J(\beta) -    e_{J'}(\beta) $ is strictly positive for $\beta \in ]-\infty;0[$, strictly negative for $\beta \in ]0;+\infty[$, and equal to 0 for $\beta=0$. Consequently, the derivative $\frac{\partial}{\partial \beta_0} E^{\infty}_{\beta_0}(\beta_B) $ is strictly negative if and only if $\beta_0\beta_B >0 $, strictly positive if and only if $\beta_0\beta_B<0$, and equal to zero if and only if $\beta_0\beta_B=0$.

\section{Local state}\label{applocstate}
In this Section we show that the local state of each spin, denoted by $\rho_{\rm Loc}$, is not a thermal state for ensembles with local spins $s$ larger or equal to $1$. Since the local state of each spin is the same we consider in the following the local state of the ``first'' spin. We denote by $p_{\rm Loc}(m_1):=~\!_1\bra s,m_1|\rho_{\rm Loc}|s,m_1\ket_1$ the population of the local state of the first spin with $m_1 \in [-s;s]$ (and $|s,m_1\ket_1$ is the eigenstate of the local operator $j_{z,1}$ associated to the eigenvalue $\h m_1$). We have
\bea
&&p_{\rm Loc}(m_1)=\nn\\
&&\sum_{m_2,...m_n=-s}^s\bra m_1,m_2,...,m_n|\rho_{\beta_0}^{\infty}(\beta_B)|m_1,m_2,...,m_n\ket.\nn\\
\eea
The local state $\rho_{\rm Loc}$ is a thermal state if and only if 
\be\label{thermalcond}
\frac{p_{\rm Loc}(m_1+1)}{p_{\rm Loc}(m_1)}=\frac{ p_{\rm Loc}(m_1)}{p_{\rm Loc}(m_1-1)}
\ee
 for any $m_1 \in [-s+1;s-1]$.
The general expression of $p_{\rm Loc}(m_1)$ is to complex to conclude on the validity or invalidity of \eqref{thermalcond}. Therefore we consider the limit $\h\omega|\beta_0|\gg1$ in which the steady state is simplified to 
\bea
 \rho^{\infty}_{\beta_0}(\beta_B)& \underset{\h\omega|\beta_0|\gg1}{\rightarrow} & \rho^{\infty}_{\beta_0=\pm\infty}(\beta_B)\nn\\
&=&\!\!\!\!\! Z_{ns}^{-1}(\beta_B)\sum_{m=-ns}^{ns} e^{-\hbar\omega m\beta_B}|ns,m\ket\bra ns,m|.\nn\\
\eea
The states $|ns,m\ket$ are Dicke states \cite{Dicke,Gross_1982} which can be expressed in terms of the local basis as
\be
|ns,m\ket = \frac{1}{\sqrt{I_m}}\sum_{m_1+m_2+...+m_n=m} |m_1,m_2,...,m_n\ket,
\ee
where $I_m=\sum_{J = |m|}^{ns} l_J$ (represents the dimension of the eigenspace of $J_z$ associated to the eigenvalue $\hbar m$, or equivalently the number of combinations of $m_1$, $m_2$,...,$m_n$ which sum up to $m$). Consequently, 
\be
\bra m_1,...,m_n|ns,m\ket\bra ns,m|m_1,...,m_n\ket = \frac{1}{I_m}\delta_{m,m_1+...+m_n}
\ee
where $\delta$ denotes the Kronecker delta (equal to 1 if $m=m_1+...+m_n$ and 0 otherwise). The expression of $p_{\rm Loc}(m_1)$ is simplified to
\bea
p_{\rm Loc}(m_1) &=& \sum_{m_2,...,m_n=s}^s Z_{ns}^{-1}(\beta_B)\sum_{m'=-ns}^{ns} e^{-\hbar\omega m'\beta_B}\nn\\
&&\times\frac{1}{I_{m'}}\delta_{m',m_1+...+m_n}\nn\\
&=& Z_{ns}^{-1}(\beta_B)\sum_{m=-(n-1)s}^{(n-1)s}~\sum_{m_2+...+m_n=m}\nn\\ &&\times\sum_{m'=-ns}^{ns} e^{-\hbar\omega m'\beta_B}\frac{1}{I_{m'}}\delta_{m',m_1+m}\nn\\
&=& Z_{ns}^{-1}(\beta_B)\!\!\!\!\!\sum_{m=-(n-1)s}^{(n-1)s}\sum_{m_2+...+m_n=m}\!\!\!\!\!\frac{e^{-\hbar\omega (m+m_1)\beta_B}}{I_{m+m_1}}\nn\\
&=& Z_{ns}^{-1}(\beta_B)\sum_{m=-(n-1)s}^{(n-1)s}K_m\frac{e^{-\hbar\omega (m+m_1)\beta_B}}{I_{m+m_1}},\nn\\
\eea
where $K_m$ denotes the number of combinations of $m_2$,...,$m_n$, summing up to $m$ (the same as $I_m$ but for combinations of $n-1$ integers). Then, the condition fro thermality \eqref{thermalcond} is equivalent to
\bea
&&p_{\rm Loc}(m_1+1)p_{\rm Loc}(m_1-1)-p_{\rm Loc}^2(m_1)=\nn\\
&&Z_{ns}^{-2}(\beta_B)\sum_{m,m'=-(n-1)s}^{(n-1)s}K_mK_{m'}e^{-\h\omega(2m_1+m+m')\beta_B}\nn\\
&&\times\Big(\frac{1}{I_{m+m_1+1}I_{m'+m_1-1}}-\frac{1}{I_{m+m_1}I_{m'+m_1}}\Big) =0.
\eea
Since this should hold for any $\beta_B$, it means that \eqref{thermalcond} is equivalent to
\be\label{f8}
\sum_{m+m'=q}\!\!\!\!\!K_mK_{m'}\Big(\frac{1}{I_{m+m_1+1}I_{m'+m_1-1}}-\frac{1}{I_{m+m_1}I_{m'+m_1}}\Big) =0
\ee
for all $q$ in $[-2(n-1)s;2(n-1)s]$. In particular, for $q=2(n-1)s$, implying that $m=m'=(n-1)s$, and for $m_1=s-1$ (remembering that we assumed $s\geq 1$), the fulfillment of \eqref{f8} leads to 
\be\label{false}
I_{ns}I_{ns-2}=I_{ns-1}^2.
\ee
However the equality \eqref{false} is not true since $I_{ns}=1$, $I_{ns-1}=n$, and $I_{ns-2}=\left(\begin{matrix} n\\ 2 \end{matrix}\right)=n(n+1)/2$ (for any  $s\geq 1$). Therefore, \eqref{thermalcond} is not satisfied and the local state $\rho_{\rm Loc}$ is not a thermal state.

\section{Steady state entropy}\label{smentropycomp}
In this section we show that $\frac{\partial}{\partial \beta_0}S[\rho_{\beta_0}^{\infty}(\beta_B)]$ is strictly positive for all $\beta_B$ and for all $\beta_0<0$ and strictly negative for all $\beta_B$ and all $\beta_0>0$. We start re-writing the entropy in the following form $S[\rho_{\beta_0}^{\infty}(\beta_B)] = \sum_{J=J_0}^{ns}l_Jp_J(\beta_0)\big[S[\rho_{J}^{\rm th}(\beta_B)]-\ln p_J(\beta_0)\big]$. Using the identity \eqref{derpj} one obtains
\bea
&&\frac{\partial}{\partial \beta_0}S[\rho_{\beta_0}^{\infty}(\beta_B)] = -\sum_{J=J_0}^{ns}l_Jp_J(\beta_0)\nn\\
&&\times\Big\{[e_J(\beta_0)-E^{\rm th}(\beta_0)+\hbar\omega ns] \big[S[\rho_{J}^{\rm th}(\beta_B)]-\ln p_J(\beta_0)\big]\nn\\
&&\hspace{0.5cm} +[e_J(\beta_0)-E^{\rm th}(\beta_0)+\hbar\omega ns]\Big\}\nn\\
 &&= -\sum_{J=J_0}^{ns}l_Jp_J(\beta_0)\Big\{[e_J(\beta_0)-E^{\rm th}(\beta_0)+\hbar\omega ns] \nn\\
&&\hspace{2.5cm}\times\big[S[\rho_{J}^{\rm th}(\beta_B)]-\ln p_J(\beta_0)\big] \Big\},\nn\\
\eea
where the simplification is due to the identity \eqref{then}. In a similar way as in Appendix \ref{derivativeE}, one can rewrite the derivative in the following symmetrical form using again the identities \eqref{then} and $\sum_{J'=J_0}^{ns} l_{J'} p_{J'}(\beta_0) = 1$,
\bea
&&\frac{\partial}{\partial \beta_0}S[\rho_{\beta_0}^{\infty}(\beta_B)] = -\sum_{J=J_0}^{ns}l_Jl_{J'}p_J(\beta_0)p_{J'}(\beta_0) \nn\\
&&\hspace{0.5cm}\times[e_J(\beta_0)-e_{J'}(\beta_0)]\big[S[\rho_{J}^{\rm th}(\beta_B)]-\ln p_J(\beta_0)\big]\nn\\
&&=-\sum_{J>J'}l_Jl_{J'}p_J(\beta_0)p_{J'}(\beta_0)[e_J(\beta_0)-e_{J'}(\beta_0)] \nn\\
&&\hspace{0.5cm}\times\big[S[\rho_{J}^{\rm th}(\beta_B)]-S[\rho_{J'}^{\rm th}(\beta_B)]-\ln p_J(\beta_0)+\ln p_{J'}(\beta_0)\big]\nn\\
&&=-\sum_{J>J'}l_Jl_{J'}p_J(\beta_0)p_{J'}(\beta_0)[e_J(\beta_0)-e_{J'}(\beta_0)] \nn\\
&&\times\Bigg[\ln\frac{Z_J(\beta_B)}{Z_{J'}(\beta_B)}+\beta_B[e_J(\beta_B)-e_{J'}(\beta_B)]- \ln\frac{Z_J(\beta_0)}{Z_{J'}(\beta_0)}\Bigg].\nn\\
\eea
Identity \eqref{ent3} was used to obtain the last line. Remembering the identity \eqref{appendiff}, one can observe that $\beta_B[e_J(\beta_B)-e_{J'}(\beta_B)] = -\beta_B\frac{\partial}{\partial \beta_B} \ln \frac{Z_J(\beta_B)}{Z_{J'}(\beta_B)}$. Finally, using $\frac{Z_J(\beta)}{Z_{J'}(\beta)} = \frac{\sinh(J+1/2)\omega\beta}{\sinh(J'+1/2)\omega\beta}$, one can show that $\frac{\partial^2}{\partial \beta_B^2} \ln \frac{Z_J(\beta_B)}{Z_{J'}(\beta_B)}$ is positive for all $\beta_B$ and $J>J'$ (i.e. $\ln \frac{Z_J(\beta_B)}{Z_{J'}(\beta_B)}$ is a convex function). Consequently, the following inequality holds for all $\beta_B$ (still with $J>J'$),
\be
\ln\frac{Z_J(\beta_B)}{Z_{J'}(\beta_B)}+\beta_B[e_J(\beta_B)-e_{J'}(\beta_B)] \leq \ln\frac{Z_J(0)}{Z_{J'}(0)}.
\ee
It follows that 
\bea
&&\ln\frac{Z_J(\beta_B)}{Z_{J'}(\beta_B)}+\beta_B[e_J(\beta_B)-e_{J'}(\beta_B)]- \ln\frac{Z_J(\beta_0)}{Z_{J'}(\beta_0)}\nn\\
&&\leq \ln\frac{Z_J(0)}{Z_{J'}(0)}-\ln\frac{Z_J(\beta_0)}{Z_{J'}(\beta_0)}\nn\\
&& < 0,
\eea
for all $\beta_0\ne 0 $ since $ \ln \frac{Z_J(\beta_0)}{Z_{J'}(\beta_0)}$ is strictly decreasing on $]-\infty;0[$ and strictly increasing on $]0;+\infty[$ (as already mentioned in Appendix \ref{derivativeE}). Finally, since for $J>J'$, $e_J(\beta_0)-e_{J'}(\beta_0)>0$ for $\beta_0<0$ and $e_J(\beta_0)-e_{J'}(\beta_0)<0$ for $\beta_0>0$, one concludes as announced in the beginning of the appendix: $\frac{\partial}{\partial \beta_0}S[\rho_{\beta_0}^{\infty}(\beta_B)]$ is strictly positive for all $\beta_B$ and for all $\beta_0<0$ and strictly negative for all $\beta_B$ and all $\beta_0>0$.

\section{Mitigation of the free energy variation}\label{appfreeenergy}
The free energy variation for collective coupling is $\Delta F^{\infty}_{\beta_0}(\beta_B):=F[\rho^{\infty}_{\beta_0}(\beta_B)]-F[\rho^{\rm th}(\beta_0)]$ while its counterpart for independent coupling is $\Delta F^{\rm th}(\beta_B):=F[\rho^{\rm th}(\beta_B)]-F[\rho^{\rm th}(\beta_0)]$. A quick calculation using \eqref{entropygen} shows that the difference of free energy variations can be expressed as
\be
\Delta F^{\rm th}(\beta_B)-\Delta F^{\infty}_{\beta_0}(\beta_B) = \frac{1}{\beta_B}\sum_{J=J_0}l_Jp_J(\beta_0)\ln\frac{p_J(\beta_B)}{p_J(\beta_0)}.
\ee
The sign of the above quantity is not so obvious since the probabilities $p_J(\beta)$ are in general non-monotonic. We therefore compute its derivative with respect to $\beta_0$ and find,
\bea
&&\frac{\partial}{\partial \beta_0}\beta_B\sum_{J=J_0}^{ns} l_J p_J(\beta_0)\ln \frac{p_J(\beta_B)}{p_J(\beta_0)} \label{d1}\\
&&=-\beta_B\sum_{J=J_0}^{ns} l_J p_J(\beta_0)[e_J(\beta_0)-E^{\rm th}(\beta_0)+\hbar\omega ns]\nn\\
&&\hspace{0.3cm}\times\ln \frac{p_J(\beta_B)}{p_J(\beta_0)}\nn\\
&&\hspace{0.3cm}+\beta_B\sum_{J=J_0}^{ns} l_J p_J(\beta_0)[e_J(\beta_0)-E^{\rm th}(\beta_0)+\hbar\omega ns]\label{d2}\\
&&=-\beta_B\sum_{J=J_0}^{ns}\sum_{J'=J_0}^{ns} l_Jl_{J'} p_J(\beta_0)p_{J'}(\beta_0)\nn\\
&&\hspace{0.3cm}\times[e_J(\beta_0)-e_{J'}(\beta_0)]\ln \frac{p_J(\beta_B)}{p_J(\beta_0)}\nn\\
&&=-\beta_B\sum_{J>J'}^{ns}l_Jl_{J'} p_J(\beta_0)p_{J'}(\beta_0)\label{d3}\\
&&\hspace{0.3cm}\times[e_J(\beta_0)-e_{J'}(\beta_0)]\left[\ln \frac{p_J(\beta_B)}{p_J(\beta_0)}-\ln \frac{p_{J'}(\beta_B)}{p_{J'}(\beta_0)}\right]\nn\\
&&=-\beta_B\sum_{J>J'}^{ns}l_Jl_{J'} p_J(\beta_0)p_{J'}(\beta_0)\label{d4}\\
&&\hspace{0.3cm}\times[e_J(\beta_0)-e_{J'}(\beta_0)]\left[\ln \frac{Z_J(\beta_B)}{Z_{J'}(\beta_B)}-\ln \frac{Z_{J}(\beta_0)}{Z_{J'}(\beta_0)}\right].\nn
\eea
where we used \eqref{derpj} from \eqref{d1} to \eqref{d2}, and the definition of $p_J(\beta_0):=Z_J(\beta_0)/Z(\beta_0)$ from \eqref{d3} to \eqref{d4}. We saw in Appendix \ref{derivativeE} that for $J>J'$, $\frac{Z_{J}(\beta)}{Z_{J'}(\beta)}$ is a strictly decreasing function on $]-\infty;0[$ and strictly increasing on $]0;+\infty[$. Therefore, using the property $Z_J(-\beta) = Z_J(\beta)$, one can conclude that, for $\beta_B>0$, $\frac{\partial}{\partial \beta_0}\beta_B\sum_{J=J_0}^{ns} l_J p_J(\beta_0)\log \frac{p_J(\beta_B)}{p_J(\beta_0)}$ is strictly negative for $\beta_0 > \beta_B$ and $-\beta_B < \beta_0 <0$, and strictly positive on $0<\beta_0< \beta_B$ and $\beta_0<-\beta_B$. Since $\beta_B\sum_{J=J_0}^{ns} l_J p_J(\beta_0)\log \frac{p_J(\beta_B)}{p_J(\beta_0)}$ is equal to zero when $\beta_0=\pm\beta_B$, we reach the conclusion that 
\be
\Delta F^{\rm th}(\beta_B)<\Delta F^{\infty}_{\beta_0}(\beta_B) 
\ee
for all $\beta_0\ne \pm\beta_B$ and for all $\beta_B>0$. Conversely, for $\beta_B<0$, 
\be
\Delta F^{\rm th}(\beta_B)>\Delta F^{\infty}_{\beta_0}(\beta_B) 
\ee
for all $\beta_0\ne \pm\beta_B$. 
Finally, since the variation of free energy is always negative for $\beta_B>0$ and always positive for $\beta_B<0$, we obtain that the absolute value of the free energy variation is always strictly smaller for collective coupling (for $\beta_0\ne\pm\beta_B$)
\be
|\Delta F^{\infty}_{\beta_0}(\beta_B) |<|\Delta F^{\rm th}(\beta_B)|.
\ee


\section{Effective amplification}\label{appeffamplification}
In this section we show that we can have $|W^{\rm coh}| >|W^{\rm inc}|$ for a large range of bath temperatures $\beta_h$, $\beta_c$, and initial temperatures $\beta_0$ of the spin ensemble. Using the expressions of $-W^{\rm coh}$ and $-W^{\rm inc}$ in \eqref{wextcoh} and \eqref{wextinc}, respectively, one can see the work extracted by indistinguishable particles is larger than the one extracted by distinguishable particles if 
\bea
E^{\infty}_{\beta_0}(\beta_h)-E^{\infty}_{\beta_0}(\lambda\beta_c) >E^{\rm th}(\beta_h)-E^{\rm th}(\lambda\beta_c),
\eea
which happens if and only if the function $E^{\rm th}(\beta)-E^{\infty}_{\beta_0}(\beta)$ is a strictly growing function of $\beta$ for at least some intervals within $[\beta_h,\lambda\beta_c]$. 
To see when it can happen, we compute the derivative, 
\bea\label{derivative}
&&\frac{\partial}{\partial \beta} [E^{\rm th}(\beta)-E^{\infty}_{\beta_0}(\beta)] \nn\\
&&= \sum_{J=J_0}^{ns} p_J(\beta_0)l_J\frac{\partial}{\partial \beta}[n e_{s}(\beta) - e_J(\beta)]\nn\\
&&=\frac{\h^2\omega^2}{4\sinh^2 x} \sum_{J=J_0}^{ns} p_J(\beta_0)l_J\nn\\
&&\times\Big[n(2s+1)^2\frac{\sinh^2x}{\sinh^2(2s+1)x}-n \nn\\
&&\hspace{2.5cm}- (2J+1)^2\frac{\sinh^2x}{\sinh^2(2J+1)x}+1\Big],\nn\\
\eea
where we defined $x:=\hbar\omega\beta/2$ for convenience. The function $\frac{\sinh^2x}{\sinh^2(2J+1)x}$ is monotonic decreasing for $x\in[0;+\infty[$ (and monotonic increasing for $x\in]-\infty;0]$), taking the value $(2J+1)^{-2}$ in $x=0$ and going to $0$ for increasing $x$. Therefore, the derivative \eqref{derivative} is always negative for $|x|\gg1$. However, for $|x| \ll 1$,  
\bea\label{derivativeexp}
&&\frac{\partial}{\partial \beta} [E^{\rm th}(\beta)-E^{\infty}_{\beta_0}(\beta)]\nn\\
&&=\frac{\h^2\omega^2}{3}\sum_{J=J_0}^{ns} p_J(\beta_0)l_J(J^2+J-ns^2-ns)\nn\\
&&+ {\cal O}(\hbar^2\omega^2\beta^2)
\eea
which can become strictly positive for $|\beta_0|$ large enough. Indeed, for $\omega|\beta_0|\gg1$, $p_J(\beta_0)$ tends to 0 for $J<ns$ and to 1 for $J=ns$ so that the above derivative \eqref{derivativeexp} is strictly positive for any $n\geq 2$ and any $s$. Furthermore, the positivity of the derivative \eqref{derivativeexp} is not limited to large value of $|\beta_0|$. One can see that for $n\gg1$, the quantity $J^2+J-ns^2-ns$ is positive for $J \in [\sqrt{ns(s+1)};ns]$, so that for $\h\omega|\beta_0| \simeq 1$ or even $\h\omega|\beta_0| \ll1$ the derivative \eqref{derivativeexp} can remain strictly positive. This shows that the work extracted by indistinguishable spins can be strictly larger than the work extracted by distinguishable spins for a large range of initial temperatures $\beta_0$ (ultimately determined by $n$ and $s$).

In the following, aiming to analyse how large can be the indistinguishability-enhanced extracted work, we consider the most favourable situation which is $\h\omega|\beta_0|\gg1$. In this limit, $E_{\beta_0}^{\infty}(\beta)$ tends to $e_{ns}(\beta)$, so that the derivative \eqref{derivative} becomes
\bea\label{der+}
&&\frac{\partial}{\partial \beta} [E^{\rm th}(\beta)-E^{\infty}_{\beta_0}(\beta)] \nn\\
&&=\frac{\h^2\omega^2}{4\sinh^2 x} \Big[n(2s+1)^2\frac{\sinh^2x}{\sinh^2(2s+1)x}-n \nn\\
&&\hspace{2.5cm}- (2ns+1)^2\frac{\sinh^2x}{\sinh^2(2ns+1)x}+1\Big],\nn\\
\eea
which is positive for all $\beta \in [0;\beta_l]$, where $\beta_l$ is strictly positive and such that $\frac{\partial}{\partial \beta} [E^{\rm th}(\beta)-E^{\infty}_{\beta_0}(\beta)]_{|\beta_l}=0 $. The analytical expression of $\beta_l$ is challenging to obtain in general, but one can estimate it graphically.   
 For any $\beta_h$ and $\lambda\beta_c$ (such that $\lambda\beta_c>\beta_h$) belonging to the interval $[0;\beta_l]$, we have $E^{\infty}_{\beta_0}(\beta_h)-E^{\infty}_{\beta_0}(\lambda\beta_c) >E^{\rm th}(\beta_h)-E^{\rm th}(\lambda\beta_c)$ and consequently $|W^{\rm coh}| >|W^{\rm inc}|$. In particular, the largest indistinguishability-induced enhancement is obtained for $\beta_h=0$ and $\lambda\beta_c=\beta_l$, leading to
 \be
  {\rm max}\left[|W^{\rm coh}| - |W^{\rm inc}|\right] =(1-\lambda)[E^{\rm th}(\beta_l) - E^{\infty}_{+}(\beta_l)].
  \ee
Thus, $\beta_l$ corresponds to the point where the curves of $E^{\rm th}(\beta_B)$ and $E^{\infty}_{+}(\beta_B)$ are the most far apart, see Figs. \ref{enVSns} and \ref{workdiff}.


\section{Role of the bath-induced coherences}\label{roleofbathinducedcoh}
In the previous study on a pair of two-level systems \cite{bathinducedcohTLS} it was possible to explicitly relates the steady state energy to the amount of bath-induced coherences in the local basis. In the present situation however such direct relation is much more complex to exhibit. Nevertheless, one can still pinpoint bath-induced coherences (between states of the local basis) as responsible for the dramatic alteration of the steady state energy. Note that we are not claiming that coherences between degenerate energy levels contribute directly to the internal energy of the spin ensemble, but instead that such coherences prevent the ensemble from reaching the thermal equilibrium energy. 
This is the aim of this Section.

As shown in Section \ref{sectionsteadys}, the steady states reached under collective dissipation are convex combinations of highly coherent states. Indeed, all eigenstates $|J,m\ket_i$ (expect $|J=ns,m=\pm ns\ket$) are {\it coherent superpositions} of the local basis states $|m_1,...,m_n\ket$. Then, expect for very particular initial conditions, namely $\beta_0=\beta_B$, the steady state $\rho_{\beta_0}^{\infty}(\beta_B)$ contains coherences (in the local basis). 
These coherences are global in the sense that locally each spin remains in a diagonal state. This can be seen in the following way. We denote by $\rho_1:={\rm Tr}_{2...n}\rho_{\beta_0}^{\infty}(\beta_B)$ the reduced density operator of the spin 1, where ${\rm Tr}_{2...n}$ stands for the partial trace over all other spins, from 2 to $n$. One can see that for $m_1 \ne m_1'$ there is no local coherence between $|m_1\ket$ and $|m_1'\ket$,
\bea
&&\bra m_1|\rho_1|m_1'\ket \nn\\
&&= \sum_{m_2,...,m_n} \bra m_1, m_2,...,m_n|\rho_{\beta_0}^{\infty}(\beta_B) |m_1',m_2,...,m_n\ket \nn\\
&&=0.
\eea
We use the fact that $\rho_{\beta_0}^{\infty}(\beta_B)$ is a mixture of states $|J,m\ket_i\bra J,m|$ which contain components only in the eigenspace $m$ (subspace spanned by the eigenstates of $J_z$ associated to the eigenvalue $m$) whereas $|m_1,m_2,...,m_n\ket$ and $|m_1',m_2,...,m_n\ket$ do not belong to 
the same eigenspaces. This implies that the coherences contained in the steady state are global and can be seen alternatively as correlations between spins since the global state of the spin ensemble cannot be written as a tensor product of local density operators. 

 At first sight it might appear contradictory that global coherences (or correlations) could be responsible for the alteration of the steady state energy -- since they do not contribute to the spins energy. How does it work? 
As a preliminary observation, the heat exchanges between the spin ensemble in a non-dark state $\rho$ and the reservoir are characterised by the apparent temperature \cite{apptemp} of the spin ensemble defined by
\be\label{defapptemp}
{\cal T} :=\omega \left(\ln\frac{{\rm Tr}J^{-}J^{+}\rho}{{\rm Tr}J^{+}J^{-}\rho}\right)^{-1}.
\ee
We recall that if the spin ensemble is in a dark state it does not interact with the bath and therefore there is no heat flow and no apparent temperature can be defined.
 Moreover, a necessary condition for the spin ensemble to be in a steady state is to have an apparent temperature equal to the bath temperature $1/\beta_B$ (otherwise the heat flow is not null). Indeed, one can verify (Appendix \ref{smapptempss}) that all states of the form \eqref{ss} have an apparent temperature equal to $1/\beta_B$ (they are all steady states of specific initial conditions). 

When the spin ensemble is initially in a thermal state at extreme temperatures $\h\omega|\beta_0| \gg1$, the distribution $p_J(\beta_0)$ tends to be concentrated in $J=ns$ (as already mentioned in Section \ref{mainenergy}), $p_{J=ns}(\beta_0)  \underset{\h\omega |\beta_0| \gg 1}{\simeq} 1$ and $p_{J<ns}(\beta_0)  \underset{\h\omega |\beta_0| \gg 0}{\simeq} 1$. In other words the steady state $\rho^{\infty}_{\beta_0}(\beta_B)$ is mostly made of the eigenstates $\{|J=ns,m\ket\}_{-ns\leq m\leq ns}$, which are Dicke states \cite{Dicke,Gross_1982} (totally symmetric states). Such states contains only positive global coherences, or correlations. As shown in \cite{apptemp}, positive correlations increase the apparent temperature of ensembles when the underlying diagonal state (in the natural basis) has a positive apparent temperature, and decrease the apparent temperature otherwise. Then, one can conclude that the positive correlations contained in the steady state $\rho^{\infty}_{\beta_0}(\beta_B)$  increase the apparent temperature of the spin ensemble when $\beta_B>0$, while they decrease the apparent temperature when $\beta_B<0$. For this reason, when $|\beta_0|>\beta_B>0$ ($-|\beta_0|<\beta_B<0$) the spin ensemble is able to reach an apparent temperature equal to $1/\beta_B$ while having lower (higher) populations of high energy levels than $\rho^{\rm th}(\beta_B)$, implying lower (higher) energy than the thermal energy $E^{\rm th}(\beta_B)$. 
To strengthen this argument we show explicitly in Appendix \ref{roleofcoh} that, for $\beta_B>0$ ($\beta_B<0$), the apparent temperature of the steady state without coherences, denoted by $\rho_{\beta_0|_D}^{\infty}(\beta_B)$, is strictly lower (larger) than the apparent temperature of $\rho_{\beta_0}^{\infty}(\beta_B)$ (equal to $1/\beta_B$), confirming that the coherences within $\rho_{\beta_0}^{\infty}(\beta_B)$ increase (decrease) the apparent temperature. Therefore, the bath-induced coherences appear as a crucial ingredient for the emergence of steady states with energy different from the thermal energy $E^{\rm th}(\beta_B)$.

We show for $\omega|\beta_0| \ll1$ that the core mechanism for the alteration of the steady state energy is the bath-induced coherences. The demonstration for arbitrary $\beta_0$ is more involved. It requires in particular to compute sums with the Clebsch-Gordan coefficients \cite{Sakurai_Book} related to the sum of $n$ spins $s$. This is a challenging task that we left for further research. Nevertheless, we can give some qualitative explanations for finite $\beta_0$. Roughly speaking, when $|\beta_0|$ decreases, the decomposition of the steady state $\rho^{\infty}_{\beta_0}(\beta_B)$ onto the global basis contains less and less components from the Dicke states $\{|J=ns,m\ket\}_{-ns\leq m\leq ns}$ and more and more from components from states of lower $J$. Such states $|J,m\ket$ with $J<ns$ contains less positive coherence than the Dicke states, and can even contain negative coherences. Therefore, when $|\beta_0|$ decreases, the coherence of the steady state together with the deviation from the thermal equilibrium energy decrease. 
When the point $|\beta_0|=|\beta_B|$ is reached, the distribution of $|J,m\ket_i\bra J,m|$ contained in the steady state becomes balanced (in terms of $J$, for fixed $m$) 
\bea
&&\sum_{J=|m|}^{ns} p_J(\beta_0=\beta_B)\frac{e^{-\hbar\omega m\beta_B}}{Z_J(\beta_B)}|J,m\ket_i\bra J,m|\nn\\
&&=\frac{e^{-\hbar\omega m\beta_B}}{Z(\beta_B)}\sum_{J=|m|}^{ns} |J,m\ket_i\bra J,m|,
\eea
 and all coherences cancel out, implying that the steady state energy is equal to the thermal equilibrium energy.
 Beyond this point, when $|\beta_0|<|\beta_B|$, the coherences within the steady state are negative. This has the opposite impact on the energy, namely the steady state energy is lowered for $\beta_B>0$ and increased for $\beta_B>0$.

\section{Apparent temperature of the steady states}\label{smapptempss}
In this Section we show that all states of the form \eqref{ss} have an apparent temperature equal to the bath temperature $1/\beta_B$. We recall that the apparent temperature of the spin ensemble in a non-dark state $\rho$ is defined by \eqref{defapptemp}, ${\cal T} :=\omega \left(\ln\frac{{\rm Tr}J^{-}J^{+}\rho}{{\rm Tr}J^{+}J^{-}\rho}\right)^{-1}$ as introduced in \cite{apptemp} (if the spin ensemble is in a dark state it does not interact with the bath and therefore no apparent temperature can be defined).  
We first show that all states $\rho_{J,i}^{\rm th}(\beta_B)$ defined in \eqref{thstji} have indeed an apparent temperature equal to $1/\beta_B$. We start with,
\bea
&&{\rm Tr}\rho_{J,i}^{\rm th}(\beta_B)J^{+}J^{-} \nn\\
&&= Z_J(\beta_B)^{-1}\sum_{m=-J}^Je^{-\hbar\omega m\beta_B}~_i\bra J,m|J^{+}J^{-}|J,m\ket_i\nn\\
&&= Z_J(\beta_B)^{-1}\sum_{m=-J}^Je^{-\hbar\omega m\beta_B}(J+m)(J-m+1)\nn\\
&&= Z_J(\beta_B)^{-1}\sum_{m=-J-1}^{J-1}e^{-\hbar\omega (m+1)\beta_B}(J+m+1)(J-m)\nn\\
&&= Z_J(\beta_B)^{-1}e^{-\hbar\omega\beta_B}\sum_{m=-J}^{J-1}e^{-\hbar\omega m\beta_B}(J+m+1)(J-m)\nn\\
&&= Z_J(\beta_B)^{-1}e^{-\hbar\omega\beta_B}\sum_{m=-J}^{J}e^{-\hbar\omega m\beta_B}(J+m+1)(J-m)\nn\\
&&=e^{-\hbar\omega \beta_B} {\rm Tr}\rho_{J,i}^{\rm th}(\beta_B)J^{-}J^{+},\label{identityapptemp}
\eea
which once inserted in the expression of the apparent temperature \eqref{defapptemp} yields an apparent temperature equal to $1/\beta_B$ for $\rho_{J,i}^{\rm th}(\beta_B)$ (if ${\rm Tr}\rho_{J,i}^{\rm th}(\beta_B)J^{+}J^{-}\ne0$, namely if $\rho_{J,i}^{\rm th}(\beta_B)$ is not a dark state). Inserting the identity \eqref{identityapptemp} in the expression of the apparent temperature for $\rho^{\infty}(\beta_B)$ we have
\bea
{\cal T}^{\infty} &:=& \hbar\omega \left(\ln\frac{{\rm Tr}J^{-}J^{+}\rho^{\infty}(\beta_B)}{{\rm Tr}J^{+}J^{-}\rho^{\infty}(\beta_B)}\right)^{-1}\nn\\
&=& \hbar \omega \left(\ln\frac{\sum_{J=J_0}^{ns}\sum_{i=1}^{l_J}p_{J,i}{\rm Tr}J^{-}J^{+}\rho_{J,i}^{\rm th}(\beta_B)}{\sum_{J=J_0}^{ns}\sum_{i=1}^{l_J}p_{J,i}{\rm Tr}J^{+}J^{-}\rho_{J,i}^{\rm th}(\beta_B)}\right)^{-1}\nn\\
=  \hbar&\omega& \left(\ln\frac{\sum_{J=J_0}^{ns}\sum_{i=1}^{l_J}p_{J,i}{\rm Tr}J^{-}J^{+}\rho_{J,i}^{\rm th}(\beta_B)}{\sum_{J=J_0}^{ns}\sum_{i=1}^{l_J}p_{J,i}e^{-\hbar\omega\beta_B}{\rm Tr}J^{-}J^{+}\rho_{J,i}^{\rm th}(\beta_B)}\right)^{-1}\nn\\
&=& 1/\beta_B,
\eea
which is the result announced in Appendix \ref{roleofbathinducedcoh}.

\section{Apparent temperature without bath-induced coherences}\label{roleofcoh}
In this Section we explicitly show that for $\h\omega|\beta_0|\gg1$, the apparent temperature of $\rho^{\infty}_{\beta_0|_D}(\beta_B)$, the steady state without coherences in the local basis (obtained from $\rho^{\infty}_{\beta_0}(\beta_B)$ by canceling all non-diagonal elements in the local basis), is strictly lower than the bath temperature $1/\beta_B$ if $\beta_B>0$, and strictly larger than $1/\beta_B$ if $<\beta_B<0$. We recall that 
\bea
 \rho^{\infty}_{\beta_0}(\beta_B)& \underset{\h\omega|\beta_0|\gg1}{\rightarrow} & \rho^{\infty}_{\beta_0=\pm\infty}(\beta_B)\nn\\
&=&\!\!\!\!\! Z_{ns}^{-1}(\beta_B)\sum_{m=-ns}^{ns} e^{-\hbar\omega m\beta_B}|ns,m\ket\bra ns,m|.\nn\\
\eea
As already mention in Appendix \ref{applocstate}, the states $|ns,m\ket$ are Dicke states \cite{Dicke,Gross_1982} which can be expressed in terms of the local basis as
\be
|ns,m\ket = \frac{1}{\sqrt{I_m}}\sum_{m_1+m_2+...+m_n=m} |m_1,m_2,...,m_n\ket,
\ee
where $I_m=\sum_{J = |m|}^{ns} l_J$. In the following we simplify slightly the notation by using $\rho_{+}^{\infty}(\beta_B):=\rho_{\beta_0=\pm \infty}^{\infty}(\beta_B)$. We have,
\bea
\rho^{\infty}_{\beta_0|_D}(\beta_B)& \underset{\h\omega|\beta_0|\gg1}{\rightarrow} &
\rho^{\infty}_{+|_D}(\beta_B):=\rho_{\beta_0=\pm \infty|_D}^{\infty}(\beta_B)\nn\\
&=&\sum_{m_1=-s}^s...\sum_{m_n=-s}^s \nn\\
&&\bra m_1,...m_n|\rho^{\infty}_{+}(\beta_B)|m_1,...,m_n\ket \nn\\
&&\times |m_1,...,m_n\ket \bra m_1,...,m_n|\nn\\
&\underset{\h\omega|\beta_0|\gg1}{=}&\sum_{m_1=-s}^s...\sum_{m_n=-s}^s \sum_{m=-ns}^{ns}\frac{e^{-\hbar\omega m\beta_B}}{Z_{ns}(\beta_B)} \nn\\
&&\bra m_1,...m_n|ns,m\ket\bra ns,m|m_1,...,m_n\ket \nn\\
&&\times |m_1,...,m_n\ket \bra m_1,...,m_n|\nn\\
&=& Z_{ns}^{-1}(\beta_B) \sum_{m=-ns}^{ns} ~\sum_{m_1+...+m_n=m} \nn\\
&&\frac{e^{-\hbar\omega m\beta_B}}{I_m}|m_1,...,m_n\ket \bra m_1,...,m_n|.\nn\\
\eea
In order to obtain the apparent temperature of the state $\rho^{\infty}_{+|_D}(\beta_B)$ we need to calculate ${\rm Tr}J^{+}J^{-}\rho^{\infty}_{+|_D}(\beta_B)$ and ${\rm Tr}J^{-}J^{+}\rho^{\infty}_{+|_D}(\beta_B)$, which can be done as follows,
\bea\label{trj+j-}
&&{\rm Tr}J^{+}J^{-}\rho^{\infty}_{+|_D}(\beta_B) =  Z_{ns}^{-1}(\beta_B) \sum_{m=-ns}^{ns}  \frac{e^{-\hbar\omega m\beta_B}}{I_m}\nn\\
&&\times\sum_{m_1+m_2+....+m_n=m} \bra m_1,m_2,...,m_n|J^{+}J^{-}|m_1,m_2,...,m_n\ket.\nn\\
\eea
We can identify 
\be
\sum_{m_1+m_2+....+m_n=m} \bra m_1,m_2,...,m_n|J^{+}J^{-}|m_1,m_2,...,m_n\ket
\ee
as the trace of $J^{+}J^{-}$ restricted to the eigenspace $m$ (eigenspace of $J_z$ associated to the eigenvalue $\hbar m$). Therefore, the trace does not depend on the choice of the basis and can be calculated alternatively in the basis $\{|J,m\ket_i\}_{|m|\leq J\leq ns, 1\leq i\leq l_J}$, yielding
\bea
&&{\rm Tr}J^{+}J^{-}\rho^{\infty}_{+|_D}(\beta_B) \nn\\
&&=  Z_{ns}^{-1}(\beta_B) \sum_{m=-ns}^{ns}  \frac{e^{-\hbar\omega m\beta_B}}{I_m}\sum_{J=|m|}^{ns} \sum_{i=1}^{l_J}~_i\bra J,m|J^{+}J^{-}|J,m\ket_i\nn\\
&&=  Z_{ns}^{-1}(\beta_B) \sum_{m=-ns}^{ns}  \frac{e^{-\hbar\omega m\beta_B}}{I_m}\sum_{J=|m|}^{ns} l_J(J+m)(J-m+1)\nn\\
&&=  Z_{ns}^{-1}(\beta_B) \sum_{J=J_0}^{ns}\sum_{m=-J}^J  \frac{e^{-\hbar\omega m\beta_B}}{I_m} l_J(J+m)(J-m+1).\nn\\
\eea
Using a similar procedure we obtain
\bea
&&{\rm Tr}J^{-}J^{+}\rho^{\infty}_{+|_D}(\beta_B) \nn\\
&&=  Z_{ns}^{-1}(\beta_B) \sum_{J=J_0}^{ns}\sum_{m=-J}^J  \frac{e^{-\hbar\omega m\beta_B}}{I_m} l_J(J-m)(J+m+1),\nn\\
\eea
which can be rewritten as (following the same steps as in \eqref{identityapptemp})
\bea
&&{\rm Tr}J^{-}J^{+}\rho^{\infty}_{+|_D}(\beta_B) \nn\\
&&=  Z_{ns}^{-1}(\beta_B) \sum_{J=J_0}^{ns}\sum_{m=-J+1}^{J+1}  \frac{e^{-\hbar\omega (m-1)\beta_B}}{I_{m-1}}\nn\\
&&\hspace{4cm}\times l_J(J-m+1)(J+m),\nn\\
&&= e^{\hbar\omega \beta_B} Z_{ns}^{-1}(\beta_B) \sum_{J=J_0}^{ns}\sum_{m=-J}^{J}  \frac{e^{-\hbar\omega m\beta_B}}{I_{m-1}}\nn\\
&&\hspace{4cm} \times l_J(J+m)(J-m+1).\nn\\
\eea
Then, one can see that unless $I_{m-1} = I_m$ for all $m$ in $[-ns+1;ns]$, ${\rm Tr}J^{-}J^{+}\rho^{\infty}_{+|_D}(\beta_B)  \ne e^{\hbar\omega\beta_B} {\rm Tr}J^{+}J^{-}\rho^{\infty}_{+|_D}(\beta_B) $. Since $l_{J=ns}=1$ and $l_{J=ns-1}=n-1$, we have that $I_{ns}=1$ and $I_{ns-1}=n$, which confirms that for any ensemble of at least 2 spins the apparent temperature ${\cal T}_D = \h\omega \left(\ln \big[{\rm Tr}J^{-}J^{+}\rho^{\infty}_{+|_D}(\beta_B)/{\rm Tr}J^{+}J^{-}\rho^{\infty}_{+|_D}(\beta_B)\big] \right)^{-1}$ of $\rho_{+|_D}^{\infty}(\beta_B)$ is different from $1/\beta_B$. We can go further and show that ${\cal T}_D$ is strictly smaller (larger) than $1/\beta_B$ for $\beta_B >0$ ($\beta_B<0$). This is equivalent to 
\bea\label{smineq}
&&e^{\hbar\omega/{\cal T}_D} = \frac{\bra J^{-}J^{+}\ket_{\rho^{\infty}_{+|_D}(\beta_B)}}{\bra J^{+}J^{-}\ket_{\rho^{\infty}_{+|_D}(\beta_B)}} > e^{\hbar\omega\beta_B}\nn\\
&&\Leftrightarrow \bra J^{-}J^{+}\ket_{\rho^{\infty}_{+|_D}(\beta_B)}-e^{\hbar\omega\beta_B}\bra J^{+}J^{-}\ket_{\rho^{\infty}_{+|_D}(\beta_B)} >0\nn\\
&&\Leftrightarrow \sum_{J=J_0}^{ns} \sum_{m=-J+1}^J e^{-\hbar\omega m\beta_B}l_J(J+m)(J-m+1)\nn\\
&&\hspace{2cm}\times\left(\frac{1}{I_{m-1}}-\frac{1}{I_m}\right) > 0.\nn\\ 
\eea
We can see explicitly that the above inequality is always verified for $\beta_B>0$ by making the transformation $m \rightarrow -m$ in the last line of \eqref{smineq}, so that the above condition becomes,
 \bea
&&\sum_{J=J_0}^{ns} \sum_{m=-J}^{J-1} e^{\hbar\omega m\beta_B}l_J(J-m)(J+m+1)\nn\\
&&\hspace{2cm}\times\left(\frac{1}{I_{-m-1}}-\frac{1}{I_{-m}}\right) > 0\nn\\
&&\Leftrightarrow \sum_{J=J_0}^{ns} \sum_{m=-J+1}^J e^{\hbar\omega (m-1)\beta_B}l_J(J-m+1)(J+m)\nn\\
&&\hspace{2cm}\times\left(\frac{1}{I_{-m}}-\frac{1}{I_{-m+1}}\right) > 0.\label{F10}
\eea
Now we sum together \eqref{smineq} and \eqref{F10}, and using the  symmetry property $I_{-m}=I_m$ one obtains the following condition
\bea\label{F11}
&&\sum_{J=J_0}^{ns} l_J \sum_{m=-J+1}^J (J+m)(J-m+1) \left(\frac{1}{I_{m-1}}-\frac{1}{I_m}\right)\nn\\
&&\times(e^{-\hbar\omega m\beta_B}-e^{\hbar\omega(m-1)\beta_B}) >0,
\eea
which is always verified for $\beta_B>0$ since $I_m = \sum_{J=|m|}^{ns} l_J$ so that $I_{m-1} = I_m+l_{m-1}> I_m$ for $m\geq1$ and $I_{m-1} < I_m=I_{m-1} +l_m$ for $m\leq 0$.

For $\beta_B<0$, the inequality \eqref{F11} is inverted so that the apparent temperature ${\cal T}_D$ is strictly larger than $1/\beta_B$. 
Note that for $\beta_B=0$, ${\cal T}_D = 1/\beta_B = \infty$.

\section{Inhomogeneities and interactions between subsystems}\label{appnoisy}
In this section we analyse the impact of small defects like homogeneities rising for instance from unequal subsystems or local disorder and variation of the subsystems' surrounding \cite{Tsyplyatyev_2009,Sun_2019}. We also consider small interactions between the subsystems. Then, the Hamiltonian of the ensemble of spins or two-level atoms is 
\be
H_A := H_0 + H_{\rm inh}+H_{\rm int}
\ee
where $H_0=\sum_k \h\omega j_{z,k}=\h\omega J_z$ is the noiseless Hamiltonian considered throughout the paper, $H_{\rm inh} = \sum_k \h\delta_k j_{z,k}$ represents the contributions from the inhomogeneities which result in different energy splitting $\h(\omega+\delta_k)$ for each subsystem $k$, and $H_{\rm int}=\sum_{k>l} \h\Omega_{k,l} (j_{+,k}j_{-,l}+j_{-,k}j_{+,l})$ is the interaction Hamiltonian between the subsystems of the ensemble, where $j_{\pm,k}:=j_{x,k}\pm ij_{y,k}$ are the local ladder operators of each subsystem. For atomic ensembles, $H_{\rm int}$ represents the Van der Waals interaction \cite{Gross_1982,Coffey_1977}.

Intuitively, one can consider that for small inhomogeneities and small interactions, the contributions of both $H_{\rm inh}$ and $H_{\rm int}$ manifests itself only for very long times. Then, if the dissipative dynamics induced by the bath happens on a smaller timescale (meaning that the bath coupling involves larger energies), the ensemble reaches its steady state before the appearance of inhomogeneities and interactions effects. Then, for larger times, inhomogeneities and interactions might affect the steady state (which is therefore not anymore the real steady state). Note that for most applications detailed in Section \ref{applications} (and in particular for the ``effective amplification'' \ref{seceffectiveamp}) the crucial point is to reach, even temporarily, the steady state $\rho_{\beta_0}^{\infty}(\eta_B)$. In the following we show rigorously that the above considerations indeed hold. 

The global evolution of the ensemble plus bath is given by the unitary evolution $U_t:=e^{-i(H_A + V+ H_B)t/\h}$. Using operator calculus formulas \cite{Feynman_1951} one can re-written the operator evolution as
\bea
U_t &=& e^{-iH_{\rm int}t/\h} e^{-\frac{i}{\h}{\cal T}\int_0^t du [H_0 + \tilde H_{\rm inh}(u) + \tilde V(u) +H_B]}\nn\\
\eea
where $\tilde {\cal O} (u): =e^{i H_{\rm int} u/\h} {\cal O} e^{-iH_{\rm int}u/\h}$ for any operator ${\cal O}$. Note that the form of the chosen interactions preserve the energy of the ensemble, $[H_0,H_{\rm int}]=0$, so that $\tilde H_0(u) = H_0$. 
Since $H_{\rm int}$ is Hermitian, it can be diagonalised. We denote by $\chi_i$ and $|\chi_i\ket$ its eigenvalues and eigenstates, respectively, so that we can re-write the interaction Hamiltonian as $H_{\rm int} = \sum_{i=1}^{s^n}\h\chi_i|\chi_i\ket\bra \chi_i|$. This leads to
\be
\tilde {\cal O}(u) = \sum_{i,j} e^{i(\chi_i-\chi_j)u}|\chi_i\ket\bra \chi_j| O_{i,j}
\ee
where $O_{i,j}:=\bra \chi_i|{\cal O}|\chi_j\ket$. Then, if we consider only times $t$ much smaller than $|\chi_i-\chi_j|^{-1}$ for all $i,j$, we can safely write $\tilde {\cal O}(t) = {\cal O}$ and therefore,
\bea\label{u1}
U_t &=& e^{-iH_{\rm int}t/\h} e^{-i [H_0 + H_{\rm inh}+ V+H_B]t/\h}.
\eea

The eigenvalues $\chi_i$ are functions of the coupling constant $\Omega_{k,l}$ so that $\chi_i$ is of the order of magnitude of the $\Omega_{k,l}$ (more precisely, of the order of magnitude of $\Omega_{k,l}$ times the average number of interacting neighbours each subsystem has). Denoting by $\Omega$ the order of magnitude of the interaction strength between subsystems, the above equality \eqref{u1} holds only for $t\ll \Omega^{-1}$.
We can repeat a similar operation with the Hamiltonian $H_{\rm inh}$,
\bea\label{u2}
U_t &=& e^{-iH_{\rm int}t/\h}e^{-iH_{\rm inh}t/\h}  e^{-\frac{i}{\h}{\cal T}\int_0^t du [H_0 + \bar V(u) +H_B]}\nn\\
\eea
 where $\bar V(u):= e^{i H_{\rm inh} u/\h} V e^{-iH_{\rm inh}u/\h}$. One can proceed in a similar way as for $H_{\rm int}$, or alternatively exploit the simple form of $H_{\rm inh}$ as follows. Having the bath coupling of the form $V =\sum_k g (j_{+,k}+j_{-,k})O_B$ (see also Section \ref{seccollective} of the main text), we obtain,
 \be
 \bar V(u) = \sum_k g(e^{i\delta_h u}j_{+,k} + e^{-i\delta_k u}j_{-,k})O_B.
 \ee
Thus, for times $t$ such that $t\ll \Omega^{-1} $ and $t\ll \delta^{-1}$ ($\delta$ denoting the order of magnitude of the inhomogeneities), we have simply
\be
U_t\underset{t\ll\Omega^{-1},\delta^{-1}}{=} e^{-iH_{\rm int}t/\h}e^{-iH_{\rm inh}t/\h}  e^{-i [H_0 + V +H_B]t/\h}.
\ee

The conclusion of these manipulations is that for times smaller than $\Omega^{-1}$ and $\delta^{-1}$ the effects of spins interactions and inhomogeneities are decoupled from the bath's action. In the rotating picture with respect to $H_{\rm int}$ and $H_{\rm inh}$ we recover the dynamics \eqref{me} considered throughout the paper. If $\Omega$ and $\delta$ are much smaller than $|\Gamma(\omega)| \simeq g^2 \tau_c$, where $\tau_c$ denotes the bath correlation time, the steady state $\rho_{\beta_0}^{\infty}(\beta_B)$ is reached before the effects of spin interactions and inhomogeneities start appearing. Similar behaviour was obtained under a rigorous mathematical analysis in \cite{Merkli_2015} for a quasi-degenerate three-level system. In particular, it was pointed out that, when the energy difference between the quasi degenerate levels is smaller than the coupling with the bath, the system reaches a quasi-stationary manifold (corresponding here to the family of states $\rho_{\beta_0}^{\infty}(\beta_B)$) before converging to the thermal equilibrium state on a long time scale (given by the inverse of the energy gap between the quasi-degenerate levels). 

Looking now at the energy of the ensemble $E_t={\rm Tr} H_0U_t \rho_0U_t^{\dag}$, since $[H_0,H_{\rm int}]=[H_0,H_{\rm inh}]=0$, we have $E_t = {\rm Tr} H_0 \rho_t^R$, where $\rho_t^R$ denotes the density operator of the ensemble in the rotating picture with respect to $H_{\rm int}$ and $H_{\rm inh}$. $E_t$ tends to $E_{\beta_0}^{\infty}(\beta_B)$ (Section \ref{mainenergy}) if $\rho_0$ is a thermal state at inverse temperature $\beta_0$. Therefore, under the conditions $\Omega \ll g^2\tau_c$ and $\delta \ll g^2\tau_c$, the mitigation and amplification effects still happen. However, for times much larger than $\Omega^{-1}$ and $\delta^{-1}$, spins interactions and inhomogeneities might start affecting these phenomena. 

Note that we did not take into account the equilibration speed-up emerging form collective effects \cite{Kloc_2019}. This implies that our rough estimate of the equilibration timescale $(g^2\tau_c)^{-1}$ is probably overestimating the actual equilibration time. This might relax the above conditions $\Omega \ll g^2\tau_c$ and $\delta \ll g^2\tau_c$ under which mitigation and amplification effects survive.

Summarising, taking into account small imperfections, the mitigation and amplification of the bath's action survive, at least temporarily, which is enough for most envisioned applications of Section \ref{applications}. For stronger imperfections, of the order of at least $g^2\tau_c$, the above treatment is not valid. Alternative methods have to be used to investigate the behaviour of the mitigation and amplification effects. This is left for future research.


\begin{thebibliography}{1} 
\bibitem{Dicke} RH. Dicke, {\it Coherence in Spontaneous Radiation Processes}, Phys. Rev. {\bf 93}, 99 (1954).
\bibitem{Gross_1982} M. Gross and S. Haroche, {\it Superradiance: An essay on the theory of collective spontaneous emission}, Physics Reports (Review Section of Physics Letters) {\bf 93}, 301-396 (1982).
\bibitem{Benatti_2003} F. Benatti, R. Floreanini, and M. Piani, {\it Environment Induced Entanglement in Markovian Dissipative Dynamics}, Phys. Rev. Lett. {\bf 91}, 070402 (2003).
\bibitem{Benatti_2010} F. Benatti, R. Floreanini, and U. Marzolino, {\it Entangling two unequal atoms through a common bath}, Phys. Rev. A {\bf 81}, 012105 (2010).
\bibitem{Passos_2018} M. H. M. Passos, W. F. Balthazar, A. Z. Khoury, M. Hor-Meyll, L. Davidovich, and J. A. O. Huguenin, {\it Experimental investigation of environment-induced entanglement using an all-optical setup}, Phys. Rev. A {\bf 97}, 022321 (2018).
\bibitem{bathinducedcohTLS}  C. L. Latune, I. Sinayskiy, F. Petruccione, {\it Energetic and entropic effects of bath-induced coherences}, Phys. Rev. A {\bf 99}, 052105 (2019).
\bibitem{Wang_2009} H. Wang, S. Liu, and J. He, {\it Thermal entanglement in two-atom cavity QED and the entangled quantum Otto engine}, Phys. Rev. E {\bf 79}, 041113 (2009).
\bibitem{Scully_2011} M. O. Scully, K. R. Chapin, K. E. Dorfman, M. B. Kim, and A. Svidzinsky, {\it Quantum heat engine power can be increased by noise-induced coherence}, Proc. Natl. Acad. Sci. {\bf 108}, 15097 (2011).
\bibitem{Gelbwaser_2015} D. Gelbwaser-Klimovsky, W. Niedenzu, P. Brumer, and G. Kurizki, {\it Power enhancement of heat engines via correlated thermalization in a three-level ``working fluid''}, Sci. Rep. {\bf 5}:14413 (2015).
\bibitem{Uzdin_2016} R. Uzdin, {\it Coherence-Induced Reversibility and Collective Operation of Quantum Heat Machines via Coherence Recycling}, Phys. Rev. Applied {\bf 6}, 024004 (2016).
\bibitem{Mehta_2017} V. Mehta and R. S. Johal, {\it Quantum Otto engine with exchange coupling in the presence of level degeneracy}, Phys. Rev. E {\bf 96}, 032110 (2017).
\bibitem{Cakmak_2017} B. {\c C}akmak, A. Manatuly, and \"{O}. E. M\"{u}stecapl\i o\v{g}lu, {\it Thermal production, protection, and heat exchange of quantum coherences}, Phys. Rev. A {\bf 96}, 032117 (2017).
\bibitem{Niedenzu_2018} W. Niedenzu and G. Kurizki, {\it Cooperative many-body enhancement of quantum thermal machine power}, New J. Phys. {\bf 20}, 113038 (2018).
\bibitem{Holubec_2018} V. Holubec and T. Novotn{\'y}, {\it Effects of noise-induced coherence on the performance of quantum absorption refrigerators}, J. Low Temp. Phys. {\bf 192}, 147-198 (2018). 
\bibitem{Hewgill_2018} A. Hewgill, A. Ferraro, and G. De Chiara, {\it Quantum correlations and thermodynamic performances of two-qubit engines with local and common baths}, Phys. Rev. A {\bf 98}, 042102 (2018).
\bibitem{Watanabe_2019} G. Watanabe, B.P. Venkatesh, P. Talkner, M.-J. Hwang, and A. del Campo, {\it Quantum Statistical Enhancement of the Collective Performance of Multiple Bosonic Engines}, arXiv:1904.07811.
\bibitem{Scully_2010} M. O. Scully, {\it Quantum Photocell: Using Quantum Coherence to Reduce Radiative Recombination and Increase Efficiency}, Phys. Rev. Lett. {\bf 104}, 207701 (2010).
\bibitem{Svidzinsky_2011} A. A. Svidzinsky, K. E. Dorfman, M. O. Scully, {\it Enhancing photovoltaic power by Fano-induced coherence}, Phys. Rev. A {\bf 84}, 053818 (2011).
\bibitem{Svidzinsky_2012} A. A. Svidzinsky, K. E. Dorfman, M. O. Scully, {\it Enhancing photocell power by noise-induced coherence}, Coherent Optical Phenomena {\bf 1}, 7-14 (2012). 
\bibitem{Dorfman_2013} K. E. Dorfman, D. V. Voronine, S. Mukamel, and M. O. Scully, {\it Photosynthetic reaction center as a quantum heat engine}, Proc. Natl. Acad. Sci. {\bf 110}, 2746–2751 (2013).
\bibitem{Creatore_2013} C. Creatore, M. A. Parker, S. Emmott, and A.W. Chin, {\it Efficient Biologically Inspired Photocell Enhanced by Delocalized Quantum States}, Phys. Rev. Lett. {\bf 111}, 253601 (2013).
\bibitem{Romero_2014} E. Romero, R. Augulis, V. I. Novoderezhkin, M. Ferretti, J. Thieme, D. Zigmantas and R. van Grondelle, {\it Quantum coherence in photosynthesis for efficient solar-energy conversion},  Nature Phys. {\bf 10}, 676-682 (2014).
\bibitem{Killoran_2015} N. Killoran, S. F. Huelga, and M. Plenio, {\it Enhancing light-harvesting power with coherent vibrational interactions: A quantum heat engine picture},  Journal of Chemical Physics {\bf 143}, 155102 (2015).
\bibitem{Xu_2015} D. Xu, C. Wang, Y. Zhao and J. Cao, {\it Polaron effects on the performance of light-harvesting systems: a quantum heat engine perspective}, New J. Phys. {\bf 18}, 023003 (2016).
\bibitem{Su_2016} S.-H. Su, C.-P. Sun, S.-W. Li, J.-C. Chen, {\it Photoelectric converters with quantum coherence}, Phys. Rev. E {\bf 93}, 052103 (2016).
\bibitem{Chen_2016} H.-B. Chen, P.-Y. Chiu, and Y.-N. Chen, {\it }, {\it Vibration-induced coherence enhancement of the performance of a biological quantum heat engine}, Phys. Rev. E {\bf 94}, 052101 (2016).
\bibitem{Romero_2017} E. Romero, V. I. Novoderezhkin and R. van Grondelle, {\it Quantum design of photosynthesis for bio-inspired solar-energy conversion}, Nature {\bf 543}, 355-365 (2017).
\bibitem{Brown_2018} W. M. Brown and E. M. Gauger, {\it Light-harvesting with guide-slide superabsorbing condensed-matter nanostructures}, arXiv:1803.08036.
\bibitem{Robentrost_2009} P. Rebentrost, M. Mohseni, I. Kassal, S. Lloyd and A. Aspuru-Guzik, {\it Environment-assisted quantum transport}, New J. Phys. {\bf 11}, 033003 (2009).
\bibitem{Ishizaki_2009} A. Ishizaki, and G. R. Fleming, {\it Unified treatment of quantum coherent and incoherent hopping dynamics in electronic energy transfer: Reduced hierarchy equation approach }, J. Chem. Phys. {\bf 130}, 234111 (2009).
\bibitem{Lloyd_2011} S. Lloyd, {\it 
Journal of Physics: Conference Series
Quantum coherence in biological systems}, Journal of Physics: Conference Series {\bf 302}, 012037 (2011).
\bibitem{Lee_2017} M. H. Lee, and A. Troisi, {\it Vibronic enhancement of excitation energy transport: Interplay between local and non-local exciton-phonon interactions}, J. Chem. Phys. {\bf 146}, 075101 (2017).
\bibitem{heatflowreversal} C. L. Latune, I. Sinayskiy, F. Petruccione, {\it Heat flow reversals without reversing the arrow of time: The role of internal quantum coherences and correlations}, Phys. Rev. Research {\bf 1}, 033097 (2019).
\bibitem{Collini_2010} E. Collini, C. Y. Wong, K. E. Wilk, P. M. G. Curmi, P. Brumer, and G. D. Scholes, {\it Coherently wired light-harvesting in photosynthetic marine algae at ambient temperature}, Nature {\bf 463}, 644-648 (2010).
\bibitem{Lambert_2013} N. Lambert, Y.-N. Chen, Y.-C. Cheng, C.-M. Li, G.-Y. Chen and F. Nori, {\it Quantum Biology}, Nat. Phys. {\bf 9}, 10-18 (2013).
\bibitem{Huelga_2013} S. F. Huelga and M. B. Plenio, {\it Vibrations, quanta and biology}, Contemporary Physics, {\bf 54}, 181–207 (2013).
\bibitem{Chin_2013} A.W. Chin J. Prior, R. Rosenbach, F. Caycedo-Soler, S. F. Huelga and M. B. Plenio, {\it The role of non-equilibrium vibrational structures in electronic coherence and recoherence in pigment–protein complexes}, Nature Phys. {\bf 9}, 113-118 (2013).
\bibitem{Krisnanda_2018} T. Krisnanda, C. Marletto, V. Vedral, M. Paternostro, and T. Paterek, {\it Probing quantum features of photosynthetic organisms}, npj Quantum Inf. {\bf 4}, 60 (2018).

\bibitem{Farre_2018} S. Juli{\`a}-Farr{\'e}, T. Salamon, A. Riera, M. N. Bera, and M. Lewenstein, {\it Bounds on Capacity and Power of Quantum Batteries}, arXiv:1811.04005.
\bibitem{Brandner_2017} K. Brandner, M. Bauer, and U. Seifert, {\it Universal Coherence-Induced Power Losses of Quantum Heat Engines in Linear Response}, Phys. Rev. Lett. {\bf 119}, 170602 (2017).
\bibitem{Uzdin_2015} R. Uzdin, A. Levy, and R. Kosloff, {\it Equivalence of Quantum Heat Machines, and Quantum-Thermodynamic Signatures}, Phys. Rev. X {\bf 5}, 031044 (2015).
\bibitem{Andolina_2018} G. M. Andolina, M. Keck, A. Mari, V. Giovannetti, and M. Polini, {\it Quantum versus classical many-body batteries}, Phys. Rev. B {\it 99}, 205437 (2019).
\bibitem{Gonzalez_2018} J. O. Gonz{\'a}lez, J. P. Palao, D. Alonso, and L. A. Correa, {\it Classical simulation of quantum-coherent thermal machines},  Phys. Rev. E {\it 99}, 062102 (2019).
\bibitem{Andolina_2019} G. M. Andolina, M. Keck, A. Mari, M. Campisi, V. Giovannetti, and M. Polini, {\it Extractable Work, the Role of Correlations, and Asymptotic Freedom in Quantum Batteries}, Phys. Rev. Lett. {\bf 122}, 047702 (2019).
\bibitem{Hovhannisyan_2013} K. V. Hovhannisyan, M. Perarnau-Llobet, M. Huber, and A. Ac{\'i}n, {\it Entanglement Generation is Not Necessary for Optimal Work Extraction}, Phys. Rev. Lett. {\bf 111}, 240401 (2013).
\bibitem{Kilgour_2018} M. Kilgour and D. Segal, {\it Coherence and decoherence in quantum absorption refrigerators}, Phys. Rev. E {\bf 98}, 012117 (2018).
\bibitem{Muller_2018} M. P. M{\"u}ller, {\it Correlating Thermal Machines and the Second Law at the Nanoscale}, Phys. Rev. X {\bf 8}, 041051 (2018).
\bibitem{Spohn_1978} Herbert Spohn, J. Math. Phys. {\bf 19}, 1227 (1978).
\bibitem{Spohn_1978b} H. Spohn and J. L. Lebowitz, Adv. Chem. Phys {\bf 38}, 109 (1978).
\bibitem{Alicki_1979} R. Alicki, J. Phys. A: Math. Gen. {\bf 12}, (1979).
\bibitem{Brandao_2013} F. G. S. L. Brand{\~a}o, M. Horodecki, J. Oppenheim, J. M. Renes, and R. W. Spekkens, {\it Resource Theory of Quantum States Out of Thermal Equilibrium}, Phys. Rev. Lett. {\bf 111}, 250404 (2013).
\bibitem{Parrondo_2009} J. M. R. Parrondo, C. Van den Broeck and R. Kawai, {\it Entropy production and the arrow of time}, New J. Phys. {\bf 11}, 073008 (2009).
\bibitem{Deffner_2011} S. Deffner and E. Lutz, {\it Nonequilibrium Entropy Production for Open Quantum Systems}, Phys. Rev. Lett. {\bf 107}, 140404 (2011).
\bibitem{Santos_2017} J. P. Santos, G. T. Landi, and M. Paternostro, {\it Wigner Entropy Production Rate}, Phys. Rev. Lett. {\bf 118}, 220601 (2017).
\bibitem{Brunelli_2018} M. Brunelli, L. Fusco, R. Landig, W. Wieczorek, J. Hoelscher-Obermaier, G. Landi, F. L. Semi{\~a}o, A. Ferraro, N. Kiesel, T. Donner, G. De Chiara, and M. Paternostro, {\it Experimental Determination of Irreversible Entropy Production in out-of-Equilibrium Mesoscopic Quantum Systems}, Phys. Rev. Lett. {\bf 121}, 160604 (2018).
\bibitem{Santos_2019} J. P. Santos, L. C. C{\'e}leri, G. T. Landi and M. Paternostro, {\it The role of quantum coherence in non-equilibrium entropy production}, npj Quantum Information {\bf 5}:23 (2019).

\bibitem{Barato_2015} A. C. Barato and U. Seifert, {\it Thermodynamic Uncertainty Relation for Biomolecular Processes}, Phys. Rev. Lett. {\bf 114}, 158101 (2015)
\bibitem{Gingrich_2016} T. R. Gingrich, J. M. Horowitz, N. Perunov, and J. L. England, {\it Dissipation Bounds All Steady-State Current Fluctuations}, Phys. Rev. Lett. {\bf 116}, 120601 (2016).
\bibitem{Pietzonka_2016} P. Pietzonka, A. C. Barato, and U. Seifert, {\it Universal bounds on current fluctuations}, Phys. Rev. E {\bf 93}, 052145 (2016).
\bibitem{Pietzonka_2018} P. Pietzonka and U. Seifert, {\it Universal Trade-Off between Power, Efficiency, and Constancy in Steady-State Heat Engines}, Phys. Rev. Lett. {\bf 120}, 190602 (2018).
\bibitem{Guarnieri_2019} G. Guarnieri, G. T. Landi, S. R. Clark, and J. Goold, {\it Thermodynamics of precision in quantum non equilibrium steady states}, arXiv:1901.10428.
\bibitem{Timpanaro_2019} A. M. Timpanaro, G. Guarnieri, J. Goold, and G. T. Landi, {\it Thermodynamic uncertainty relations from exchange fluctuation theorems}, arxiv:1904.07574.
\bibitem{Su_2019} S. Su, W. Shen, J. Du, and J. Chen, {\it Thermodynamic coupling rule for far-from-equilibrium systems}, arXiv:1904.04113.
\bibitem{apptemp} C. L. Latune, I. Sinayskiy, F. Petruccione, {\it Apparent temperature: demystifying the relation between quantum coherence, correlations, and heat flows}, Quantum Sci. Technol. {\bf 4}, 025005 (2019).
\bibitem{Aberg_2014} J. Aberg, {\it Catalytic Coherence}, Phys. Rev. Lett. {\bf 113}, 150402 (2014).
\bibitem{Ghosh_2017} A. Ghosh, C. L. Latune, L. Davidovich, and G. Kurizki, {\it Catalysis of heat-to-work conversion in quantum machines}, Proc. Natl. Acad. Sci. {\bf 114}, 12156–12161 (2017).
\bibitem{Wood_2014} C. J. Wood, T. W. Borneman, and D. G. Cory, {\it Cavity Cooling of an Ensemble Spin System}, Phys. Rev. Lett. {\bf 112}, 050501 (2014).
\bibitem{Wood_2016}  C. J. Wood and D. G. Cory, {\it Cavity cooling to the ground state of an ensemble quantum system}, Phys. Rev. A {\bf 93}, 023414 (2016).
\bibitem{Hama_2018} Y. Hama, W. J. Munro, and K. Nemoto, {\it Relaxation to Negative Temperatures in Double Domain Systems}, Phys. Rev. Lett. {\bf 120}, 060403 (2018).
\bibitem{Raimond_1982} J. M. Raimond, P. Goy, M. Gross, C. Fabre, and S. Haroche, {\it Collective Absorption of Blackbody Radiation by Rydberg Atoms in a Cavity: An Experiment on Bose Statistics and Brownian Motion}, Phys. Rev. Lett. {\bf 49}, 117 (1982).
\bibitem{Devoe_1996} R. G. DeVoe and R. G. Brewer, Phys. Rev. Lett. {\bf 76}, 2049 (1996).
\bibitem{Barberena_2019} D. Barberena, R. J. Lewis-Swan, J. K. Thompson, and A. Maria Rey, Phys. Rev. A {\bf 99}, 053411 (2019).
\bibitem{Petruccione_Book} H. Breuer and F. Petruccione, {\it Theory of Open Quantum Systems}, (Oxford, Oxford, 2002).
\bibitem{Cohen_Book} C. Cohen-Tannoudji, J. Dupont-Roc, and G. Grynberg, {\it Processus d'interaction entre photons et atomes}, (EDP Science/CNRS {\'E}ditions, Paris, 2001).
\bibitem{Alicki_2014} R. Alicki, {\it Comment on "Nanoscale Heat Engine Beyond the Carnot Limit"}, arXiv:1401.7865.
\bibitem{Alicki_2015} R. Alicki and D. Gelbwaser-Klimovsky, {\it Non-equilibrium quantum heat machines}, New J. Phys. {\bf 17}, 115012 (2015).
\bibitem{Assis_2018} R. J. de  Assis, T. M. de Mendonça, C. J. Villas-Boas, A. M. de Souza, R. S. Sarthour, I. S. Oliveira, N. G. de Almeida, {\it Efficiency of a quantum Otto heat engine operating under a reservoir at effective negative temperatures}, Phys. Rev. Lett. {\bf 122}, 240602 (2019).
\bibitem{Kosloff_2019} R. Kosloff, {\it Quantum thermodynamics and open-systems modeling}, J. Chem. Phys. {\bf 150}, 204105 (2019).
\bibitem{Brunner_2012} N. Brunner, N. Linden, S. Popescu, and P. Skrzypczyk, {\it Virtual qubits, virtual temperatures, and the foundations of thermodynamics}, Phys. Rev. E {\bf 85}, 051117 (2012).
\bibitem{autonomousmachines} C. L. Latune, I. Sinayskiy, F. Petruccione, {\it Quantum coherence, many-body correlations, and non-thermal effects for autonomous thermal machines}, Scientific Reports {\bf 9}:3191 (2018). 
\bibitem{Sakurai_Book} J. J. Sakurai, {\it Modern Quantum Mechanics} (Addison Wesley, Reading, MA, 1993).
\bibitem{Mpemba_1969} E. B. Mpemba and D. G. Osborne, {\it Cool?}, Phys. Educ. {\bf 4}, 172 (1969).
\bibitem{Lasanta_2017} A. Lasanta, F. Vega Reyes, A. Prados, and A. Santos, {\it When the Hotter Cools More Quickly: Mpemba Effect in Granular Fluids}, Phys. Rev. Lett. {\bf 119}, 148001 (2017).
\bibitem{Lu_2017} Z. Lu and O. Raz, {\it Nonequilibrium thermodynamics of the Markovian Mpemba effect and its inverse}, Proc. Natl. Acad. Sci. {\bf 114}, 5083 (2017).
\bibitem{Nielsen_Book} M. A. Nielsen and I. L. Chuang, {\it Quantum Computation and Quantum Information} (10th Anniversary Edition, Cambridge University Press, 2010).
\bibitem{Batalhao_2018} T. B. Batalh{\~a}o, S. Gherardini, J. P. Santos, G. T. Landi, and M. Paternostro, {\it Characterizing irreversibility in open quantum systems}, arXiv:1806.08441.
\bibitem{Strasberg_2019} P. Strasberg and M. Esposito, {\it Non-Markovianity and negative entropy production rates}, Phys. Rev. E {\bf 99}, 012120 (2019). 
\bibitem{Campaioli_2017} F. Campaioli, F. A. Pollock, F. C. Binder, L. C\'eleri, J. Goold, S. Vinjanampathy, and K. Modi, {]\it Enhancing the Charging Power of Quantum Batteries}, Phys. Rev. Lett. {\bf 118}, 150601 (2017).
\bibitem{Ferraro_2018} D. Ferraro, M. Campisi, G. M. Andolina, V. Pellegrini, and M. Polini, {\it High-Power Collective Charging of a Solid-State Quantum Battery}, Phys. Rev. Lett. {\bf 120}, 117702 (2018).
\bibitem{Campaioli_2018} F. Campaioli, F. A. Pollock, and S. Vinjanampathy, {\it Quantum Batteries - Review Chapter}, arXiv:1805.05507.
\bibitem{Altintas_2014} F. Altintas, A. \"{U}. C. Hardal, and \"{O}. E. M\"{u}stecapl\i o\v{g}lu, {\it Quantum correlated heat engine with spin squeezing}, Phys. Rev. E {\bf 90}, 032102 (2014).
\bibitem{Barrios_2017} G. Alvarado Barrios, F. Albarr{\'a}n-Arriagada, F. A. C{\'a}rdenas-L{\'o}pez, G. Romero, and J. C. Retamal, {\it Role of quantum correlations in light-matter quantum heat engines}, Phys. Rev. A {\bf 96}, 052119 (2017).
\bibitem{Hardal_2018} A. \"{U}. C. Hardal, M. Paternostro, and \"{O}. E. M\"{u}stecapl\i o\v{g}lu, {\it Phase-space interference in extensive and nonextensive quantum heat engines}, Phys. Rev. E {\bf 97}, 042127 (2018).
\bibitem{Gelbwaser_2019}  D. Gelbwaser-Klimovsky, W. Kopylov, and G. Schaller, {\it Cooperative efficiency boost for quantum heat engines}, Phys. Rev. A {\bf 99}, 022129 (2019).
\bibitem{Feldmann_2006} T. Feldmann and R. Kosloff, {\it Quantum lubrication: Suppression of friction in a first-principles four-stroke heat engine}, Phys. Rev. E {\bf 73}, 025107 (2006).
\bibitem{Plastina_2014} F. Plastina, A. Alecce, T. J. G. Apollaro, G. Falcone, G. Francica, F. Galve, N. Lo Gullo, and R. Zambrini, {\it Irreversible Work and Inner Friction in Quantum Thermodynamic Processes}, Phys. Rev. Lett. {\bf 113}, 260601 (2014).
\bibitem{Holubec_2018b} V. Holubec and A. Ryabov, {\it Cycling Tames Power Fluctuations near Optimum Efficiency}, Phys. Rev. Lett. {\bf 121}, 120601 (2018).
\bibitem{Scully_2002} M. O. Scully, {\it Quantum Afterburner: Improving the Efficiency of an Ideal Heat Engine}, Phys. Rev. Lett. {\bf 88}, 050602 (2002).
\bibitem{Quan_2007} H. T. Quan, Y.-x. Liu, C. P. Sun, and F. Nori, {\it Quantum thermodynamic cycles and quantum heat engines}, Phys. Rev. E {\bf 76}, 031105 (2007).
\bibitem{Jaramillo_2016} J. Jaramillo, M. Beau, and A. del Campo. {\it Quantum supremacy of many-particle thermal machines}, New J. Phys. {\bf 18}, 075019 (2016).
\bibitem{Campisi_2016} M. Campisi, R. Fazio, {\it The power of a critical heat engine}, Nature Comm. {\bf 7}:11895 (2016).
\bibitem{Holubec_2017} V. Holubec, and A. Ryabov, {\it Work and power fluctuations in a critical heat engine}, Phys. Rev. E {\bf 96}, 030102 (2017).
\bibitem{Ma_2017} Y.-H. Ma, S.-H. Su, and C.-P. Sun, {\it Quantum thermodynamic cycle with quantum phase transition}, Phys. Rev. E {\bf 96}, 022143 (2017).
\bibitem{Kloc_2019} M. Kloc, P. Cejnar, and G. Schaller, {\it Collective performance of a finite-time quantum Otto cycle}, arXiv:1905.08692. 
\bibitem{Altintas_2015} F. Altintas, A. \"{U}. C. Hardal, and \"{O}. E. M\"{u}stecapl\i o\v{g}lu, {\it Rabi model as a quantum coherent heat engine: From quantum biology to superconducting circuits}, Phys. Rev. A {\bf 91}, 023816 (2015).
\bibitem{Nefedkin_2016} N. E. Nefedkin, E. S. Andrianov, A. A. Zyablovsky, A. A. Pukhov, A. V. Dorofeenko, A. P. Vinogradov, and A. A. Lisyansky, {\it Superradiance of a subwavelength array of classical nonlinear emitters}, Opt. Express {\bf 24}, 3464-3478 (2016).
\bibitem{Tsyplyatyev_2009} O. Tsyplyatyev and D. Loss, {\it Dynamics of the inhomogeneous Dicke model for a single-boson mode coupled to a bath of nonidentical spin-1/2 systems}, Phys. Rev. A {\bf 80}, 023803 (2009).
\bibitem{Sun_2019} C. Sun, V. Y. Chernyak, A. Piryatinski, N. A. Sinitsyn, {\it Cooperative Light Emission in the Presence of Strong Inhomogeneous Broadening}, arXiv:1906.11412.
\bibitem{Coffey_1977} B. Coffey, and R. Friedberg, {\it Effect of short-range Coulomb interaction on cooperative spontaneous emission}, Phys. Rev. A {\bf 17}, 1033 (1977). 
\bibitem{Boykin_2002} P. O. Boykin, T. Mor, V. Roychowdhury, F. Vatan, and R. Vrijen, {\it Algorithmic cooling and scalable nmr quantum computers}, Proc. Ntl Ac. Sci., {\bf 99}(6):3388–3393 (2002).
\bibitem{Alhambra_2019} {\'A}. M. Alhambra, M. Lostaglio, and C. Perry, {\it Heat-Bath Algorithmic Cooling with optimal thermalization strategies}, Quantum {\bf 3}, 188 (2019).
\bibitem{Merkli_2015} M. Merkli, H. Song, and G. P. Berman, J. Phys. A: Math. Theor. {\bf 48}, 275304 (2015).
\bibitem{Feynman_1951} R. P. Feynman, {\it An Operator Calculus Having Applications in Quantum Electrodynamics}, Phys. Rev. {\bf 84}, 108 (1951).  



















%



%
%

%
%
%

%
%
%
%
%


\end{thebibliography}
\end{document}